\documentclass[11pt,preprint]{aastex}
\usepackage[normalem]{ulem}
\usepackage{mathptmx}
\usepackage{lscape}
\usepackage{longtable}
\usepackage{color}
\usepackage{float}
\usepackage[usenames,dvipsnames,svgnames,table]{xcolor}

\def  \eg          {{\rm e.g.}}

\def\msun{\rm M_{\sun}}

\def\micron{$\mu$m}

\def\mbh{\rm M_{\bullet}}

\def\lbol{L_{\rm bol}}
\def  \ergs        {\hbox{erg s$^{-1}$}}              % erg/sec
\def  \kms         {\hbox{km s$^{-1}$}}          % kilometers per sec

\newcommand{\etal}{et al.}

\def  \Ha          {\ifmmode {\rm H}\alpha \else H$\alpha$\fi}
\def  \Hb          {\ifmmode {\rm H}\beta \else H$\beta$\fi}
\def  \CIIIb       {\ifmmode {\rm C}\,{\sc iii]}\,\lambda1909
                     \else C\,{\sc iii]}\,$\lambda1909$\fi}
\newcommand{\hbeta}{\rm H{$\beta$}}

\newcommand{\civ}{\rm CIV}
\newcommand{\mgii}{\rm MgII} %Mg {\rm II}
\newcommand{\feii}{\rm FeII}
\newcommand{\oiii}{[OIII]}
\newcommand{\oiiiab}{[OIII]\,$\lambda\lambda$4959,\,5007}

\newcommand{\oiiia}{[O{\sevenrm\,III}]\,$\lambda$4959}
\newcommand{\oiiib}{[O{\sevenrm\,III}]\,$\lambda$5007}

 \font\sevenrm=cmr7 scaled 1000

\usepackage{CJK}

\begin{document}
\begin{CJK}{GB}{gbsn}

\shorttitle{Spectral Catalog}
\shortauthors{Dai et al.}
\slugcomment{prepared for ApJ: \today}
\title{Mid-Infrared Selected Quasars I: Virial Black Hole Mass and Eddington Ratios\footnote{Observations reported here were obtained at the MMT Observatory, a joint facility of the Smithsonian Institution and the University of Arizona. }}

\author{Y. Sophia Dai ({´÷êÅ})\altaffilmark{1,2}, 
Martin Elvis\altaffilmark{1},
Jacqueline Bergeron\altaffilmark{3},
Giovanni G. Fazio\altaffilmark{1},
Jia-Sheng Huang\altaffilmark{1}, 
Belinda J. Wilkes \altaffilmark{1},
Christopher N. A.  Willmer \altaffilmark{4},
Alain Omont\altaffilmark{3},
and Casey Papovich\altaffilmark{5}
\altaffiltext{1}{Harvard-Smithsonian Center for Astrophysics, 60 Garden Street,
Cambridge, MA 02138, USA}
\altaffiltext{2}{Caltech/IPAC, 1200 E California Blvd, Pasadena, CA 91125, USA}
%\altaffiltext{2}{Boston College, 140 Commonwealth Ave, Chestnut Hill, MA 02467, USA}
\altaffiltext{3}{CNRS, UMR7095, Institut d$'$Astrophysique de Paris, F-75014, Paris, France}
\altaffiltext{4}{Steward Observatory, University of Arizona, 933 North Cherry Avenue, Tucson, AZ 85721, USA}
\altaffiltext{5}{Department of Physics and Astronomy, Texas A\&M University, College Station, TX 77843, USA} 
\email{ydai$@$caltech.edu}}

\begin{abstract}
We provide a catalog of 391 mid-infrared-selected (MIR, 24\,$\mu$m) broad-emission-line (BEL, type 1) 
quasars in the 22 deg$^2$ SWIRE Lockman Hole field.
This quasar sample is selected in the MIR from Spitzer MIPS with $S_{\rm 24} > 400\mu$Jy,
jointly with an optical magnitude limit of r (AB) $<$ 22.5 for broad line identification.
The catalog is based on MMT and SDSS spectroscopy to select BEL quasars,
extends the SDSS coverage to fainter magnitudes and lower redshifts,
and recovers a more complete quasar population.
The MIR-selected quasar sample peaks at $z\sim$1.4, 
and recovers a significant and constant (20\%) fraction of 
extended objects with SDSS photometry across magnitudes,
which was not included in the SDSS quasar survey
dominated by point sources.
This sample also recovers a significant population of $z < 3$ quasars at $i > 19.1$.
We then investigate the continuum luminosity and line profiles of these MIR quasars, 
and estimate their virial black hole masses
and the Eddington ratios.
The SMBH mass shows evidence of downsizing,
though the Eddington ratios remain constant at $1 < z < 4$. 
Compared to point sources in the same redshift range,
extended sources at $z < 1$
show systematically lower Eddington ratios.
The catalog and spectra are publicly available online.
%This sample contains dust-rich quasars 
%constructed to facilitate the systematic studies 
%of the supermassive black hole (SMBH)-host correlation,
%which have been hampered by the small sample sizes.
%Six BEL objects at $i > 19.1$, not in the SDSS DR7 quasar catalog,  
%were also recovered as quasars with SDSS spectroscopy. 
%This leads to a revised estimate of the SDSS 
%completeness to be $(59 \pm\ 10)\%$ at $i < 19.1$ for both extended and point sources ,
%and (13 $\pm$ 7)\% at 19.1 $< i <$ 20.2 for point sources.

\end{abstract}


\section{INTRODUCTION}

The apparent connection between supermassive black holes (SMBHs) and their host galaxies
has been explained by a variety of theories. 
In the merger driven model, the collision of dust-rich galaxies drives gas inflows,
 fueling both starbursts and buried quasars until feedback disperses
the gas and dust, allowing the quasar to be briefly visible as a 
bright optical source \citep[e.g.][]{sanders88, hopkins06}. 
Instead of physical coupling between the BH and host galaxy,
the central-limit-theorem can be used to explain
the linear SMBH mass and bulge mass correlation 
by the hierarchical assembly 
of BH and stellar mass \citep{peng07,jahnke11}.
Alternatively, the cold flow model \citep[e.g.][]{dekel09,bournaud11,dimatteo12}
introduces inflowing cosmological cold gas streams rather than collisions to fuel the star formation and
quasar, and better explains the clumpy disks observed for high-$z$ galaxies. 
Observationally, a SMBH-host connection is 
supported by the discovery of correlations
of the SMBH mass with bulge luminosity, mass,
and velocity dispersion, especially with bulges and ellipticals 
 \citep[e.g.][]{kormendy95, ferrarese00, kormendy13}.
However, despite tremendous progress on the 
demographic studies of SMBHs,
whether or how the SMBH
regulates the formation and evolution of their hosts
via the possible `feedback' process
is still under debate. 
One sign of such feedback may be the ongoing star formation
observed for host galaxies of active galactic nuclei (AGNs) and quasars,
and vice versa, starbursts are found to 
host buried AGNs \citep{kauffmann03, shi09, dai12}. 
Based on the similar star formation rate (SFR)
observed for galaxies with and without an active galactic nucleus (AGN),
recent studies suggest that the SMBH-host correlation results from the gas availability,
instead of major interaction between the SMBH growth and host star formation
\citep[e.g.][]{goulding14, lilly13}.
In this paper, we present a mid-infrared (MIR) selection to
effectively select quasar candidates 
with dusty nuclear material in a 
disk/wind or torus geometry \citep[e.g.][`torus' hereafter]{elvis00,antonucci93}.
This selection is relatively unaffected by obscuration.
%in the peak of SMBH accretion -- quasars,
%which also reside in a dust-rich host,
%where joint star formation and AGN activities are likely to be present.

In the high redshift ($z > 0.5$) universe, 
it is hard to observe both broad-line (type 1) quasars and their
host galaxies simultaneously. 
The quasar glare usually outshines the host galaxy at optical wavelengths,
and the host has a small angular size.  
In large optical surveys, 
the focus has been 
on broad-emission-line (BEL) quasars \citep[e.g.][S11]{rich06a, shen11}, 
or `blue' quasars, 
which are biased towards optically-unobscured (Type 1) objects
with limited information about the host galaxy.
Studies on the cosmic history of quasars
show an evolution over redshifts,
with a quasar peak appearing at $z \sim$ 1.5 \citep[e.g.][]{hms05, silverman08}. 
At longer infrared (IR) wavelengths, 
where thermal emission from dust is dominant,
quasars have characteristic power-law shaped MIR SEDs,
and are selected by different color wedges in the Spitzer IRAC \citep{fazio04}
and Wide-field Infrared Survey Explorer \citep[][WISE]{wright10} bands 
\citep{lacy04, sajina05, stern05, stern12, donley12}.

Recent surveys in the IR have detected optically obscured (type 2),
dust-reddened quasars \citep[e.g.][]{rich03, rich09, polletta06, glikman12, lacy13}.
These quasars are 
marked by having reddened UV-optical SEDs resulting from dust absorption. 
At different redshift and luminosity ranges, 
quasars are reported to have an obscuration fraction from
20\% to over 50\% \citep{lacy02, glikman04,glikman07, urrutia09, juneau13, lacy13}.
In the merger-driven model,
these quasars are in an early transitional phase,
and are in the process of expelling 
their dusty environment before becoming `normal' blue quasars (type 1).
This IR-luminous phase also evolves with time,
and was more common at high $z$ \citep[e.g.][]{caputi07, serjeant10}.
Optical studies of quasar and host systems
%especially in reddened and dusty quasars,
are challenged 
by the high contrast between the bright point-source quasar and starlight.
Infrared-selected quasars 
are good candidates to study the SMBH-host connection,
as they are not biased against dusty hosts.

In this paper, we present a catalog of 391
MIR-selected BEL objects 
in the $\sim$22 deg$^2$ Lockman Hole - Spitzer Wide-area InfraRed Extragalactic 
Survey (LHS) Field \citep[SWIRE,][]{lonsdale03}. 
As will be pointed out in \S\ref{sec:bel}, since all of the objects have BEL features,
and the majority also qualify the classical Seyfert / quasar luminosity separation ($M_{\rm B} < -23$),
hereafter we will simply refer to these BEL objects as quasars.
Combining the mid-IR (MIR) 24\,$\mu$m flux-limit
and optical identification has been 
demonstrated to be an effective way of selecting quasars (with a 13\% detection 
rate in \citet{papovich06}). 
This MIR selection was designed to be biased towards dusty systems,
where ample hot dust exists in the nuclear region
with higher likelihood of tracing remnant or ongoing star formation (cool dust).
The spectroscopic sample used in this work comprises new observations
taken with the Hectospec at the MMT 
of the wide-angle SWIRE field and of a smaller MIPS GTO field, 
and spectra obtained by the Sloan Digital Sky Survey (SDSS) within
the Lockman Hole footprint.
We hope that this sample will provide a new test bed to study
the SMBH self-regulation or AGN feedback when the system has not 
relaxed to equilibrium, if such effects do exist.
In \S\ref{sec:sample} we review the sample selection and 
introduce the spectroscopic data and the MIR additions to the SDSS quasar catalog; 
followed by the spectral measurements in \S\ref{sec:spec};
in \S\ref{sec:virial} and \S\ref{sec:lbol}, we describe the virial black hole mass and bolometric luminosity estimates; 
we then follow with the spectral catalog (\S\ref{sec:cata}), discussion (\S\ref{sec:discussion}) and 
the summary (\S\ref{sec:sum}).
Throughout the paper, we assume a concordance cosmology with $H_0$=70 km$/$s Mpc$^{-1}$, $\Omega_{\rm M}$=0.3, and
$\Omega_{\Lambda}$=0.7.  
All magnitudes are in AB system except where otherwise noted.

\section{THE SAMPLE}
\label{sec:sample}
\subsection{MIR MIPS 24\,\micron\ Selection}
\label{sec:s24}
%One approach to facilitate the systematic studies 
%of the SMBH-host correlation is to select a sample 
%whose SMBH and host properties are well defined. 
The combined MIR 24\,\micron\ and optical
selection for this survey
was designed to detect objects with luminous torus / nucleus
and not biased against dusty hosts.
%We decided to select objects that have luminous
%torus / nuclei and not biased against dusty hosts with 
%MIR 24\,\micron\  selection and optical identification.
The MIR selection allows detection of hot dust (a few hundred K) 
at the redshifts z$\sim$1.5; %, ensuring bright torus with ample dust;
while optical follow-up spectroscopically identified the BEL objects ,
confirming their unobscured (type 1) quasar nature.
This MIR selection also allows far-infrared (FIR) cross-match to look for
cool dust for SMBH-host studies, as demonstrated in \citet{dai12}.

\begin{figure}[h]
\begin{center}
\includegraphics[scale=1,angle=0]{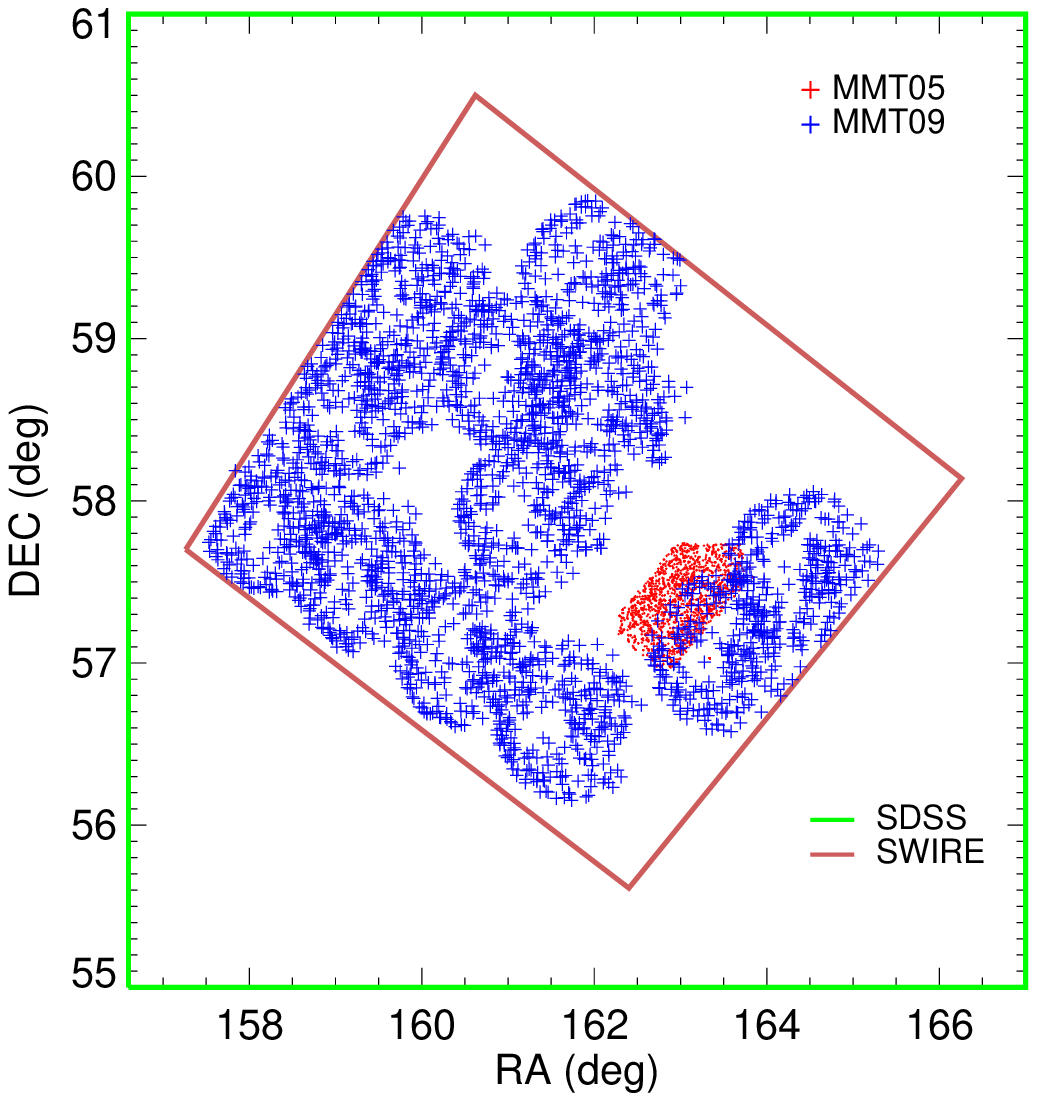} %{coverage.eps}  %
\end{center}
\figcaption{Spectroscopic targets and survey coverage in the Lockman Hole-SWIRE (LHS) field: Green square,  {\it SDSS photometry, covering the whole field}; Brown square, {\it SWIRE}; Red pluses, {\it MMT-Hectospec (2005)}; Blue pluses, {\it MMT-Hectospec (2009)}. 
The hole at the center of each MMT configuration is due to the spacing limitation of the 300 fibers in the Hectospec instrument.
\label{fig:coverage}} 
\end{figure}

We select Spitzer MIPS \citep{rieke04} 24 $\mu$m sources from the SWIRE survey in 
the $\sim$ 22 deg$^{2}$ Lockman Hole - SWIRE (LHS) field
centered at RA$=$10:46:48, DEC$=$57:54:00 \citep{lonsdale03}. 
The SDSS imaging also covers 
the LHS region to $r = 22.2$ at 95 \% detection repeatability, 
but can go as deep as $r = 23$. 
All magnitudes are taken from the SDSS photoObj catalog in DR7,
already corrected for Galactic extinction according to \citet{schlegel98}. 
They are the SDSS approximate AB system \citep{oke83,fukugita96,smith02}. 
SDSS has astrometric uncertainties $< 0.1 \arcsec$ on average \footnote{http://www.sdss.org/dr7}.
In Fig.~\ref{fig:coverage} we show the SWIRE and SDSS coverages in the LHS field.

We first apply a 24 $\mu$m flux limit of 400 $\mu$Jy ($\sim$ 8\,$\sigma$),  % 
which yields a sample of 23,\,402 objects. 
The completeness at 400\,$\mu$Jy for SWIRE-MIPS catalog is $\sim$ 90\% \citep{shupe08}.
The confusion limit due to extragalactic sources for MIPS 24\,$\mu$m band is 56\,$\mu$Jy \citep{dole04},
so source confusion is not an issue in this sample.
The errors in position for these sources are between 0.2$\arcsec$--0.4$\arcsec$, 
and the effective beam size (FWHM) of MIPS at 24\,$\mu$m is $\sim 6 \arcsec$
\footnote{http://irsa.ipac.caltech.edu/data/SPITZER/docs/mips/mipsinstrumenthandbook/}. 

\begin{figure}[h]
\begin{center}
\includegraphics[scale=0.6,angle=0]{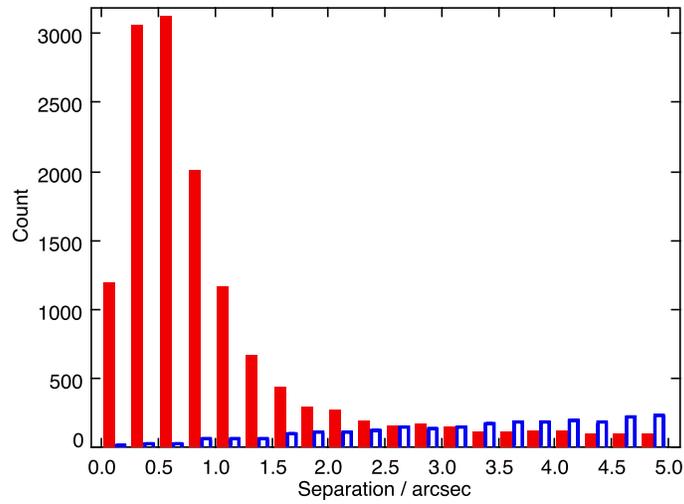} %{offset2.eps} %fig2a
\end{center}
\figcaption{Histogram of the offset between SWIRE and SDSS positions (red). 
In blue is the same histogram for mismatches after a random position offset ($<$ 10\,$\arcsec$ radius).
The random association rate within 5\,$\arcsec$
is $\sim$18\%, % (2,467 out of 15,549 matches),
but declines to $\sim$6\% within 2.5\,$\arcsec$.
A matching radius of 2.5\,$\arcsec$ was used to maximize 
the matching counts while minimizing the random associations.
\label{fig:offset}}
\end{figure}

We then match the 24\,$\mu$m flux limited sources to the SDSS DR7 catalog.
We determine an association radius
of 2.5\,$\arcsec$ to maximize the matching number counts 
while at the same time minimizing the cases of random association (Fig.~\ref{fig:offset}).
We first match the SWIRE and the SDSS r band catalogs.
Then we offset the SWIRE position by a random number 
within 10\,$\arcsec$ radius,
and match them to the SDSS r band catalog.
The association radius is determined by comparing the random association rate
at different radii.
The random association rate within 5\,$\arcsec$
is $\sim$18\% (2,467 out of 14,069 matches),
but declines to $\sim$6\% within 2.5\,$\arcsec$.
Beyond 2.5\,$\arcsec$ radius 
there are $> 50\%$ random associations.
The estimated total number of false associations within 2.5\,$\arcsec$
is %2467 (25\%) after the position offset, and 
868 (6\%).
Adopting the association radius of 2.5\,$\arcsec$ we find 14,069 matches.
Of these 87$\%$ (12,255) 24\,$\mu$m sources also satisfy $r < $ 22.5.
This r limit allows follow-up optical spectroscopic observations with the MMT. 

\begin{figure}[h]
\begin{center}
\includegraphics[scale=0.85,angle=0]{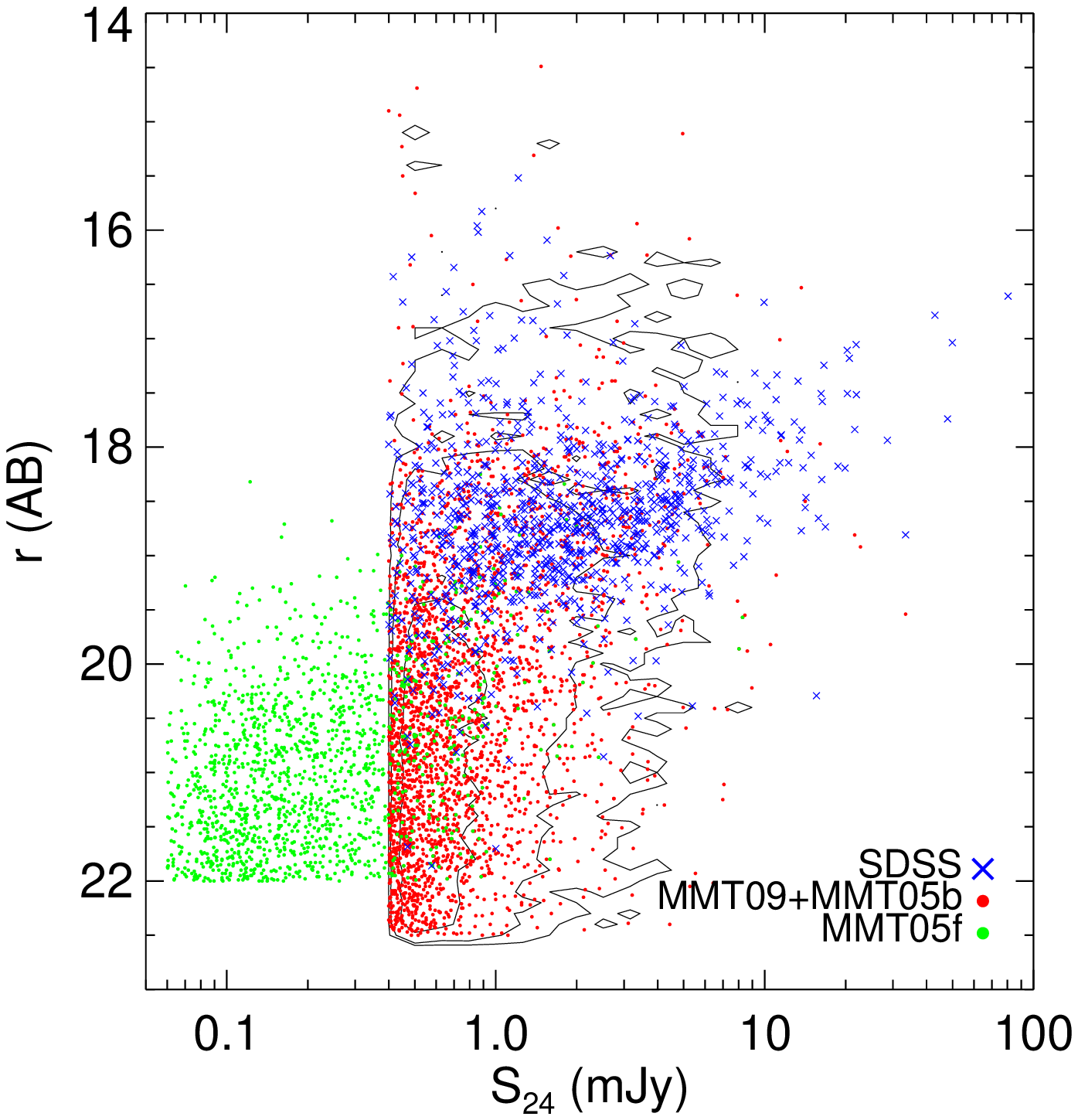} %{f24-r-cov.eps} %fig3a
\includegraphics[scale=0.8,angle=0]{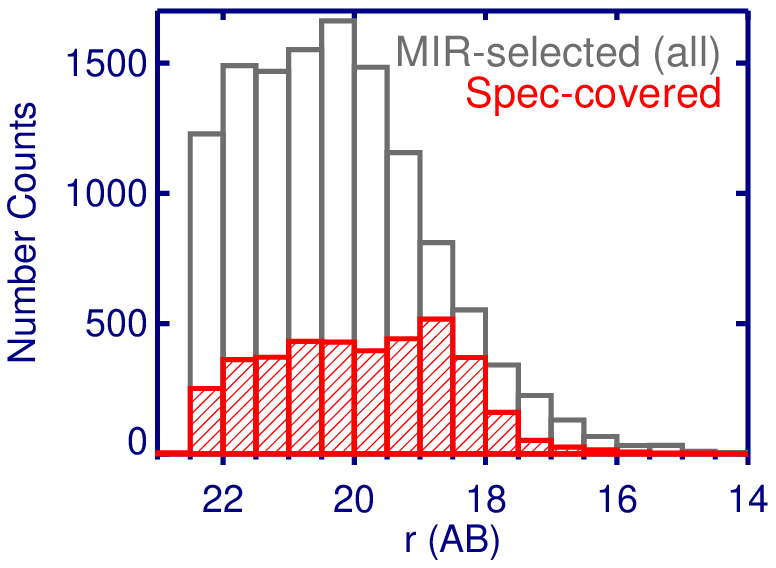} %{r-s.eps}
\includegraphics[scale=0.8,angle=0]{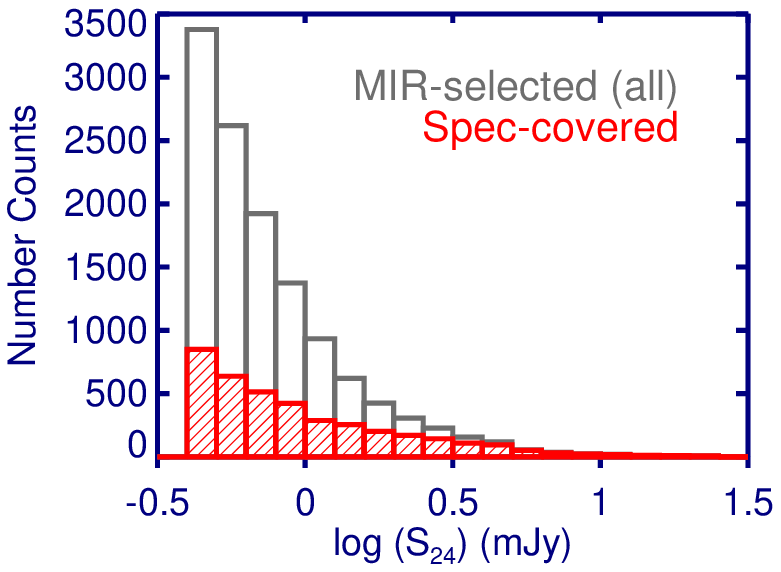} %{f24-s.eps}
\end{center}
\figcaption{ (Top): Spectroscopic coverage in the $S_{\rm 24} - r$ band space for MIR-selected targets: 
Blue crosses,  {\it 854 SDSS DR7 targets}; 
Red, {\it 2,485 MMT09 plus 273 MMT05b targets}; 
Green, {\it 902 MMT05f targets}. 
The contours in the background are the 12,255 MIR-selected targets 
that satisfy the $S_{\rm 24} > 400\mu$Jy (\& $r <$ 22.5) limits.
(Bottom): The r band magnitude and 24\,\micron\ flux distribution 
for all MIR-selected targets (black), 
and the spectroscopic covered objects (Spec-covered, red).
\label{fig:f24rcov}} 
\end{figure}

The optical spectroscopic survey consists of four parts (Fig.~\ref{fig:f24rcov}, Table.~\ref{tab:sample}): 
(1) SDSS DR7; (2) MMT 2009 survey (MMT09);  and (3) MMT 2005 bright targets (MMT05b).
These three subsamples include the MIR-selected targets as described above.
A fourth subsample comes from MMT 2005 observations for MIPS deep 
targets (\S\ref{sec:mmt05}): % that goes to $S_{\rm 24} = 60\,\mu {\rm Jy}$ 
(4) MMT 2005 faint targets (MMT05f) ($60\,\mu {\rm Jy} \leq S_{\rm 24} < 400\,\mu {\rm Jy}$),
kept only for comparison purpose.
Table~\ref{tab:obslog} summarizes the MMT covered observations.

\subsection{SDSS spectroscopy}
\label{sec:sdss}
In order to minimize the need for new spectroscopy,
we downloaded and analyzed the existing SDSS spectroscopy of LHS MIPS 24$\mu$m targets directly
from the SDSS DR7 SkyServer\footnote{http://cas.sdss.org/dr7}. 
The SDSS spectra have a resolving power of R\,$\sim$1800--2200,
with a wavelength coverage of 3800--9200 $\AA$.
In this study, we use the `1d' calibrated spectra from the DR7 Data Archive Server\footnote{http://das.sdss.org/spectro}, 
stored in logarithmic pixel scale of 10$^{-4}$. 
The redshifts given in SDSS DR7 SpecObj catalog were determined by
the $spectro1d$ package \citep{stoughton02}.
We made a SQL search (with a 5 degree radius, 
$r <$ 22.5)
in the SDSS DR7 SpecObj catalog$\footnote{http://cas.sdss.org/astrodr7/en/tools/search/sql.asp}$
and found 2,978 objects. 
Spectra for all SDSS objects with redshifts 
in the LHS field were downloaded, 
irrespective of their SDSS classification.
We matched these sources with the SWIRE MIPS 24 $\mu$m catalog,  %2.5" separation
and excluded 2,019 SDSS targets not detected by SWIRE,
and 38 SDSS targets with $S_{24} < $400\,$\mu$Jy.
Within the remaining 921 qualified spectra,
we only retain, for BEL identification, the 854 objects (93\,\%) with a redshift 
confidence $\ge$ 0.9. 

\subsection{MMT 2009 Spectroscopy}
\label{sec:mmt09}
\begin{figure}[ht]
\begin{center}
\includegraphics[scale=0.6,angle=90]{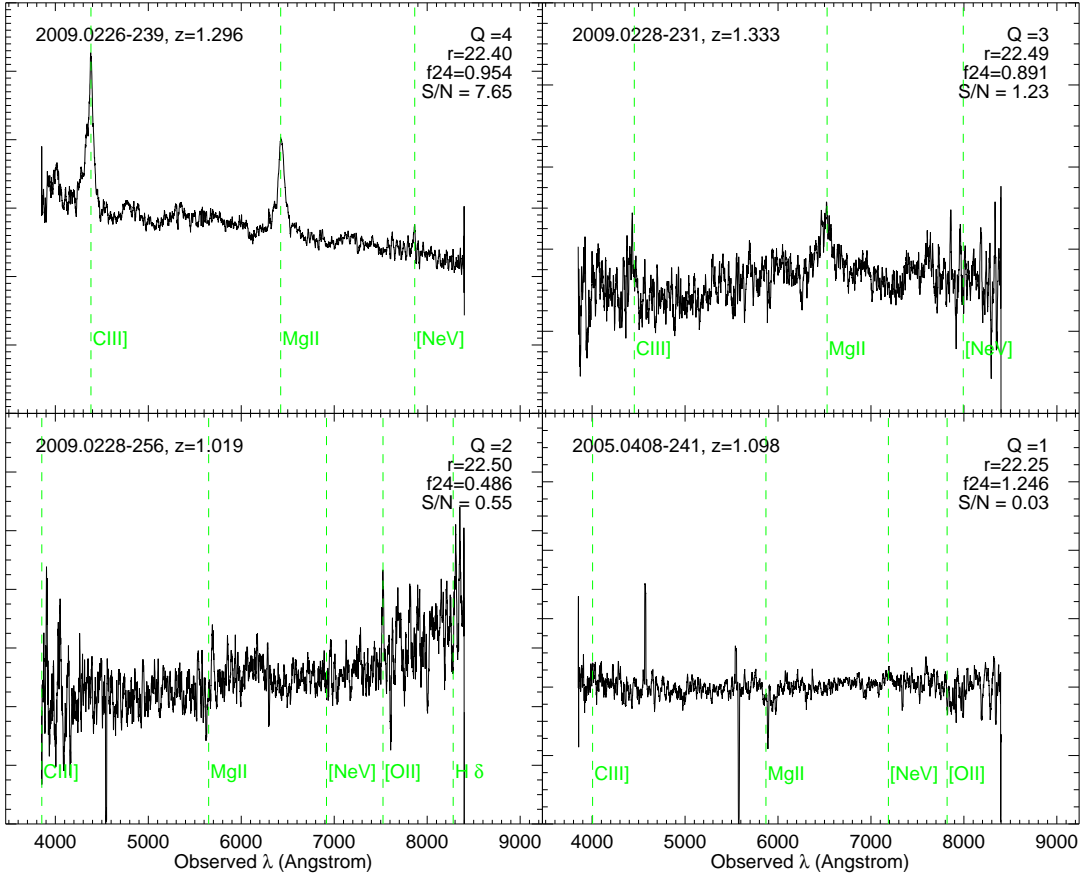} %{zqual.eps} %fig4
\end{center}
\figcaption{Examples of spectra with different redshift quality flags from Q $=$ 4 (probability $> 95\%$ of being correct) to 
Q $=$1 (no features recognized), following the same
procedure as in the DEEP2 survey \citep{willmer06, newman13}.
Q values are listed in the top right corner;
also shown in the top right corner are the SDSS r band magnitude (AB),  the MIPS 24\,$\mu$m flux in mJy, 
and the median S/N per pixel in the plotted region 
(Note this is different from the median S/N of the emission line region, 
which is usually of a higher value).
Typical quasar lines are marked in green. 
The redshifts given for the Q $=$1 and Q $=$ 2 examples
are generated from the code
or after visual check, 
and are not reliable values. 
Only spectra with Q $\ge$ 3 are kept in this work (See also \S~\ref{sec:datareduction}).
\label{fig:zqual}}
\end{figure}

Hectospec is a 300 fiber spectrometer with a 1$^{\circ}$ diameter field of view (FOV) 
mounted on the MMT \citep{fab05,mink07}. 
The combination of a wide field with a large aperture
makes Hectospec well-suited to cover extended areas such as the LHS.
Hectospec covers a wavelength range of 3650-9200\,${\AA}$ with a 6\,${\AA}$ resolution
 (1.2\,${\AA}$ pixel$^{-1}$, R$=$600--1500).
The primary spectroscopic data specific to this study were taken in 2009 (MMT09, PI: Huang)
over 11 dark photometric nights with good seeing ($<$ 2\,\arcsec)
with 12 FOVs. 
The MMT data cover a total area of $\sim$\,12 deg$^2$ (50\% of LHS field).
An ongoing MMT project (PI: Dai) is complementing the 2009
observations by targeting unobserved areas within the LHS.
But the new project adopts a different selection 
that emphasizes Herschel \citep{pilbratt10} targets,
to favor objects with cool dust ($< 60K$) that traces the host star formation. 
These data will be published in a forthcoming paper (Dai \etal, in preparation).
In Fig.~\ref{fig:coverage} the spectroscopic targets in the 12 fields observed in 2009
are marked as blue pluses.
At the center of each MMT FOV,
an area with fewer targets can be noticed.
This is due to the spacing limitations of Hectospec,
whose 300 fibers cannot be crossed or placed less than 50\,$\arcsec$ from one another.
The $\sim$ 3000 spectroscopic targets were selected from the 11,\,401 MIPS and r-band flux limited 
catalog from \S\ref{sec:s24} (after excluding the 854 SDSS objects from \S\ref{sec:sdss}).
Brighter 24\,$\mu$m sources were given higher priority (See Fig.~\ref{fig:f24rcov}), 
and fibers were configured to cover as much of the LHS field as possible.
Hectospec gives a clear BEL detection (median S/N per pixel $>$ 5 ) for 
a $r = 22.4$ quasar in a 1.5 hour %5$\times$20\,min 
exposure (e.g. Fig.~\ref{fig:zqual}, LHS-2009.0226-239). 
Hence 1.5 hour exposures were used as the standard.
Spectra for 2913 objects were recorded in 2009.
The optical spectroscopic completeness in the 12 MMT09 Fields is 33\% 
for $S_{\rm 24} >$ 400 $\mu$Jy  objects,
with an average overlap of 0.08 deg$^2$ between different configurations.
After taking into account the objects missing due to fiber placement limitations,
the completeness of MMT09 sample drops to $\sim$30\%, 
and will be used in the following discussions ($\S~\ref{sec:completeness}.$)

\subsection{MMT 2005 MIPS-deep Spectroscopy}
\label{sec:mmt05}
This spectroscopic sample is extended to include 273 MMT spectra from 
an earlier 2005 deep survey (MMT05) across eight, highly overlapping FOVs 
in the LHS. 
The 2005 data cover a much smaller ($\sim$ 0.5 deg$^2$) region (PI: Papovich).
The MMT 2005 survey applied a deeper 24\,\micron\ flux limit 
of $S_{\rm 24} >$ 60 $\mu$Jy, near to the MIPS confusion limit \citep{rieke04}. 
Only r $< 22$ targets were selected in the 2005 observations.
MMT05 recorded 1,481 spectra.
Of these, 273 objects also satisfy the bright MMT09 limit ($S_{\rm 24} >$ 400 $\mu$Jy) 
and are included in this sample. 
We call this the MMT05b (bright) sample. 
The remaining 1,208 objects with fainter flux (60 $< S_{\rm 24} <$ 400 $\mu$Jy) were also kept
for comparison purposes.
This sample is designated MMT05f (faint). 

The highly overlapped MMT05 FOVs lead to an optical spectroscopy completeness of 66\%
for 24 $\mu$m bright targets ($S_{\rm 24} >$ 400 $\mu$Jy) in the 0.5 deg$^2$ area.
This higher completeness comes at the cost of lower efficiency, 
with an average overlap of 0.94 deg$^2$,
and a drop from 242 targets per FOV in MMT09 to 185 targets per FOV in MMT05 observations,
which encouraged the adoption of the MMT09 strategy. 
 
\subsection{Spectral Data Reduction}
\label{sec:datareduction}
The SDSS spectra and redshifts are used directly from the 
DR7 SpecObj catalog without further reduction.
The MMT Hectospec data (MMT09, MMT05b and MMT05f) were 
reduced using the HSRED pipeline \citep[][http://mmto.org/~rcool/hsred/index.html]{cool08}, 
which is based on the SDSS pipeline. 
HSRED extracts one dimensional (1d) spectra, subtracts the sky and then 
flux-calibrates the spectra. 
The flux-calibration is done using
spectra of 6-10 stars selected to have SDSS colors of F stars that
are observed simultaneously with the main galaxy and quasar sample.
The flux calibration correction is obtained combining the 
extinction-corrected SDSS photometry of these stars with Kurucz (1993) model fits \citep{cool08}. 
These stellar spectra are also used to remove the  telluric lines.
The spectral range covered by Hectospec allows detecting one or more
typical emission lines present in the spectra of quasars and galaxies
(\civ, \mgii, H$\beta$, [O$_{\rm III}$], H$\alpha$) for galaxies to $z\sim\,1$,
and quasars to $z\sim$4.5.
The redshifts measured by HSRED also use code adapted from SDSS
and use the same templates as SDSS.
All spectra were visually inspected for validation as described below. 

A redshift quality flag is assigned to each spectrum, following the same
procedure used for the DEEP2 survey \citep{willmer06, newman13}, where redshift
qualities range from Q $=$ 4 (probability P $> 95\%$ of being correct), 3 (90$\% <$
P $< 95\%$), 2 (P $< 90\%$) and 1 (no features recognized). 
Q $=$ 2 spectra are assigned to objects for which only a single feature is
detected, but cannot be identified without ambiguity. The Q $=$ 3 spectra have
more than one spectral feature identified, but tend to have low S/N;
typical confidence levels for these objects  is $\sim$ 90\% for the DEEP2
galaxies. Finally, Q $=$ 4 objects have 2 or more spectral features with
reasonable to high S/N. The confidence level of these redshifts is
typically $>$ 95\%. 
Because of the larger spectral range covered by HECTOSPEC (3800--9500 \AA) 
relative to DEEP2 (5000--9500\AA), 
we expect that the quoted confidence levels are the conservative
limits for our spectra.

Fig.~\ref{fig:zqual} shows examples of objects in each redshift quality category. 
In this study, 
as for the 854 SDSS spectra,
only spectra of Q = 3 and 4 were used.
This yields a total of 2,485 MMT09 spectra ($\sim$90\% of all the recorded spectra);
and 1,175 MMT05 spectra ($\sim$80\%). 
All of the 273 MMT05b subsample satisfy the redshift quality filter.

\begin{table}[htdp]
\caption{Optical spectroscopic sample summary.}
\begin{center}
\begin{tabular}{lccccccc}
\hline
 &Source & r$_{AB}$ & S$_{\rm 24} (\mu$Jy)  & N$_{spec}$ & N$_{quasar}$  & Covered deg$^2$ & Detection rate \\
\hline
(1) & SDSS & $<$ 22.5 & $>$400 & 854* &  138* & 22&16.2\%  \\   %117 in overlapped region & 96 with zconf > 0.9 
(2) & MMT09 & $<$ 22.5 & $>$400 & 2485 & 226 & 11 & 9.1 \%  \\   % 2913 w/no z cut
(3) & MMT05b & $<$ 22 & $>$400 & 273 & 27 & 0.5 & 9.9\%  \\     %1175
\hline
& Total & $<$ 22.5 & $>$400 & 3612 & 391 & 22  & 10.8\% \\ 
\hline
(4) & MMT05f & $<$ 22 & $60 < S_{\rm 24} < 400$ & 902   &  17 & 0.5 & 1.9\% \\  %902
\hline
\\
\end{tabular}
\end{center}
*: In the full $\sim$\ 22 deg$^2$ LHS field. The numbers of spectra and quasars in the 
$\sim$\ 12 deg$^2$ MMT-covered regions are 622 and 96, respectively. % (detection rate of 15.4\%)
\label{tab:sample}
\end{table}

 \begin{table}[htdp]
\caption{Observation log for MMT spectroscopic survey.}
\begin{center}
\begin{tabular}{lccccccc}
\hline
\hline
Instrument  & Telescope & RA & Dec  & Exposure   &Observation Date  \\ %&  Air & Comments\\
 && (J2000) & (J2000) & hours & \\
\hline
Hectospec & MMT Observatory & $+10:39:48.3$ & $+59:16: 56$ & 1.5  & 2009.0319 \\ %& 1.33 & \\
Hectospec & MMT Observatory & $+10:48:48.9$ & $+58:31: 58$ & 1.5  & 2009.0318 \\ %& 1.35  & very thin clouds, good seeing\\
Hectospec & MMT Observatory & $+10:33:26.2$ & $+57:55: 05$ & 1.5  & 2009.0317 \\ %& 1.32 & \\
Hectospec & MMT Observatory & $+10:45:21.6$ & $+57:53: 55$ & 1.5  & 2009.0301 \\ %& 1.18 & \\
Hectospec & MMT Observatory & $+10:37:35.1$ & $+57:32: 53$ & 1.2  & 2009.0228 \\ %& 1.18 & light clouds, mostly clear\\
Hectospec & MMT Observatory & $+10:39:48.3$ & $+59:16: 56$ & 1.5  & 2009.0227 \\ %& 1.00 & \\
Hectospec & MMT Observatory & $+10:42:20.4$ & $+57:05: 15$ & 1.5  & 2009.0226 \\ %& 1.16 & \\
Hectospec & MMT Observatory & $+10:37:12.5$ & $+58:38: 24$ & 1.5  & 2009.0223 \\ %& 1.22 & \\
Hectospec & MMT Observatory & $+10:54:13.1$ & $+57:03: 57$ & 1.5  & 2009.0222  \\ %& 1.50 & \\
Hectospec & MMT Observatory & $+10:57:45.5$ & $+57:34: 04$ & 1.5  & 2009.0222  \\ %& 1.21 & \\
Hectospec & MMT Observatory & $+10:44:31.2$ & $+58:46: 14$ & 1.5  & 2009.0220 \\ %& 1.19 & \\
Hectospec & MMT Observatory & $+10:48:29.3$ & $+59:22: 27$ & 1.5  & 2009.0131 \\ %& 1.23 & \\
Hectospec & MMT Observatory & $+10:52:31.3$ & $+57:24: 15$ & 1.3  & 2005.0410 \\ %& 1.28 & \\
Hectospec & MMT Observatory & $+10:51:02.9$ & $+57:22: 17$ & 0.6  & 2005.0409 \\ %& 1.22 & \\
Hectospec & MMT Observatory & $+10:51:46.1$ & $+57:26: 32$ & 0.3  & 2005.0408 \\ %& 1.32 & heavy clouds at end of exposure\\
Hectospec & MMT Observatory & $+10:51:02.9$ & $+57:22: 17$ & 1.7  & 2005.0405 \\ %& 1.10 & \\
Hectospec & MMT Observatory & $+10:52:09.7$ & $+57:27: 49$ & 1.0  & 2005.0310 \\ %& 1.13 & \\
Hectospec & MMT Observatory & $+10:51:42.2$ & $+57:28: 01$ & 1.0  & 2005.0308 \\ %& 1.13 & \\
Hectospec & MMT Observatory & $+10:51:42.2$ & $+57:28: 01$ & 1.0  & 2005.0304 \\ %& 1.13 & \\
Hectospec & MMT Observatory & $+10:52:03.8$ & $+57:26:22$ & 1.0  & 2005.0308 \\ %& 1.24 & \\
Hectospec & MMT Observatory & $+10:52:18.8$ & $+57:21: 53$ & 1.0 & 2005.0308 \\ %& 1.13 & \\
\hline
\hline
\end{tabular}
\end{center}
\label{tab:obslog}
\end{table}

To summarize, we have a total of 3612 spectra of MIR-selected objects
with r $<$ 22.5 observed by MMT-Hectospec
or chosen from the SDSS SpecObj catalog 
with a redshift confidence of $>$ 90\% (Table~\ref{tab:sample}). 

\subsection{Broad Line Object Identification}
\label{sec:bel}
\begin{figure}[ht]
\begin{center}
\includegraphics[scale=0.9]{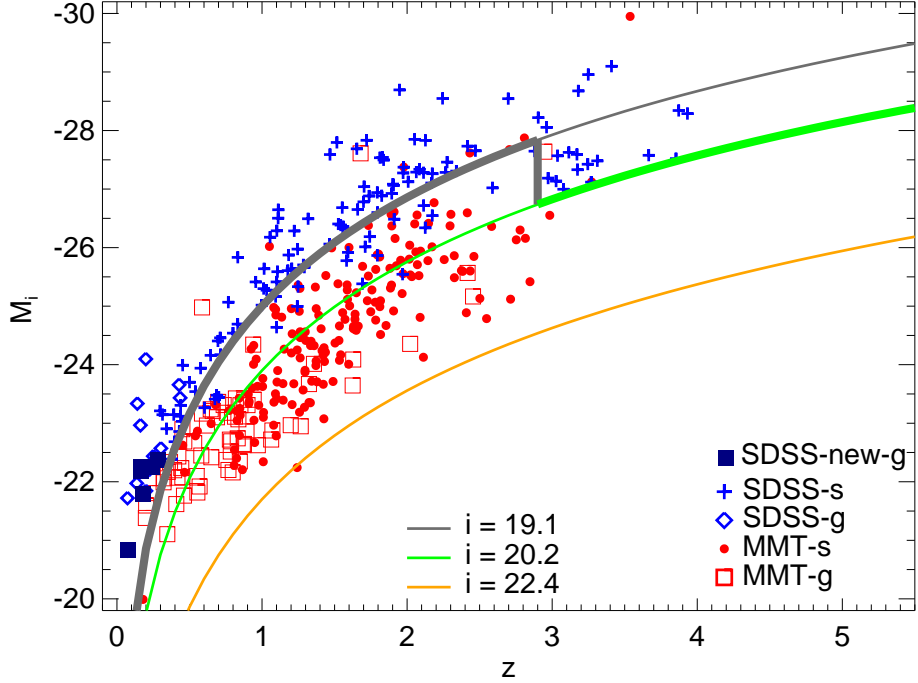} %{Mi-type.eps}%fig5
\end{center}
\figcaption{Distribution of MIR-selected quasars in the luminosity-redshift plane. 
Luminosity is indicated by i band absolute magnitude, $M_{i}$, 
calculated from the SDSS photometric magnitude. %(petroMag_i).
In blue are the SDSS-identified quasars, with blue pluses for point sources 
and blue diamonds for extended sources;
and the red are the MMT-identified quasars, with red dots for point sources
and red squares for extended sources.
Whether an object is extended (galaxy-like, `g') or a point source (star-like, `s')
is defined by the extendedness of the SDSS photometry, see \S~\ref{sec:type}.
The solid navy blue squares mark the 6 quasars newly identified with SDSS spectra
that were not included the SDSS DR7 quasar catalog.
The curves shows the two magnitude ranges of SDSS selections at  $i = 19.1$ (grey) and $i = 20.2$ (green),
and thick curves are the limiting $z$ dependent magnitudes SDSS used for 
the sample of spectroscopic targets (See \S~\ref{sec:type}). 
The orange curve shows the equivalent i band magnitude of the MMT-Hectospec limit ($r = 22.4$).
A total of 93 new quasars have been identified by MMT (red dots and squares) at $i < 20.2$,
of which 80 fall between $19.1 < i < 20.2$ at $z <$ 3,
a region SDSS did not cover in the uniform color selection (See also \S~\ref{sec:completeness}). 
The number densities of $z < 3$ quasars is $\sim$ 10\,deg$^{-2}$ at $i < 19.1$,
slightly higher than the $\sim$9\,deg$^{-2}$ at $19.1 < i < 20.2$.
\label{fig:Mi}}
\end{figure}

The 3612 reduced 1d SDSS and Hectospec spectra were first 
fitted using our IDL program adopted from the S11 procedure.
This program fits a polynomial continuum ($S_{\rm continuum} = A_{\rm \lambda} \times (\frac{\lambda}{\lambda_0})^{\alpha_{\rm \lambda}}$) 
and a Gaussian around
the redshifted CIV, MgII, and H$\beta$ regions based on the HSRED or SpecObj redshifts (See also \S\ref{sec:spec}). 
Objects are kept as quasar candidates 
if they have at least one BEL (FWHM $> 1000\,\kms$, \citet{schneider07})
in the secure spectral ranges with limited atmospheric extinction and instrument errors: 
3850--8400 ${\AA}$ \citep{fab08} for MMT targets,
and 3850--9000 ${\AA}$ \citep{stoughton02} for SDSS targets.
Outside of these ranges the spectra
start to be bounded by sky-subtraction errors and therefore not reliable.
The MMT range is from \citet{fab08},
chosen to be most consistent ($< 5\%$) with SDSS,
after comparing the optical spectra taken from SDSS and MMT
of the same targets.
The IDL program identifies 236 MMT09, 28 MMT05b, and 132 SDSS BEL objects,
all of which have an emission line equivalent width (EW) greater than 6.  
Given our flux limit ($r < 22.5$), 
the majority of the BEL quasars (83\% with $Mi < -23$) also satisfy the $M_{\rm B} < -23$,
the quasar definition in \citet{sg83} (Fig.~\ref{fig:Mi}).
Since the SDSS quasar definition is also based on the BEL features (\citet{schneider07})
in the following text we will simply refer to these BEL objects as quasars.

\begin{figure}[ht]
\begin{center}
\includegraphics[scale=0.5]{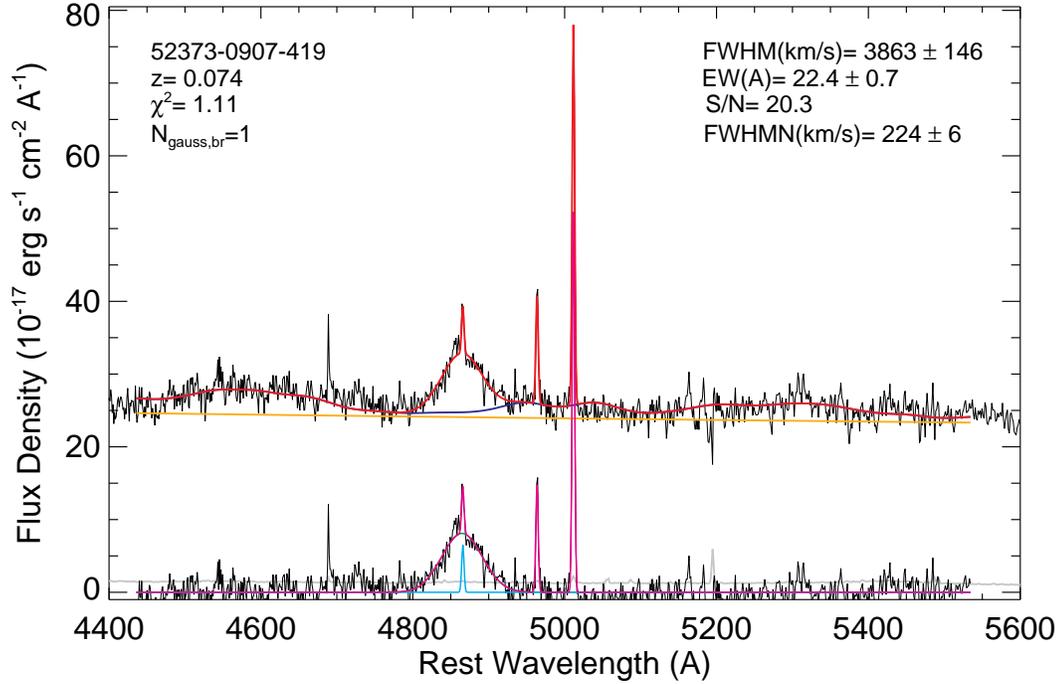} %{test-Hb-52373-0907-419.eps} %fig6
\end{center}
\figcaption{
One example of the 6 newly identified SDSS quasar zoomed in at the \hbeta\ emission 
line region  ([4400, 5600]\,\AA). 
The two sets of plots show in the top the original spectrum 
(black), 
the estimated continuum (orange), the scaled Fe
template (dark blue) and the final fitted composite spectrum (red).
The lower plot shows the continuum- and Fe-template-subtracted
spectrum (black),
 in addition to the variance spectrum (grey), 
the narrow-line emission component (cyan) and the wide-line component (green). 
The latter is mostly subsumed by the composite narrow+wide emission line
spectrum shown in magenta. 
%In $orange$ and $blue$ are the continuum and the Fe template -- the latter  
%covered by the composite spectra in $red$ except for the emission line region ([4700, 5100]\,\AA).
%In $green$ and $cyan$ are the broad and narrow Gaussian components for the \hbeta\ emission line,
%while in $magenta$ is the composite spectra for the emission line only.
The top left corner shows the ID, redshift, fitting $\chi ^2$ and number of Gaussian component used,
and top right is the fitting results (FWHM, EW, signal-to-noise, and the FWHM of the narrow line)
in the emission line region.
\label{fig:hb6}}
\end{figure}

As a check, we visually examined all 3,612 spectra from both the MMT and SDSS surveys.
This process removes 22 MMT09, 1 MMT05b,  and 5 SDSS objects 
that were erroneously identified as quasar due to bad fits.
This process also adds 12 MMT09 and 11 SDSS objects, 
but no MMT05b objects
were missed due to a poor fit by the IDL program.
Of the 11 SDSS objects,
%five were included in the SDSS DR7 quasar catalog,
6 were not included in 
%identified as quasars by the SDSS pipeline and thus not in 
the SDSS DR7 quasar catalog. 
All of the 6 new objects are confirmed as quasars with 
a broad \hbeta\,\,emission line (Fig.~\ref{fig:hb6} shows one example). 
We will explore the possible reasons why they were missed in the SDSS DR7 catalog 
in \S~\ref{sec:i191}.
Special objects with interesting features -- broad absorption Line (BAL) and narrow absorption line (NAL) quasars
--- are also flagged (See Section.~\ref{sec:discussion}). 

Combining the IDL fit and eye check, 
we identify 226 quasars from the MMT09, 
27 from the MMT05b, 
and 138 from the SDSS DR7 SpecObj catalogs. 
This adds up to a total of 391 MIR-selected quasars in the LHS field.
For comparison, we also scanned the 902 fainter ($S_{\rm 24\,\mu m} < 400\mu$Jy) 
objects from MMT05f survey and identified 17 BEL objects (one was added after eye check).
Table~\ref{tab:sample} summarizes the quasar numbers in each subsample. 
The fraction of MIR quasars in the MMT09 subsample is 
9.1\%, and 9.9\% in the MMT05b subsample,
yielding an average detection rate of 9.2\%. 
After including the SDSS quasars selected through color-color selection,
the total detection rate for this MIR quasar sample in LHS field is 10.8\%.
%,i.e. out of the 3612 spectra that satisfy the $S_{\rm 24\,\mu m} > 400\mu$Jy
%and $r < 22.5$ flux limits, 391 are identified as quasars.
If only considering the MMT and SDSS overlapping area, the quasar detection rate 
 is an almost identical 10.9\%.
These detection rates are marginally lower than the $13\pm3$\% reported in \citet{papovich06},
where a higher 24 \micron\ flux limit ($S_{\rm 24\,\mu m} > 1$\,mJy) was applied.

\begin{figure}[ht]
\begin{center}
\includegraphics[scale=0.8]{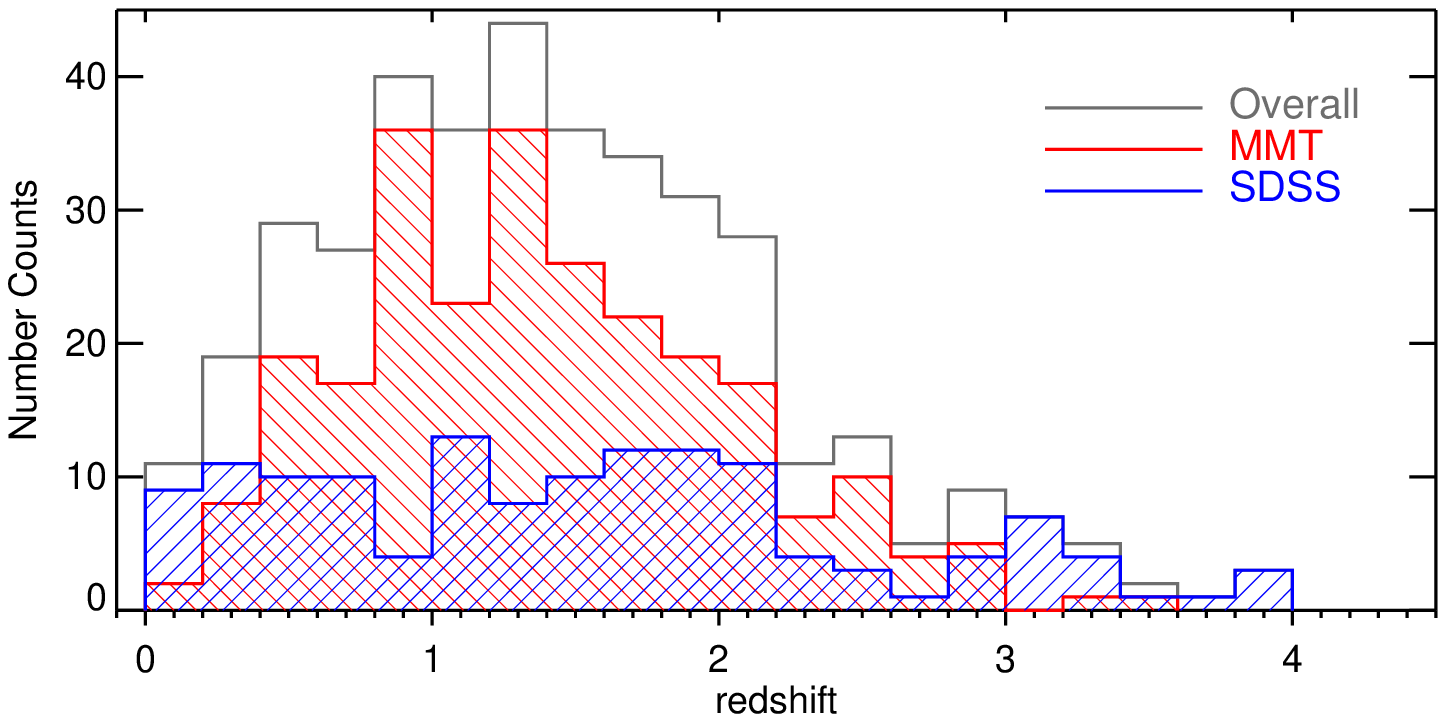} %{zdist-long2.eps} %fig7a
\includegraphics[scale=0.8]{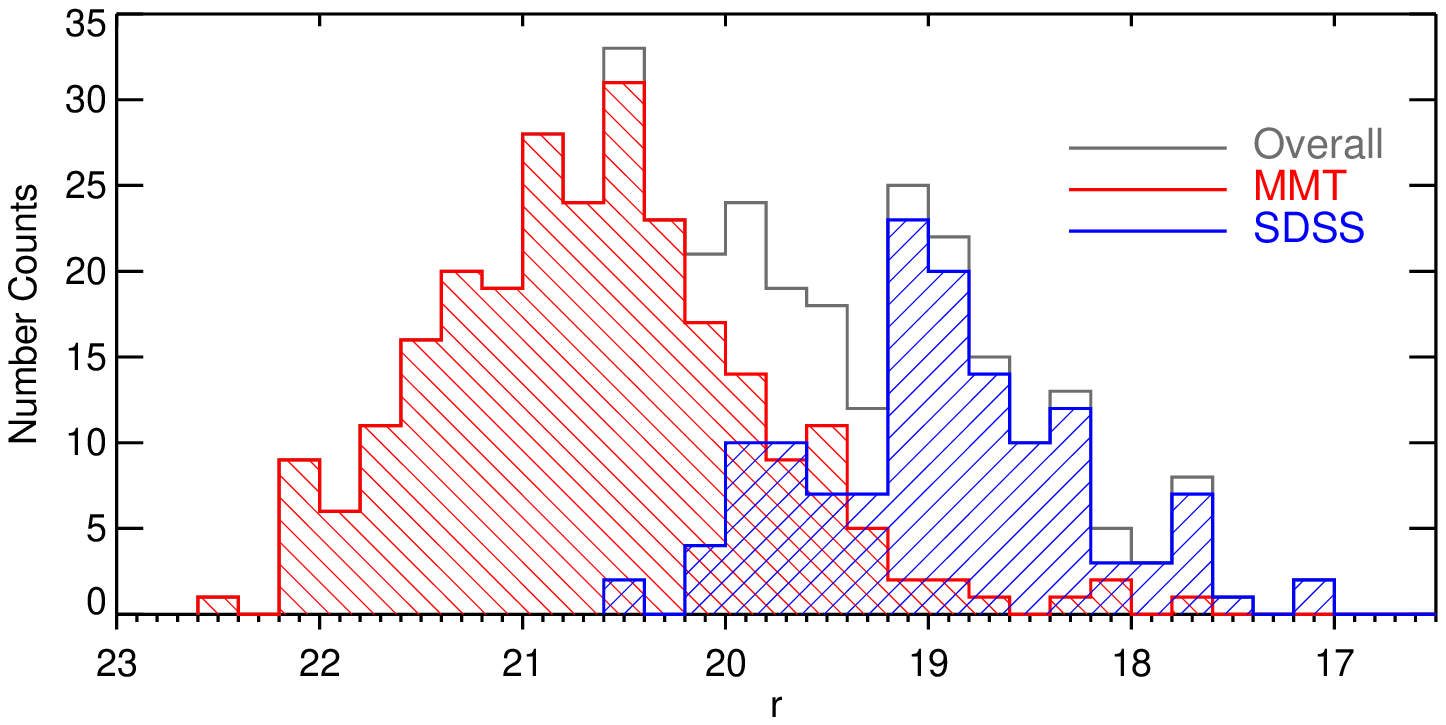} %{rdist-long2.eps} %fig7b
\includegraphics[scale=0.8]{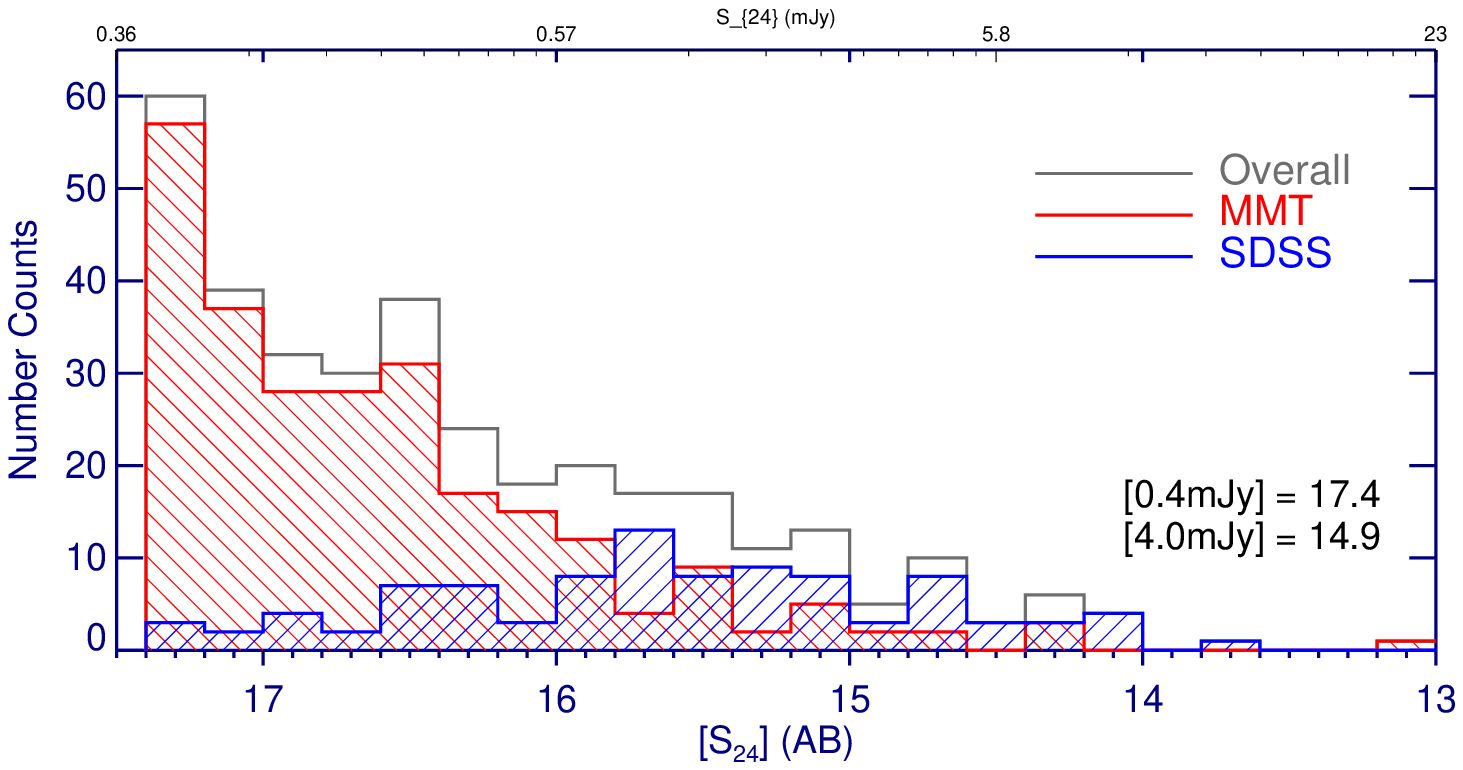} %{f24-long-r.eps} %fig7c
\end{center}
\figcaption{(top) The redshift distribution for the full MIR-selected quasars catalog (grey), 
the 253 MMT subsample (red), and the 138 SDSS subsample (blue).
Middle and bottom panels show the same color coded distribution of SDSS r band magnitude 
and 
the 24 $\mu$m flux (converted to AB magnitude for presentation purpose, 
flux conversion examples are given in the bottom right).
\label{fig:sp}}
\end{figure}

To study the overall properties of the MIR-selected quasars,
we plot the redshift ($z$, top), 
r band magnitude ($r$, middle), 
and 24 $\mu$m flux ([$S_{24}$], bottom) distributions (Fig.~\ref{fig:sp}).
The sample has a redshift range of $0.07 < z < 3.93$, with a median redshift of 1.3.
A K-S test shows significant difference (p $\ll 0.1$) between the SDSS and MMT subsamples
in all three parameters ($z,\, S_{24}$, \,and $r$).
The SDSS quasars have two $z$ peaks at 1$< z < $2 and at $z \sim 3.2$,
with an overall median $z = 1.5$.
The reason for the double peaks is because of the two main color selection criteria ($ugri\ \& \ griz$) 
applied in SDSS for low-$z$ ($z < 2.9$) and high-$z$ ($z > 2.9$) quasars (\S\ref{sec:type}).
The MMT, on the other hand, has a roughly Gaussian redshift distribution
with a peak at $z\sim$1.3.
The MIR-selected quasars
are clearly not homogeneously distributed across redshifts.
The SDSS subsample has overall brighter $r$ and $S_{24}$ 
than the MMT subsample, 
and overlaps significantly with the bright end of the MMT quasars.
These differences are due to the SDSS quasar algorithm,
which has a limit at $i = 20.2$,
about 2 magnitudes brighter than the MMT selection 
($r = 22.5$\footnote{Using \citet{rich06b} mean SDSS quasar template, 
$r = 22.5$ is equivalent to $i = 22.4$ at $z\sim1.5$.}).
 The MMT-Hectospec survey
intentionally dropped SDSS targets with existing spectra,
leaving the MMT targets biased towards the faint end. 
The combination of the MMT and SDSS provides a better way
to examine the completeness of MIR-selected quasars at $S_{24} \ge 400\mu$Jy.

\begin{figure}[ht]
\begin{center}
\includegraphics[scale=1.0]{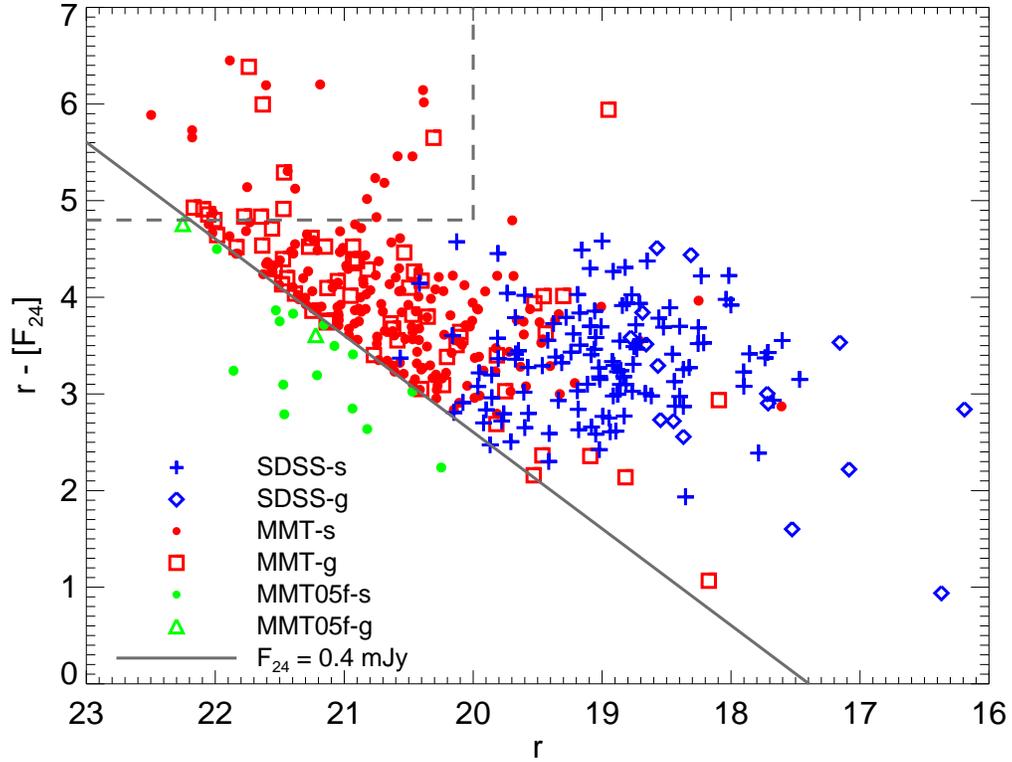} %{f24-composite-type.eps} %fig8
\end{center}
\figcaption{The comparison of different subsamples in terms of their MIR to optical [r - [F$_{\rm 24}$]]colors: 
SDSS (blue), MMT (red), and MMT05-faint (green). 
In grey we mark the 24\,$\mu$m flux limit of 400$\mu$Jy. 
Objects with point source morphology are the blue pluses (SDSS), red dots (MMT), and green dots (MMT05f);
extended sources are marked by blue diamonds (SDSS), red squares (MMT), and green triangles (MMT05f).
The morphologies are defined by the extendedness of the SDSS optical photometry, see \S~\ref{sec:type}.
The different subsamples show similar MIR to optical colors within the range of [2, 4.8].  
At $r > 20$, a very red population of MIR-selected quasars emerges (dashed region, $r > 20$, $r - [S_{24}] > 4.8$),
comprising a small fraction of 29 objects (14\%) out of the 212 $r > 20$ MIR-selected quasars. 
\label{fig:f24-r}}
\end{figure}

In Fig.~\ref{fig:f24-r} we compare the optical to MIR colors
against $r$ magnitude for
the MMT and SDSS subsamples.
The MIR-selected MMT subsamples are redder
in $r - [S_{24}] $ colors than the SDSS subsample,
with median $[r - [S_{24}]]$ values of 4.0 for MMT09, 3.9 for MMT05b,
and 3.3 for SDSS.
Though separable by $S_{24}$ flux, 
the 17 MMT05f quasars ($60 < S_{24} < 400\,\mu$Jy) show similar 
$r - [S_{24}]$ colors to the SDSS subsample,
but are bluer (median $r - [S_{24}] = 3.4$)
than the MMT subsamples.
A K-S test gives a probability of 0.975 of the MMT05f and SDSS objects,
indicating identical distributions.
Instead, the K-S test probability is $< 0.001$ between MMT05f and 
the brighter MMT subsamples (MMT09, MMT05b),
indicating significant difference in the optical-IR color $r - [S_{24}]$. 
At $r > 20$, we also notice a very red population ($r - [S_{24}] > 4.8$) of 
MIR-selected quasars (inside the dashed line, Fig.~\ref{fig:f24-r}). 
The emerge of such population may simply be a result of 
the fainter magnitudes MMT sample covers,
though this red population is still rare,
which comprises 14\% of the $r > 20$ MIR-selected quasars (32 out of 218). 
The absolute i band magnitude ($M_i$) for the red objects has a mean $M_i$ of  -23.6,  
one dex lower than the mean for the whole MIR-selected population ($M_i = -24.7$).

We further examine the SDSS and MMT subsamples
in the luminosity-redshift space (Fig.~\ref{fig:Mi}).
The majority (66\%) of the newly identified MMT quasars are fainter than 
the SDSS magnitude cut of $i = 20.2$.
A total of 93 MMT quasars also meet the SDSS magnitude limit ($i < 20.2$),
which almost doubles the number of SDSS quasars in this region.
One MMT source (2009.0131-268) at $z =$ 3.537 
has an extremely high luminosity at $M_i = -29.97$. 
Such high luminosity is also rare in the SDSS catalog, only 82 quasars (0.078\%) 
in the 105,\,783 SDSS DR7 quasars are at $M_i$ brighter than -29.9.
This quasar has consistent magnitudes at $i = 17.5$ in modeled, fiber- and Petrosian SDSS magnitudes, 
but was missed in the SDSS DR7 quasar catalog for unknown reasons. %(See also \S\ref{sec:completeness}).
The number densities of $z < 3$ quasars is $\sim$ 10\,deg$^{-2}$ at $i < 19.1$,
slightly higher than the $\sim$9\,deg$^{-2}$ at $19.1 < i < 20.2$.
The majority (78) of the new quasars are at $19.1 \leq i < 20.2$ and 
$z < 3$,  a region the SDSS selection deliberately 
avoided to ensure the selection of high $z$ targets in
 their $griz$ colors selection.
 At first glance this appears to be a major challenge to
the SDSS's claim of 90\% completeness to $i_{\rm AB} = 20.2$. 
In the following section we will explore the reasons for 
this inconsistency.

\subsection{MIR Additions to the SDSS Quasar Selection}
\label{sec:completeness}
%{\bf (This whole section has been rewritten.)}

The MIR-selected quasars are BEL (type 1) objects satisfying the joint
limits of $r < 22.5$ in the optical and $S_{\rm 24} > 400\mu$Jy in the MIR.
The limit of $r < 22.5$ is roughly equivalent 
to $i < 22.4$ at $z\sim1.5$ according to the \citet{rich06b} SED template.
In Fig.~\ref{fig:Mi}, 93 new quasars have been identified by the MMT spectroscopy above
the SDSS DR7 quasar sample limit ($i \leq 20.2$), 
of which 87 also satisfy the SDSS magnitude limit of M$_{i} < -22$.  
Another 6 quasars are identified by
re-examining the SDSS spectra. %, 2 of which would have been rejected due to faint M$_{i} (> -22$). 
In this section we study why these objects were missed 
by the SDSS quasar catalog,
and which additional objects the MIR selection 
is adding to the overall quasar population.
%All of these objects are brighter
%than the , but were not included
%in the DR7 catalog. 

\subsubsection{Comparing the Selection Criteria}
\label{sec:type}
The SDSS spectroscopic targets are selected primarily via
color-selection with the SDSS photometry \citep[][R02]{rich02a},
which includes the two main low-$z$ $ugri$, high-$z$ $griz$ color selections,
and a few other selections in the color-color or color-magnitude space:  
a mid-$z$ (2.5 $< z < $3),  two high-$z$,
UVX, and $ugr$ outlier inclusion regions. 
The 2 main uniform color selections
correspond to the two magnitude cuts at $i \leq 19.1 (ugri)$
and $ 19.1 < i \leq 20.2 (griz)$,
with the latter designed to recover high $z$ ($z > 3$) targets only
---certain conditions are set to exclude low-$z$ objects.
In both magnitude bins, 
SDSS rejected targets that fell in the color boxes of white dwarfs, A stars,
M star and white dwarf pairs.  
The SDSS selection also excludes objects in the 2 $\sigma$ wide region around the stellar locus, 
with an exception for low-$z$ resolved AGNs \citep{schneider10}. 
Therefore only in the brighter $i \le 19.1$ bin would extended sources be included,
while at fainter magnitudes ($i > 19.1$), all SDSS targets are point sources.
Secondary SDSS targets came from the FIRST radio source catalog \citep{white97} 
and ROSAT X-ray sources \citep{anderson03}. 
Color-color selected SDSS targets were qualified as quasars 
if they were spectroscopically confirmed as BEL objects or have interesting absorption 
features \citep{schneider10}.

The exclusion of extended sources in the high$-z$ $griz$ color selection 
was achieved via the SDSS star-galaxy morphology separation. 
This separation is based on comparing the small 
point-spread function (PSF) magnitude and 
the larger exponential or de Vaucouleurs magnitude resulting from their different apertures. 
Objects for which the difference between the point-spread function (PSF) and 
the modeled (exponential or de Vaucouleurs profiles) magnitudes is greater
than 0.145 mag are classified as extended (`galaxy-like', type $=$ 3, R02); 
otherwise they are classified as point-source (`star-like', type $=$ 6, R02) .

The MMT targets in the MIR quasar sample, on the other hand,
are only selected based on the 24\micron\ flux limit and
r band magnitude cuts, before they are optically identified as BEL objects.
The SDSS quasar selection criteria, 
are necessarily much more complicated
given the large sky density of $i < 20.2$ objects (\S~\ref{sec:bel}).
As a result, the quasar detection rate is higher for the SDSS spectra ($\sim$ 16\%),
than in the MMT spectra ($\sim$10\%, Table.~\ref{tab:sample}).

Table~\ref{tab:completeness} summarizes the number counts 
in 3 different $i$ magnitude bins and SDSS photometric types for the SDSS and MMT quasars in this sample.
We found a constant fraction of 20\% of `extended' MIR-selected MMT quasars
in all magnitude bins, 
with the majority ($>$ 80\%)) at lower $z$ ($ <$ 1) and 
luminosity (log\,$\lbol <$ 45.5 \ergs, see also \S~\ref{sec:lbol}). 
These extended objects were automatically rejected in the SDSS selection at $i > 19.1$. 
A second significant MIR addition comes from the fainter sources
in the MMT surveys: 
a total of 160 objects are found at $i > 20.2$, which SDSS did not cover.

\subsubsection{MIR Additions to the SDSS Completeness}
\label{sec:i191}
In this section we compare the colors and photometric morphologies 
of the SDSS and MMT identified quasars in the 3 different magnitude bins. 

The SDSS uniform color selections have an estimated completeness based on
simulated quasars, to be over 90\% at $0 < z < 5.3$
down to $i = 20.2$ (See also Table 6 in R02). 
This is an average completeness for previously known quasars,
and applies to $i < 19.1$ quasars at $z < 2.5$,
and to $i < 20.2$ quasars at $ 3.5 < z < 5.3$.
A later calibration of the completeness of the SDSS DR5 quasar survey
gives an end-to-end completeness of $\sim$89\% \citep{vandenberk05},
which was confirmed in the SDSS DR5 quasar paper
as ``close to complete" for 0.7$< z < 1.0 $ and $1.9 < z < 2.1$
at log ($\lbol) (\ergs) >$45.9 and $> $46.6, respectively \citep{rich06a, shen08}.
%The uniformly selected quasar is observed at $z=2.9$, 
%which instead of being empty, 
%still covers quasars discovered before the SDSS era (See Fig.1, S11).

\begin{figure}[ht]
\begin{center}
\includegraphics[scale=0.9]{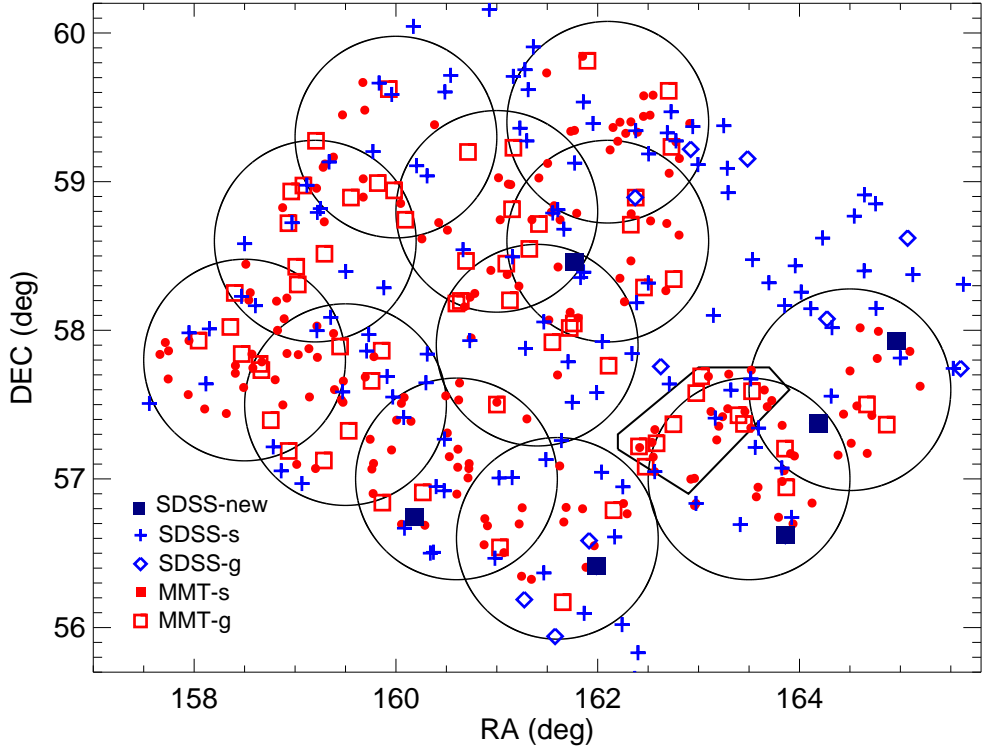} %{cov22.eps} %fig9
\end{center}
\figcaption{
Quasar distribution in the LHS field:  %MMT covered 
{\it Black circles}, the 12 MMT09 FOV;
 {\it Black polygon}, contour for the 8 MMT05 FOV;
{\it Blue pluses}, point-source quasars identified with SDSS spectra (SDSS-s); 
{\it Blue diamonds}, extended quasars identified with SDSS spectra (SDSS-g); 
{\it Red dots},  point-source quasars identified with MMT spectra (MMT-s);
{\it Red squares},  extended quasars identified with MMT spectra (MMT-g). 
{\it Navy squares}, the 6 extended quasars identified with SDSS spectra but 
not included in the SDSS DR7 quasar catalog (SDSS-new).
For definition of the photometric morphology see \S~\ref{sec:completeness}.
\label{fig:cov2}}
\end{figure}

The distribution of quasars in the $\sim$ 22 deg$^2$ LHS field is plotted in Fig.~\ref{fig:cov2}.
For a fair comparison, we focus only 
on the $\sim$12\ deg$^2$ MMT covered region (within the circles and black polygon).
There are a total of 96 SDSS quasars in the overlapping region (Table~\ref{tab:completeness}). 
Of these, 61 are uniformly color selected (uniform flag $=$ 1), 
and 29 by considering radio, X-ray, or other inclusion criteria (uniform flag $=$ 0).
None of the SDSS quasars fall into 
the high-$z$ selected SDSS ``QSO\_\ HiZ" branch (uniform flag $=$ 2).
As mentioned in \S\ref{sec:bel}, 
after re-inspecting of SDSS spectra
we identified 6 additional quasars not included in the SDSS DR7 quasar catalog.
They are represented as dark blue squares in Fig.~\ref{fig:Mi}.
There are 62 SDSS quasars at $i  \leq\ 19.1$, 
27 quasars at $19.1 < i \leq\ 20.2$,
and 1 at $i > 20.2$. 
MMT observations identify an additional 13 MMT09 and 6 SDSS quasars at $i  \leq\ 19.1$, 
of which 10 MMT09 and 4 SDSS objects qualify the SDSS M$_{i} = -22$ limit. 
At $19.1 < i \leq\ 20.2$, 73 MMT09 and 7 MMT05b quasars are
added, of which 70 MMT09 and 7 MMT05b also satisfy M$_{i} < -22$ (Fig.~\ref{fig:Mi} and Table~\ref{tab:completeness}).  

\begin{figure}[ht]
\begin{center}
\includegraphics[scale=0.6]{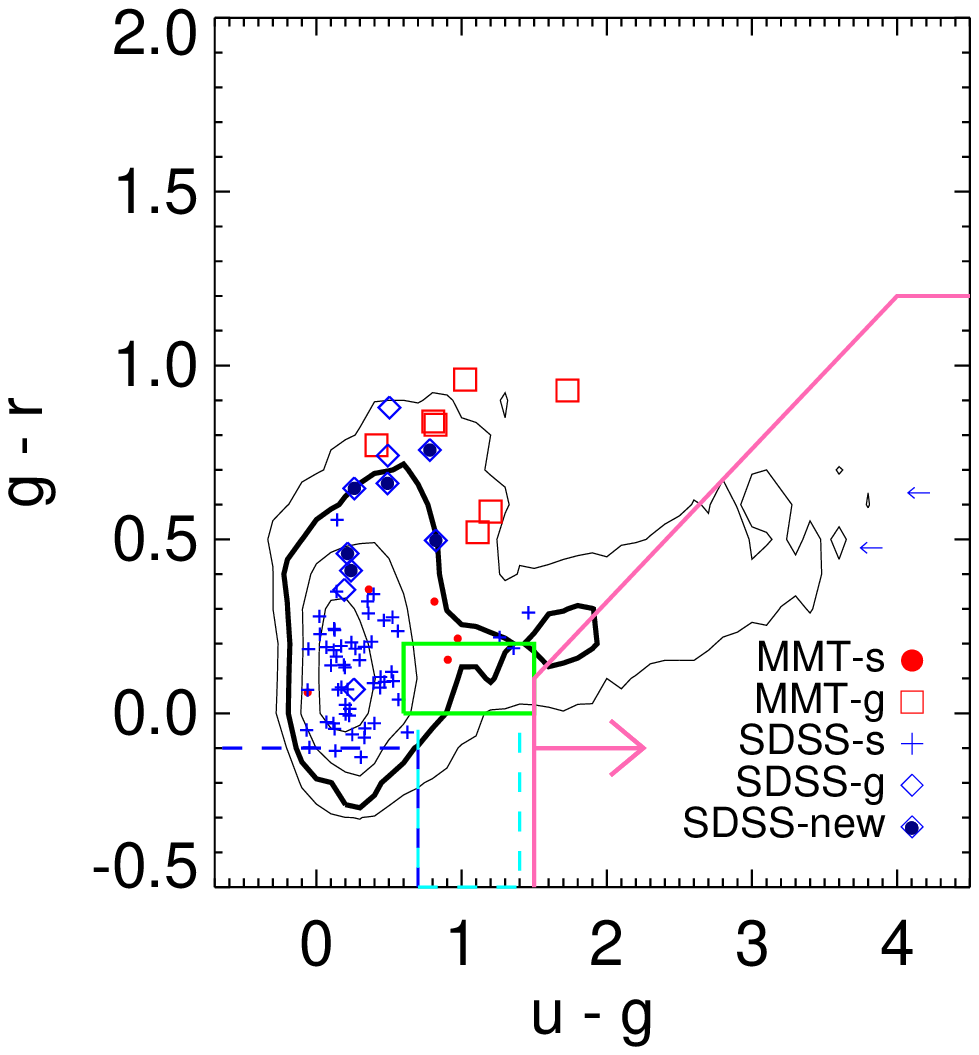} %{color1bt.eps}  %fig10
\includegraphics[scale=0.6]{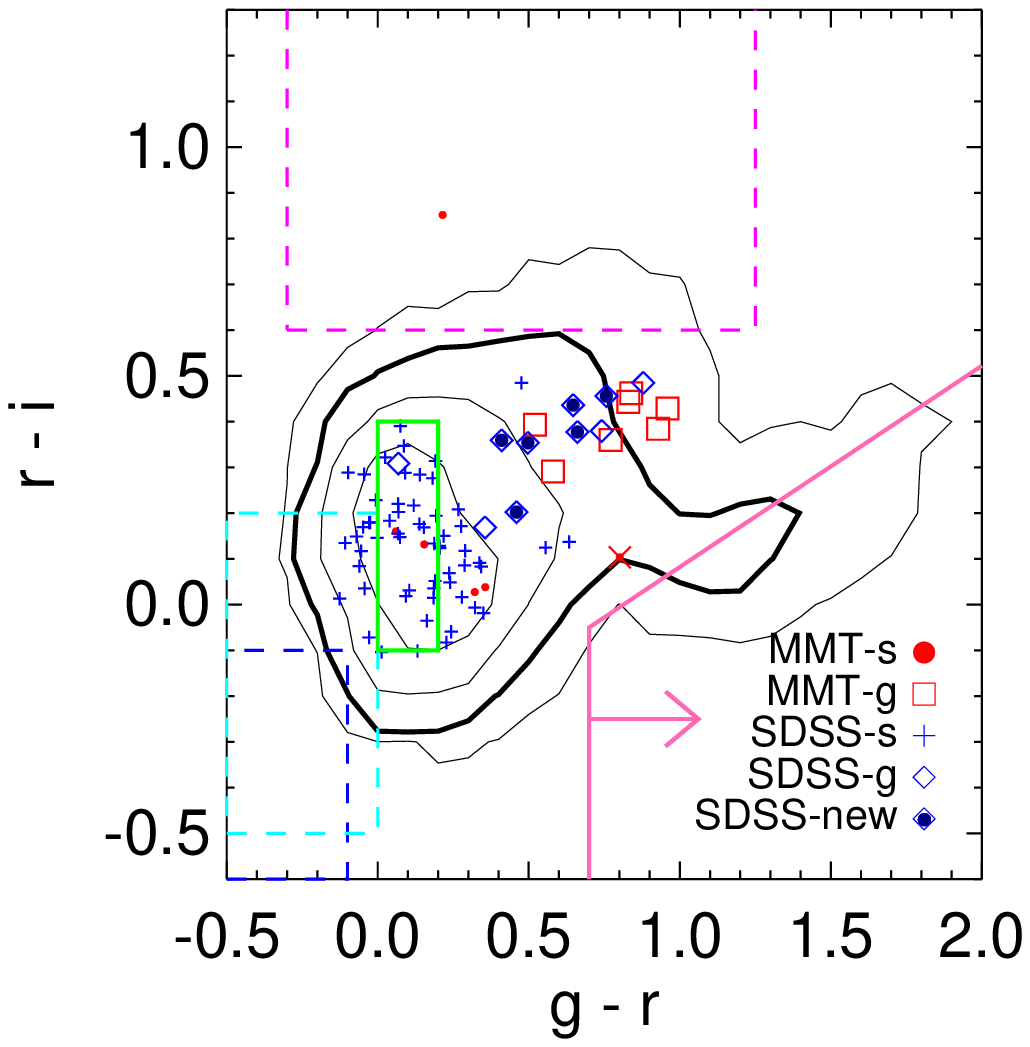} %{color2bt.eps}
\includegraphics[scale=0.6]{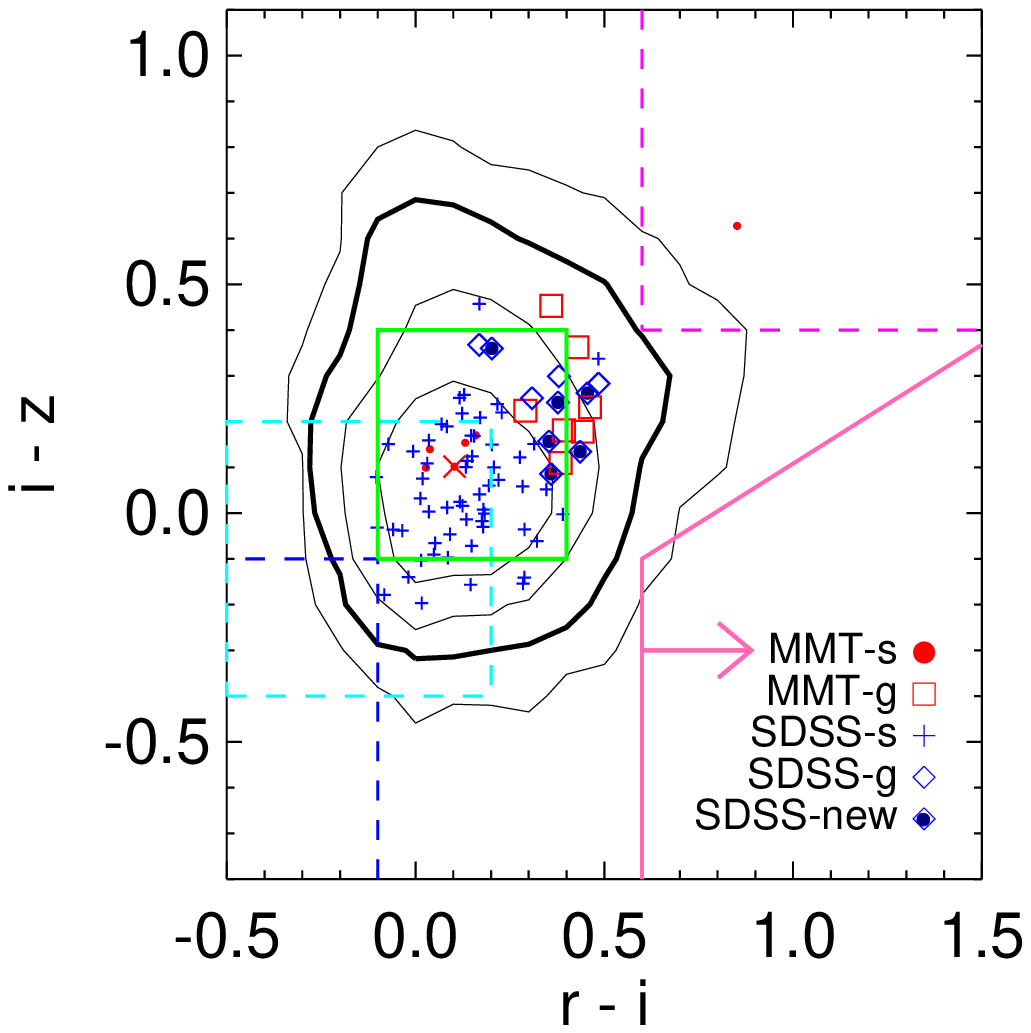} %{color3bt.eps}
\includegraphics[scale=0.6]{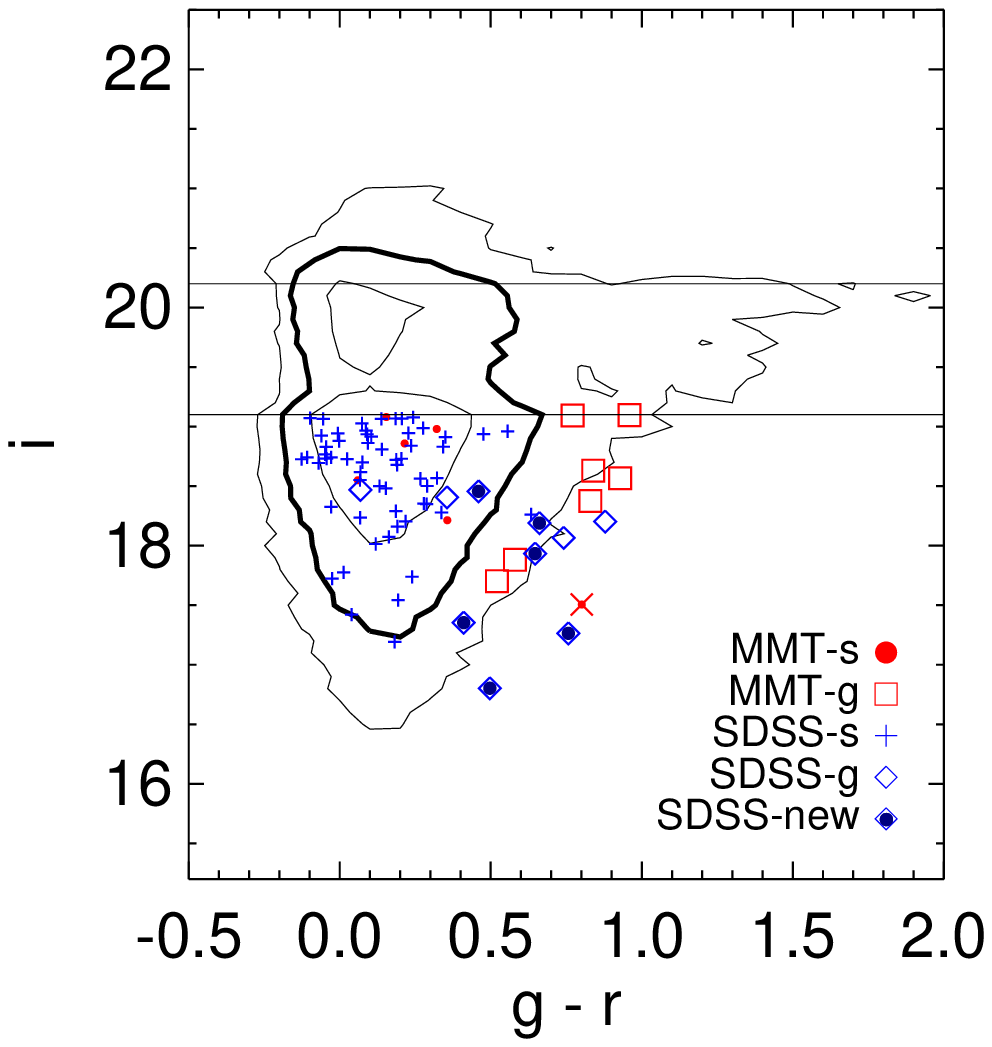} %{color4bt.eps}
\end{center}
\figcaption{Location of SDSS ($blue$) and MMT ($red$) identified quasars in the 
SDSS colors and magnitudes diagrams for objects at $i \leq 19.1$.
Symbols signals their SDSS photometric classification:
{\it blue pluses} and {\it red dots} for SDSS and MMT point source (`-s', `star like');
{\it blue diamonds} and {\it red squares} for extended sources (`-g', `galaxy like').
The {\it blue filled diamonds} are the 6 newly-identified SDSS BEL objects,
all of them have extended photometry.
The bright MMT source (`2009.0131-268') is marked with a {\it red cross} in the center. 
Contours mark the distribution of the 110,509 SDSS DR7 quasars from \citet{shen11}
at number densities of 20, 100, 500, and 1,000 per 0.1 magnitude or color bin. 
The contour level of 100 objects per 0.1 magnitude bin
is highlighted as a thick line to guide the eye.
For objects with $< 3\sigma$ detections in either band,
an upper/lower limit is used in the color-color plots.
Dashed boxes are the different SDSS exclusion regions:
{\it blue} for white dwarfs; 
$cyan$ for A stars; 
$magenta$ for M stars $+$ white dwarfs.
Solid boxes are: 
$green$, the mid-$z$ inclusion regions; 
{\it solid magenta with an arrow}, high-$z$ inclusion regions (unique in each panel, see \citet{rich02a}).
The black lines in the `g-r' vs `i' panel shows the two SDSS magnitude cuts at $i = 19.1$ and $i = 20.2$.
About half of the MMT subsample are extended sources, mostly covered by the outmost contour level of 20 objects per bin,
and 1 point MMT source falls into the SDSS M star and white dwarf exclusion region.
\label{fig:colors1}}
\end{figure}

We first examine the bright magnitude bin of $i \leq 19.1$,
where the SDSS $ugri$ color selection is optimized for low $z$ ($ z < 2.9$)
quasar selection
and includes both extended and point sources. 
At $i \leq 19.1$, 15\% of the SDSS quasars are extended (`galaxy-like', see \S\ref{sec:type}),
while in the MMT additions, 
$\sim$50\% are extended (Table~\ref{tab:completeness}).
In Figure.~\ref{fig:colors1}, we compare the MIR-selected MMT and SDSS quasars at $i \leq 19.1$ 
in the 4 color-color and color-magnitude spaces.
The majority of both MMT and SDSS samples fall inside 
the contours of 100 or more (thick curve) SDSS DR7 quasars per 0.1 magnitude bin.
Only 4 of the 62 previously-identified SDSS quasars are extended
 (`SDSS-g', marked as open blue diamonds in Fig.~\ref{fig:colors1}).
All of the 6 newly identified SDSS BEL objects 
(blue filled square) are extended. 
They were possibly rejected in the SDSS selection 
for being extended with blue $u-g$ colors (as indicated by vectors in Fig.4 of R02).

In the bright $ i \leq 19.1$ bin, 
9 of the 13 new MMT09 detections satisfy the SDSS selections, 
including 4 point sources and 5 extended sources at $z < 2.6$  (Figure.~\ref{fig:colors1}, Table~\ref{tab:completeness}). 
%with 4 point sources and 5 extended objects.
The remaining 4 MMT quasars would have been rejected in the SDSS selection, 
since 3 are fainter than $M_{\rm i} = -22$,
and one point source falls in the SDSS M star $+$ white dwarf 
exclusion region (marked by magenta dashed lines in Fig.~\ref{fig:colors1}, See also Table~\ref{tab:exclusion}).
Despite lying at the edge of the bulk of the SDSS contours,
all of the 9 new MMT objects have photometries that meet the 5$\sigma$ and error $< 0.2$ 
requirement of the SDSS selection (R02). 
After adjusting for the MMT optical spectroscopy completeness 
(30\% for MMT09, 66\% for MMT05b),
the overall completeness of the SDSS selection at $i < 19.1$ is $(67 \pm 8)\%$ (Fig.~\ref{fig:incomp}),
about 20\% lower than the simulated 90\% from R02.
%The completeness for point sources is $(76 \pm 11)\%$.
Errors are Poisson estimates based on the inverse square root of total number of objects.

\begin{figure}[ht]
\begin{center}
\includegraphics[scale=0.6]{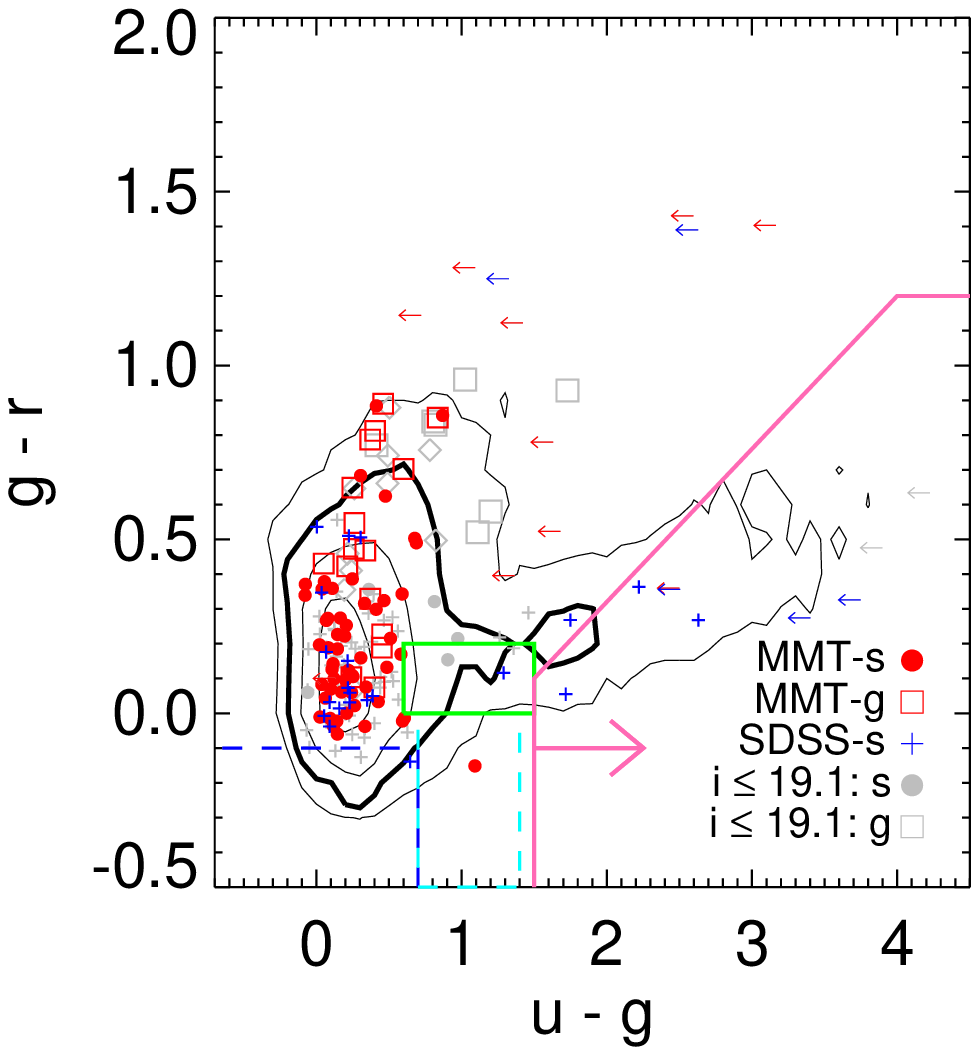} %{color1mt.eps}  %fig11
\includegraphics[scale=0.6]{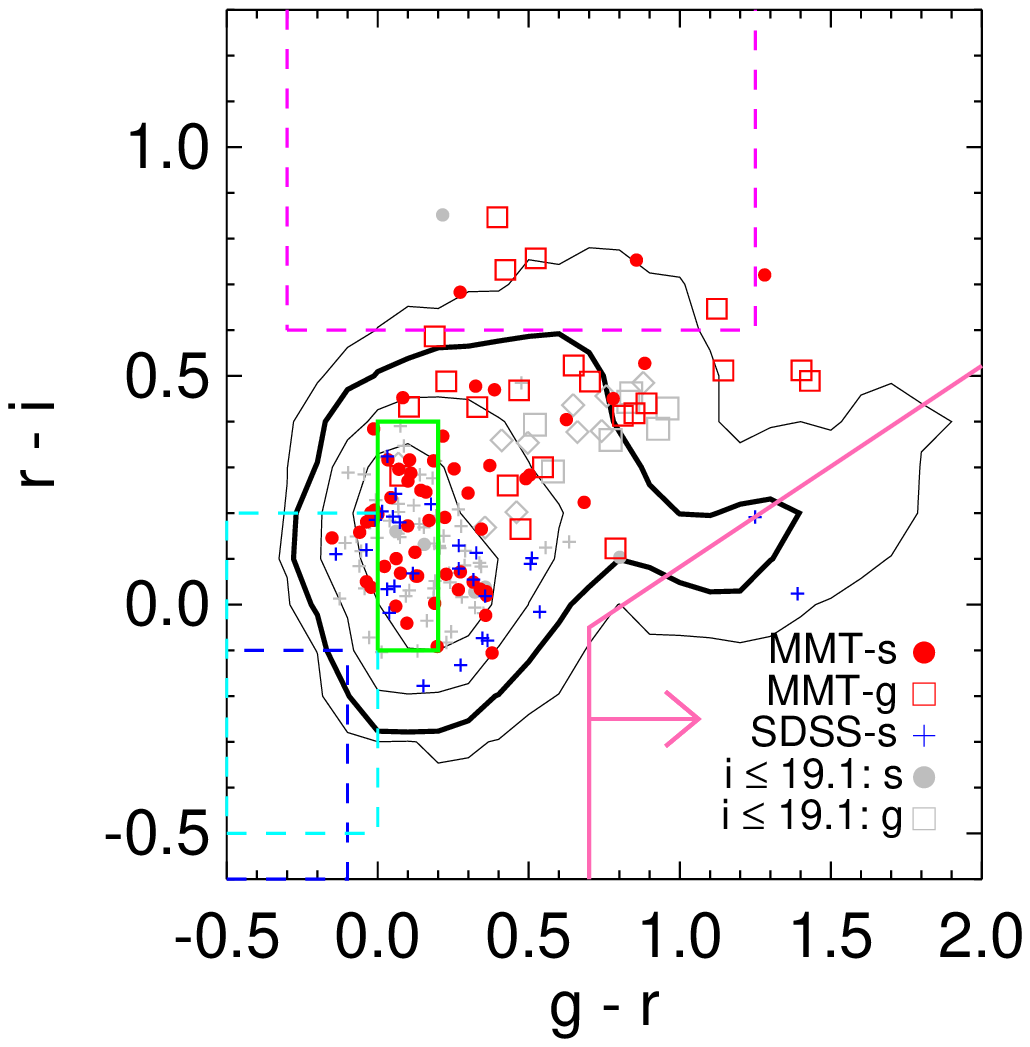} %{color2mt.eps}
\includegraphics[scale=0.6]{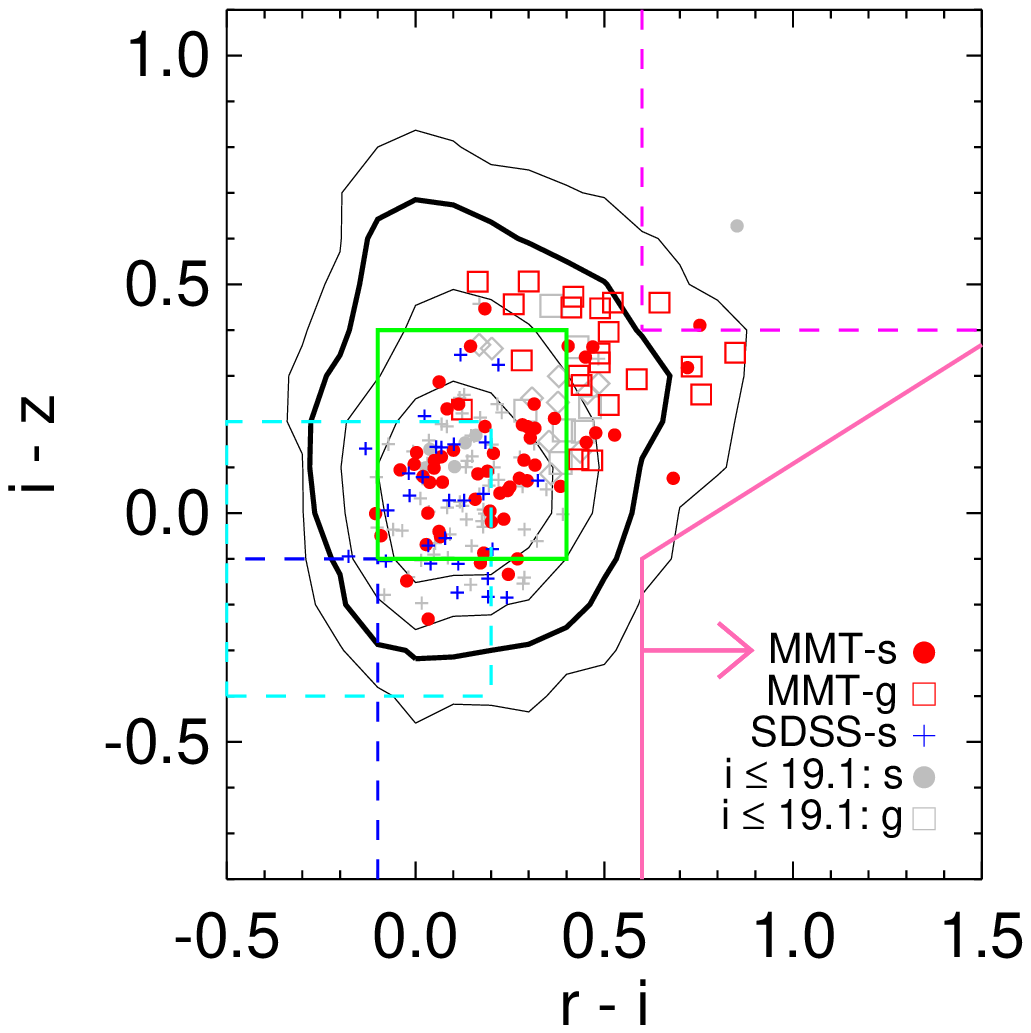} %{color3mt.eps}
\includegraphics[scale=0.6]{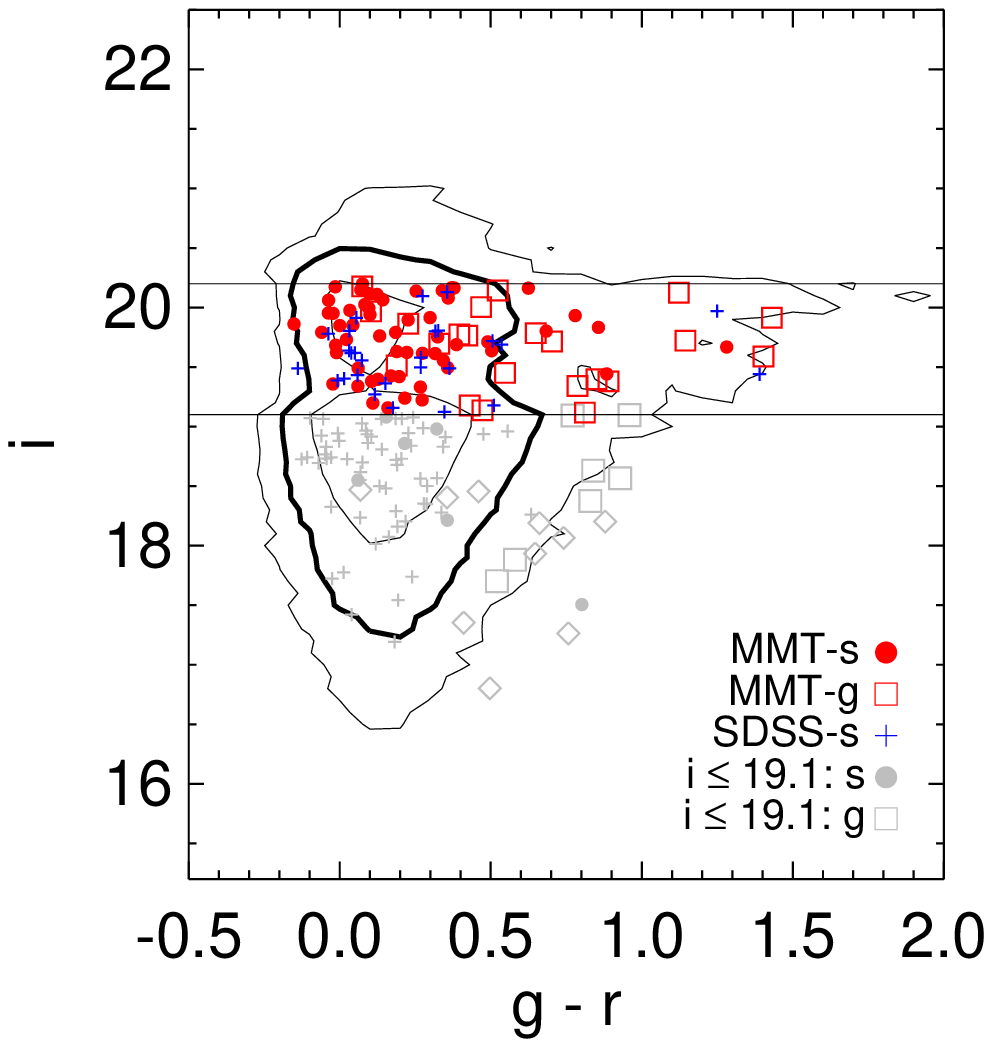} %{color4mt.eps}
\end{center}
\figcaption{Colors and magnitudes of SDSS ($blue$) and MMT ($red$) identified quasars at
$19.1 < i \leq 20.2$ (including all redshifts).
See Fig.~\ref{fig:colors1} for explanation of symbols and lines. 
In $grey$ are the brighter objects from Fig.~\ref{fig:colors1},
with point sources in dots and extended sources in squares and diamonds.
The MMT-subsample show a high fraction of extended sources (28\%). 
A total of 11 MMT quasars falls in the exclusion regions:
6 in the M star and white dwarf exclusion region,
of which 2 are point sources;
5 in the A star exclusion region,
all of which are point sources.
\label{fig:colors2}}
\end{figure}

In the fainter $19.1 < i \leq 20.2$ bin, 
SDSS applied different $griz$ color cuts to select high $z$ ($>$ 2.9), point source targets.
In this magnitude bin, MMT discovered 80 new objects (73 MMT09 and 7 MMT05b), 
the majority of which are at $z < 2.9$,
and are outside of the SDSS selected $z$ regions (R02). 
Of the 2 MMT objects that qualify the SDSS $z$ cut,
only one is a point source and could have be added to the SDSS completeness analysis. 
Therefore, it is still valid to consider the SDSS selection complete to $\sim$90\% at $z > 2.9$  (Table~\ref{tab:completeness}). 
Most ($>$90\%) of the low $z$ MMT quasars lie within the contours defined by
the SDSS DR7 quasars and satisfy the SDSS color-color selections,
though $\sim$30\% of them are extended and would have been rejected 
had SDSS explored this low $z$ regime (Fig.~\ref{fig:colors2}).

%A total of 11 MMT sources would have been rejected as they reside in 
%the SDSS exclusion zones (marked by dashed lines in Fig.~\ref{fig:colors2}, See also Table~\ref{tab:exclusion}):
%6 in the M star $+$ white dwarf exclusion region, 
%of which 4 are extended sources;
%5 in the A star exclusion zone and all are point sources.  % 0 in A star zone%
%SDSS employed additional $griz$ cuts to 
%remove ``faint, low-redshift'' quasars (Eq(1) from R02),
%which covers 4 MMT objects. 
%The remaining 45 MMT point source quasars should qualify for
%the SDSS color-color selection but were not selected by SDSS
%for unknown reasons.

\begin{figure}[ht]
\begin{center}
\includegraphics[scale=0.6]{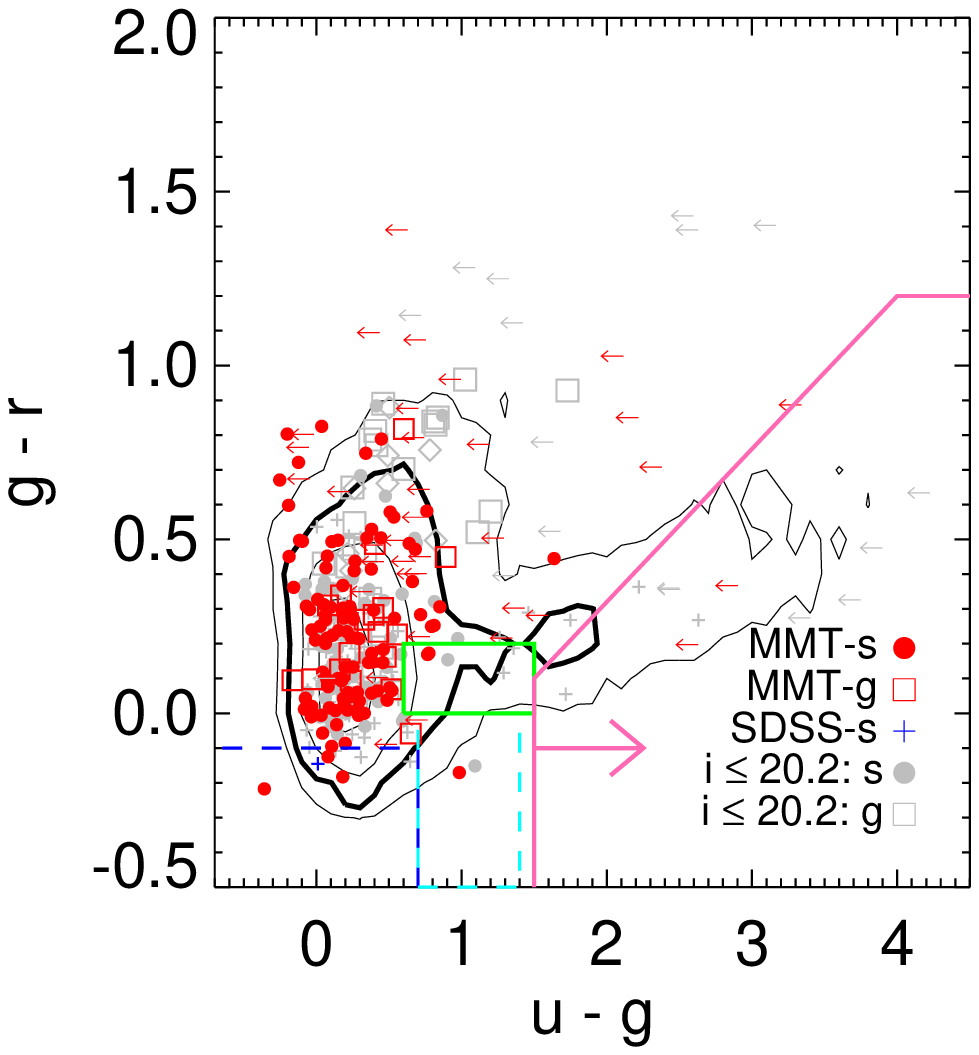} %{color1ft.eps} %fig12
\includegraphics[scale=0.6]{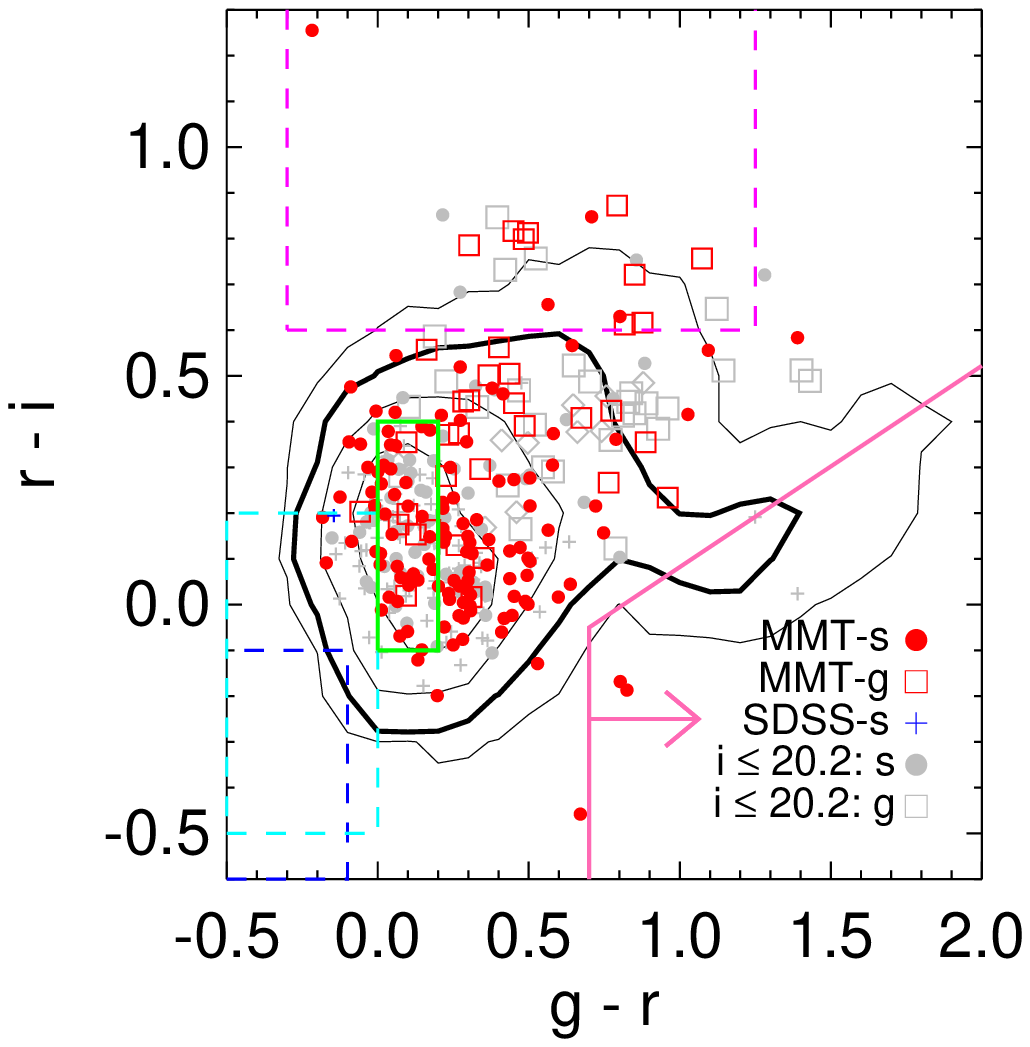} %{color2ft.eps}
\includegraphics[scale=0.6]{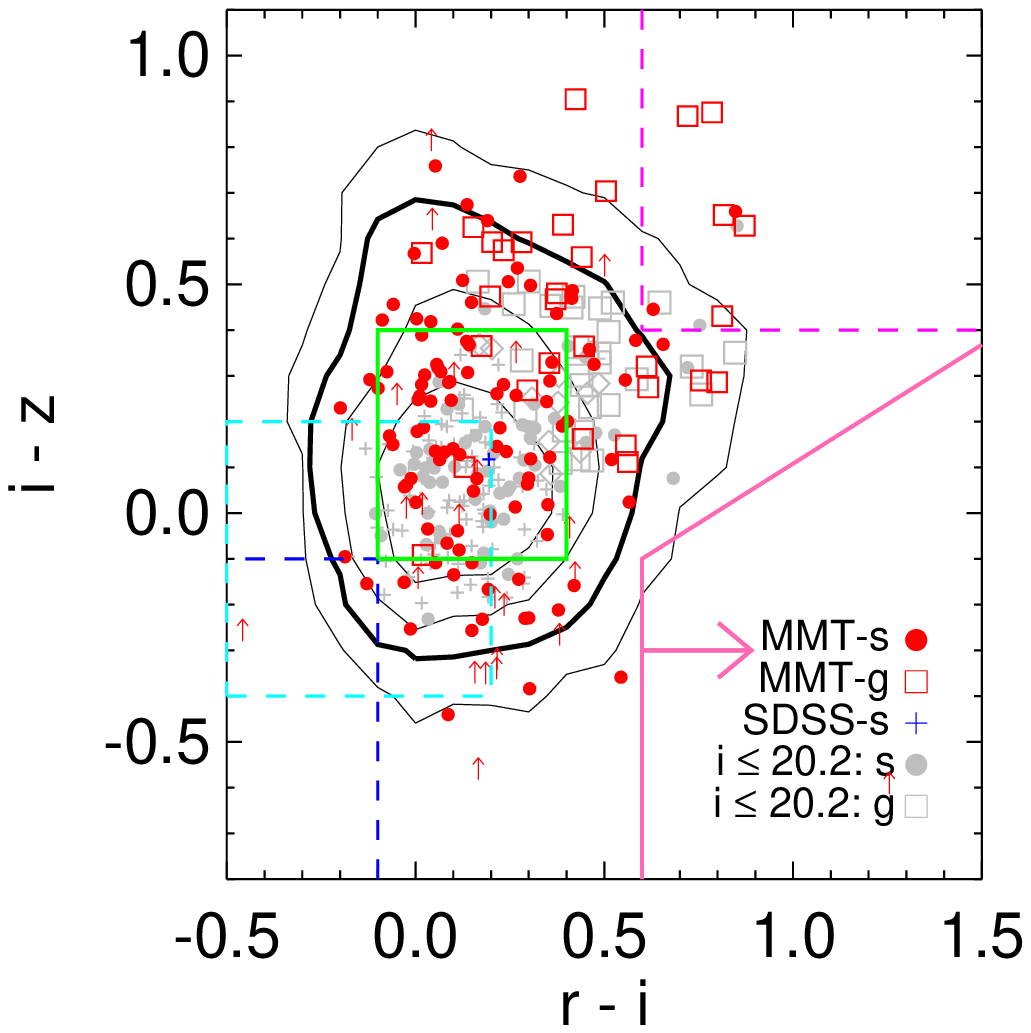} %{color3ft.eps}
\includegraphics[scale=0.6]{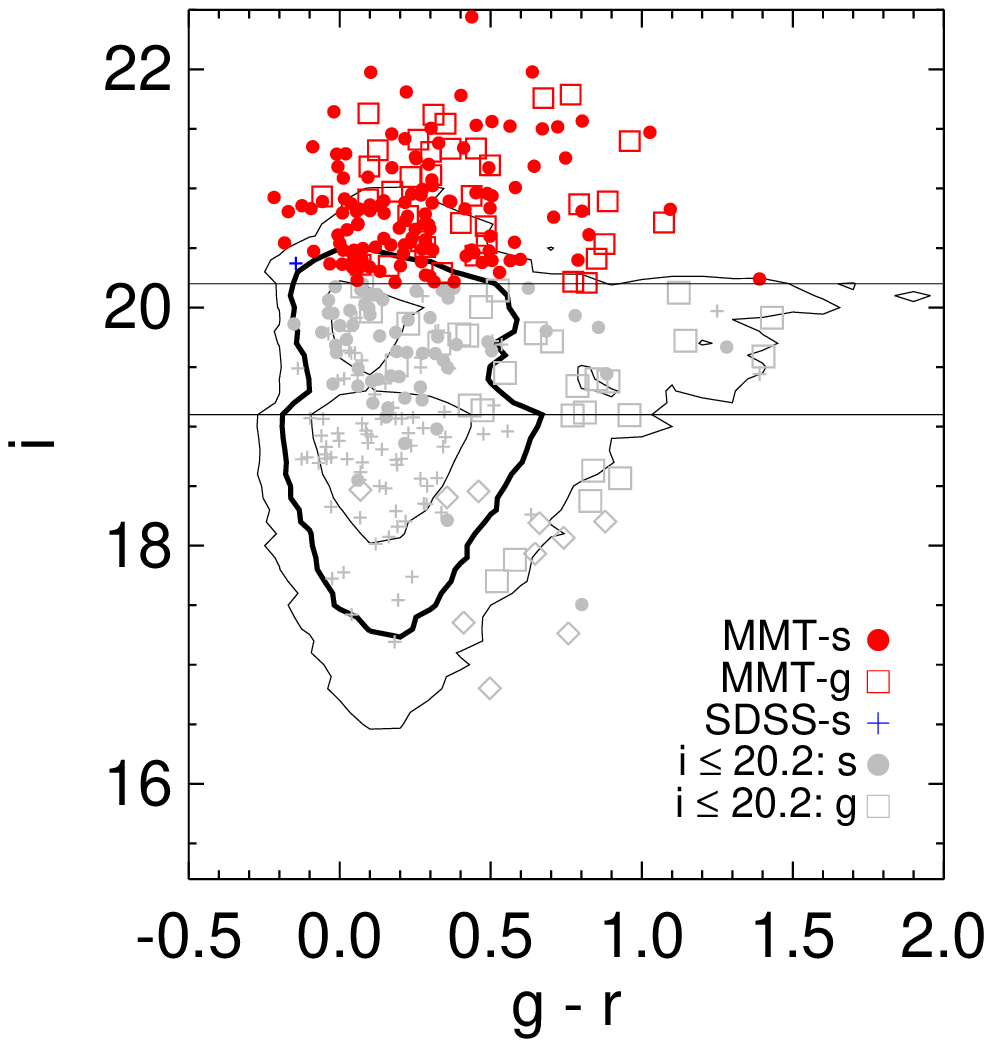} %{color4ft.eps}
\end{center}
\figcaption{Colors and magnitudes of SDSS ($blue$) and MMT ($red$) identified quasars at
the faint end of $i > 20.2$.
See Fig.~\ref{fig:colors1} and Fig.~\ref{fig:colors2} for for explanation of symbols and lines. 
In $grey$ are the brighter objects at $i \leq 20.2$.
%MMT quasars dominate this magnitude regime 
%as SDSS selection was not as deep. 
The MMT-subsample at $i > 20.2$ consists of
faint sources not covered by SDSS; 
and also shows a high fraction of extended sources (23\%).
A total of 25 MMT quasars falls in the exclusion regions:
4 point sources and 9 extended objects in the M star and white dwarf exclusion region;
9 point sources in the A star exclusion region;
%(plus another 4 upper limits);
2 point sources in the white dwarf exclusion region;
and 1 point source in the white dwarf and A star overlapping region. 
The remaining 133 objects also qualify the SDSS color selection without $z$ 
or morphological cuts.
\label{fig:colors3}}
\end{figure}

\begin{figure}[ht]
\begin{center}
\includegraphics[scale=0.6]{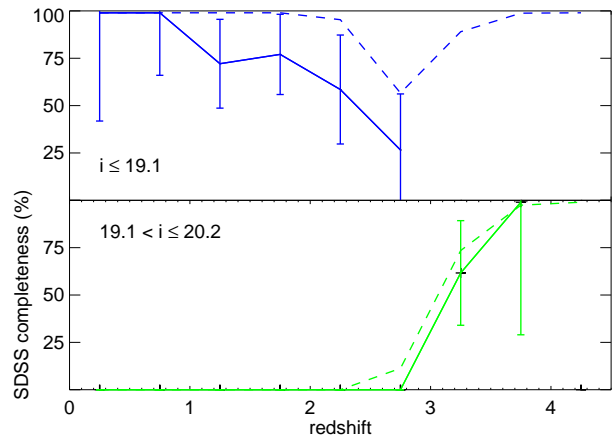} %{comp-num-all-new.eps} %fig13
\end{center}
\figcaption{
%{$top$} Number counts of SDSS ($blue$) and MMT ($red$) identified quasars at
%$i \leq 19.1$ and $19.1 < i \leq 20.2$ used for completeness analysis.
%In $grey$ is the overall distribution of quasars in each redshift bin.
%Both point source and extended sources are included.
%{$bottom$} 
The observed SDSS quasar selection completeness after including qualifying MIR MMT quasars. 
Dashed lines plots the 
simulated completeness for SDSS quasars at a 0.5 redshift bin (Table 6, \citet{rich02a}).
The SDSS completeness at $i \leq 19.1$ drops from an average 90\%
to ($67\pm 8$)\% for the MIR-selected quasars,
but is comparable at 19.1 $< i \leq$20.2.
\label{fig:incomp}}
\end{figure}

In the faintest end ($i > 20.2$), 
which is below the SDSS quasar selection magnitude limit, 
only one SDSS quasar was included
in the DR7 catalog (`52411-0947-531', not color-color selected, uniform flag $= `0$'). 
All of the 160 MMT quasars are newly identified objects.
If compared to SDSS quasars at brighter ends ($ i \leq 20.2$), 
the fainter targets show a large scatter in all colors (Fig.~\ref{fig:colors3}),
including 25 MMT sources in the SDSS exclusion 
zones (marked by dashed lines in the first 3 panels of Fig.~\ref{fig:colors3}, Table~\ref{tab:exclusion}):
13 in the M star $+$ white dwarf exclusion region, 
of which 9 are extended sources;
9 in the A star exclusion zone,  and all are point sources;
2 in the white dwarf exclusion zone,  and both are point sources;
and 1 point source in the white dwarf and A star overlapped exclusion region.
Two other extended objects failed the $M_{\rm i}$ cut. 
All of the remaining 133 targets satisfy the SDSS magnitude and $griz$ or $ugri$ color selections
but not the $z$ or point-source constraints (Table~\ref{tab:completeness}).
As at brighter magnitudes, a significant fraction (22\%) of the MIR quasars are extended, 
of which $\sim$ 70\% (25/36) lie at $z < 1$.

In Fig.~\ref{fig:incomp}, we present the measured completeness 
of the SDSS quasar selections 
as a function of redshift,
only taking into consideration the MMT objects that would otherwise
satisfy the SDSS magnitude ($M_{i} < -22$), 
redshift ($z < $2.9 at $i  < 19.1$, and $z >$2.9 at $19.1 \leq i < 20.2$), 
color ($ugri$ at $z <$2.9 and $griz$ at $z >$2.9), 
and morphology (point source only at $i > 19.1$) requirements: 
9 at $i \leq 19.1$ and one at $19.1 < i \leq 20.2$ (Table~\ref{tab:completeness}).
SDSS quasar selection is close to complete at $19.1 < i \leq 20.2$ and $z > 2.9$,
but is overestimated by $\sim$ 20\% at $i < 19.1$ and $z < 2.9$.
The modified SDSS completeness is summarized in Table~\ref{tab:comp}.
These values are corrected
for the spectroscopic completeness of the MMT survey 
---numbers of MMT09 quasars are multiplied by 3.3,
and by 1.5 for MMT05b objects. 
%The priorities assigned to high $S_{\rm 24}$ objects 
%will not greatly affect the number of undetected quasars 
%from spectroscopic completeness
%due to the similar detection rate in MMT09 and MMT05b spectra (Table.~\ref{tab:sample}). 
The corrections could be overestimated 
given the higher priority assigned to brighter 24\,\micron\ objects,
though unlikely by a significant number, 
as similar detection rates are found between MMT09 (30\% complete, 9.1\% detection rate) and 
the more complete MMT05b survey (66\% complete, 9.9\% detection rate). 

\begin{table}[htdp]
\caption{Number counts of MIR-selected quasars identified using SDSS and MMT spectra in the overlapping regions.
Second line in each magnitude bin shows the number of new objects that also satisfy the SDSS selections (see also \S~\ref{sec:i191}).}
\begin{center}
\begin{tabular}{lccccccc}
\hline
\hline
magnitude & $N_{\rm SDSS}$ & ext & point & $N_{\rm MMT}$ & ext & point & total ext\\
\hline
$ i \leq 19.1$           & (62$+$6)* & (4$+$6)* (15\%) & 58 & 13 & 7 (54\%) & 6 & 17 (21\%) \\  %05: 0+0
 && & & (9) & (5) & (4) & \\
$19.1 < i \leq 20.2$ & 27 & 0 & 27  & 80  & 22 (28\%) & 58  & 22 (20\%) \\  % 05: 5 + 2
 & && & (1) & (0) & (1) & \\
 $ i > 20.2 $              & 1  &0 & 1     &160 & 36 (23\%) & 124 & 36 (22\%) \\  % 05: 4 + 16
 & & && (133)$\dagger$ & (25) & (108) & \\
 \hline
Total                       & 96  & 10* (10\%) & 86 (90\%) & 253  & 65 (24\%) & 188 (76\%) & 75 (21\%) \\
\hline
\hline 
\end{tabular}
\end{center} 
Notes: Classification of the `extended' (ext) and `point-source' (point) morphological types are 
based on the SDSS photometry (\S~\ref{sec:type}).
%: object whose difference between
%the piont-spread function (PSF) and either the exponential of de Vaucouleurs magnitudes is greater
%than 0.145 mag are classified as extended, or galaxy-like (SDSS photometric morphology type $=$ 3, `g'). 
%Otherwise they are classified as point-source, or star-like (SDSS photometric morphology type $=$ 6, `s') \citep{rich02a}.
Throughout all magnitude bins, a constant 20\% of the MIR quasars are extended sources.
*: Six (6) are the newly-identified BEL objects with SDSS spectra not in the SDSS DR7 quasar catalog, 
all of which are extended.
$\dagger$: For the objects which would satisfy the SDSS selection at brighter magnitudes, 
no redshifts or point-source cut was applied. 
\label{tab:completeness}
\end{table}

\begin{table}[htdp]
\caption{Number counts of MIR-selected MMT quasars that fall in the SDSS exclusion zone.}
\begin{center}
\begin{tabular}{lccc}
\hline
\hline
SDSS Exclusion & $N_{\rm tot}$ & point & ext \\
\hline
 M star $+$White Dwarf     & 20 &13 & 7\\
 A star                 		 & 14 &0  & 14\\
 White Dwarf 			 & 3   & 0  &  3\\
\hline
\hline
\end{tabular}
\end{center} 
Notes: Numbers are accumulated values, for break-down in each magnitude bins, see 
figure captions of Fig.~\ref{fig:colors1},\ref{fig:colors2},\ref{fig:colors3}.
`Point' and `Ext' referred to point-like sources and extended sources, respectively.
\label{tab:exclusion}
\end{table}

\subsubsection{What makes a complete quasar sample?}
Several factors contribute to the MIR additions to the quasar population
and the biases in the SDSS quasar selection. 
Table~\ref{tab:completeness} summarizes the 
number counts in the 2 magnitude bins
in which SDSS carried out their completeness analysis.
At $i  \leq\ 19.1$, and $19.1 < i \leq\ 20.2$, 
the MMT surveys add 13 and 80 additional
quasars to the SDSS quasar catalog. 
Careful comparison reduces the numbers to 
9 and 1 quasars that also qualify the SDSS selection (Table~\ref{tab:completeness}). 
If we assume a homogeneous number density across all redshifts (R02), 
we find the SDSS completeness is overestimated by an average 20\% 
in $i < 19.1$ quasars at $z < 2.9$ (reported to be $>$ 90\% in R02),
but is comparable to the reported 90\%  for $i > 20.2$ quasars at $3.5 < z < 5.3$ (Fig.~\ref{fig:incomp}). 
This completeness assumption is however not physical,
given the known cosmic evolution of quasar number density \citep{hms05,silverman08},
and therefore should be used with caution.
Other MIR selected samples, e.g. \citet{lacy13}, 
did not show the completeness mismatch found in this paper.
This is because color selections or wedges, both in optical and MIR, 
favor the power-law shaped SEDs \citep{vandenberk01, rich02a,lacy04,stern05,donley12},
and are biased against significant host galaxy contributions, 
the presence of strong emission lines (e.g. PAH),
and other factors such as accretion rates \citep{ogle06} and LINERs \citep{sturm06}.
In contrast, the MIR flux limit applied in this sample, 
selects everything above the corresponding luminosity,
and therefore is not biased against dusty host galaxies or other above mentioned factors.
In the whole 22 deg$^2$ LHS field, 
only 6 quasars in the SDSS catalog were rejected
because of fainter MIR fluxes.
The MIR flux-limited sample provides a complementary way to 
examine the quasar population as a whole, 
being more complete than the color selections.
Of the MIR flux-limited quasars presented in this paper, 
the SDSS selection only recovers 58\% and 10\% of the total population 
at $i \leq 19.1$ and $19.1 < i \leq 20.2$, respectively.

%The modified SDSS completeness resembles the shape from the 
%simulation results in \citet{rich02a},
%but is overestimated at $ i \leq 19.1$at $z < 3$ 
%by a constant 20-30\%.  
%A similar completeness shape is also observed at $ 19.1 < i \leq 20.2$, 
%though a spike is seen at $2.5 < z < 3$, 
%due to the small number statistics (1 MMT and 2 SDSS objects) in this redshift bin
%---the modified SDSS completeness 
%will be 70\% is at $i \leq 19.1$,
%and 38\% at $19.1< i \leq 20.2$ (Fig.~\ref{fig:incomp}, bottom).

A significant fraction (50\% at $i \leq 19.1$, and 28\% at $19.1 < i \leq 20.2$) 
of the newly identified MMT quasars are extended sources (Table.~\ref{tab:completeness}). 
%which were excluded in the SDSS quasar selection at $i > 19.1$. 
SDSS chose not to include extended sources at $i > 19.1$ 
to avoid the contamination of very red, extended objects. 
Their choice was based on the observation that at $z \ge 0.6$, 
the majority of quasars are point sources.
This point-source only selection turns out to be conservative as 70\% of extended targets at $i > 19.1$
have a redshift higher than 0.6. %, though 80\% of them are at $z < 1.5$.
Regardless of apparent magnitude,
a constant fraction of 20\% MIR quasars turn out to be extended sources (Table.~\ref{tab:completeness}),
though the majority (80\%) are of relatively low $z$ and luminosities ($z < 1.5$, log$(L_{\rm bol}) <$ 45.5 $\ergs$, 
Fig.~\ref{fig:Mi}, see also Sec 5, Fig.~\ref{fig:ledd1}).

Another MIR addition to the sample arises from the 
SDSS cut of low $z$ sources in the $19.1 < i \leq 20.2$ bin (Fig.~\ref{fig:Mi}). 
Because of this redshift cut,  a significant number of quasars are missed from the sample, 
as the number density of $z < 3$ quasars 
at $19.1 < i < 20.2$ is $\sim$ 24\,deg$^{-2}$ (corrected for spectroscopic completeness),
more than doubles the $\sim$ 10\,deg$^{-2}$ found at $i < 19.1$. 
% i < 19.1, overlapped, 77 total, 12 mmt09  --> 9.54, w/o MMT05 --> same
% 19.1 < i < 20.2, overlapped,  96 total, 6 mmt05, 73 mmt09 --> 24.49, w/o MMT05 --> 23.66
% i >  20.2, overlapped, 161 total, 20 mmt05, 140 mmt09 --> 45.27, w/o MMT05 --> 42.52
Since the MMT09 survey is 30\% complete (\S\ref{sec:mmt09}) 
and MMT05 66\% complete (\S\ref{sec:mmt05}),
on top of the 80 newly identified MMT quasars, 
roughly %30 quasars may remain undetected at $i  \leq\ 19.1$,
174 quasars may remain undetected at $19.1 < i \leq\ 20.2$. 
The majority (90\%) of the MMT quasars that fall in this $z < 3$ region
also satisfy the SDSS color selections. 

The third MIR addition is the extension to faint targets ($i > 20.2$)  (Table.~\ref{tab:completeness}). 
The faint MIR quasars almost doubled the number of known quasars in this field,
and the majority (80\%) also satisfy the SDSS color selections. 
The completeness corrected number density of $z < 3$ quasars at $i > 20.2$
is $\sim$ 45\,deg$^{-2}$.

Finally, since the MIR selection does not avoid specific color areas, 
such as the SDSS exclusion regions of white dwarfs, M stars, 
and A stars, 
a total of 37 MMT quasars have been recovered (Table~\ref{tab:exclusion}). 
They contribute to $\sim$10\% of the total MIR quasar population.
This is the fourth MIR addition to the SDSS quasar selection criteria.

\begin{table}[htdp]
\caption{Observed SDSS completeness of MIR-selected quasars.}
\begin{center}
\begin{tabular}{lcccccccc}
\hline
\hline
Apparent magnitude &  &&&Redshifts &&&& \\
                                & 0-0.5        & 0.5-1          & 1-1.5          & 1.5-2         & 2-2.5        & 2.5-3            & 3-3.5 & 3.5-4 \\
\hline
%$ i \leq 19.1$           &   100.0   &   100.0   &   72.8   &   77.8   &   59.1   &   26.8   &   100.0  & ... \\%
$ i \leq 19.1$           &   100.0   &   100.0   &   72.8   &   77.8   &   59.1   &   26.8   &   ...  & ... \\%
                               &   ( 100.0    &  100.0    &  100.0    &  100.0  &   96.3   &   57.2    &  89.9   &   99.8 )\\
%$19.1 < i \leq 20.2$ &    10.9   &   8.7   &   8.8  &   6.1  &   4.6    &  100.0   &  62.3  & 100.0\\ 
$19.1 < i \leq 20.2$ &    ...   &   ...   &  ...  &  ...  &   ...   &  ...   &  62.3  & 100.0\\ 
			 &   (      0.0    &  0.0  &    0.0  &    0.0  &    0.0  &    11.4   &   74.2  &    98.4  ) \\
%$ i > 20.2 $              & 0.0   &   0.0   &   0.0   &   1.1  &   0.0   &   0.0   &   ...   & ... \\
%			  &   ( 0.0   &   0.0   &   0.0  &    0.0  &    0.0  &    10.0   &   64.2  &    97.0  ) \\
\hline
\hline
\end{tabular}
\end{center} 
Notes: Numbers are in percentage. 
In parenthesis is the SDSS simulated completeness from Table 6 in \citet{rich02a}.
\label{tab:comp}
\end{table}

\begin{table}[htdp]
\caption{Number of Gaussians used in fits with and without an F-test.}
\begin{center}
\begin{tabular}{lccc}
\hline
Emission line & $N_{\rm Gaussian}$ & $N_{\rm obj}$  & $N_{\rm obj}$ \\
 & & without F-test & with F-test \\
\hline
\civ & 1 & 30 (21\%) & 34 (24\%)\\
(143)      & 2 & 33 (23\%)& 66 (46\%)\\ 
      & 3 & 80 (56\%) & 43 (30\%)\\
\hline
\mgii & 1 & 75 (26\%) & 201 (71\%)\\
(285)        & 2 & 50 (18\%)  & 77 (27\%)\\ 
        & 3 & 160 (56\%) & 7 (2\%)\\
\hline
\hbeta& 1 & 8 (10\%)    & 70 (94\%)\\
(75)          & 2 & 66 (88\%)  & 4 (5\%)\\ 
          & 3 & 1 (2\%)      & 1 (1\%)\\
\hline
\end{tabular}
\end{center} 
\label{tab:ftest}
\end{table}

\begin{table}[htdp]
\caption{Wavelength ranges used for spectral measurements in rest frame.}
\begin{center}
\begin{tabular}{lccccc}
\hline
Emission line & redshift range & Continuum ($\AA$) & Fe Template & Emission ($\AA$) \\
\hline
\civ & 1.63 $< z < $ 4.39 & [1445, 1465] \& [1700, 1705] & ...  & [1500,1600] \\
\mgii & 0.43 $< z < $ 2.10 & [2200, 2700] \& [2900, 3090] & VW01 & [2700, 2900] \\
\hbeta & $z < 0.76$ & [4435, 4700] \& [5100, 5535] & BG92 & [4700, 5100] \\
\hline
\end{tabular}
\end{center}
Notes: The redshift ranges are the MMT \& SDSS accessible ranges based on their 
secure spectral ranges (See \S~\ref{sec:bel}).
VW01, \citet{vandw01}; BG92, \citet{bg92}. 
\label{tab:fitregion}
\end{table}

\section{MEASUREMENTS of SPECTRA}
\label{sec:spec}

Different virial SMBH mass ($\mbh$) estimators have used different line width parameters,
with either FWHM (`full-width-half-maximum', in $\kms$) or 
line dispersion, i.e., the second moment of the emission-line profile.
FWHM is easier and more straightforward to measure,
but can be easily overestimated in cases of line blending or extended wings.
Line dispersion ($\sigma_l$), on the other hand, 
has relatively lower uncertainties,
but may be overestimated for specific line profiles.
Unfortunately, both parameters are affected by measurement errors,
and can provide unreliable estimates for 
low S/N ($<$10) spectra \citep{denney13}.
This problem can be circumvented via model fits,
and Gaussian functions are widely used to fit the BELs. 
All the BH mass estimators we use later  (MC04, VP06, VO09, and S11) are 
based on either one or both the FWHM and $\sigma_l$ of the emission line.
The line dispersion $\sigma_{l}$ is arguably 
more reliable,
given its better consistency between different lines \citep{park13,denney13},
and its better scaling to the 
widely used empirical $\mbh - \sigma_{*}$ relation \citep{tremaine02}.
Because the line broadening can be due to several components,
a straightforward measurement of $\sigma_{l}$ is complicated, 
and for this work we decided to use the FWHM of the continuum subtracted emission line 
as the line width proxy.
For a Gaussian, the FWHM has a simple correlation with $\sigma_l$, 
as FWHM $= 2 \sqrt{2ln2} \sigma_l$, or 2.35 $\sigma_l$. 
If only one Gaussian is used then 
the FWHM and $\sigma_l$ will be linearly correlated.
If multiple Gaussians are used, 
the $\sigma_l$ will give a higher equivalent value
than the dominant FWHM.
We do provide the $\sigma_l$ measurements in the machine-readable table.
%(for example see Table~\ref{tab:param}, column 10).

\begin{figure}[ht]
\begin{center}
\includegraphics[scale=0.35]{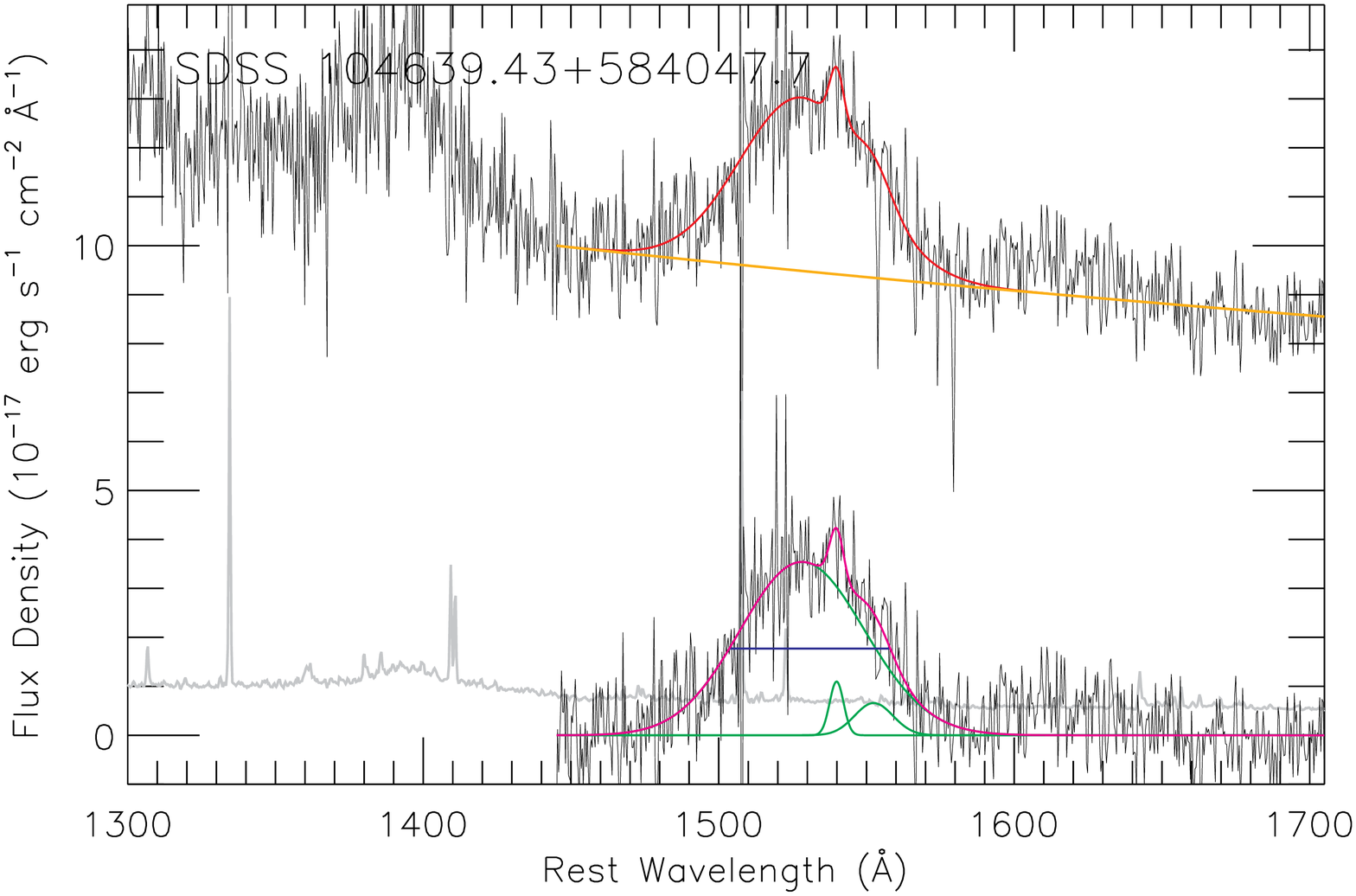} %{shen-civ-0948-563.eps} %fig14
\includegraphics[scale=0.35]{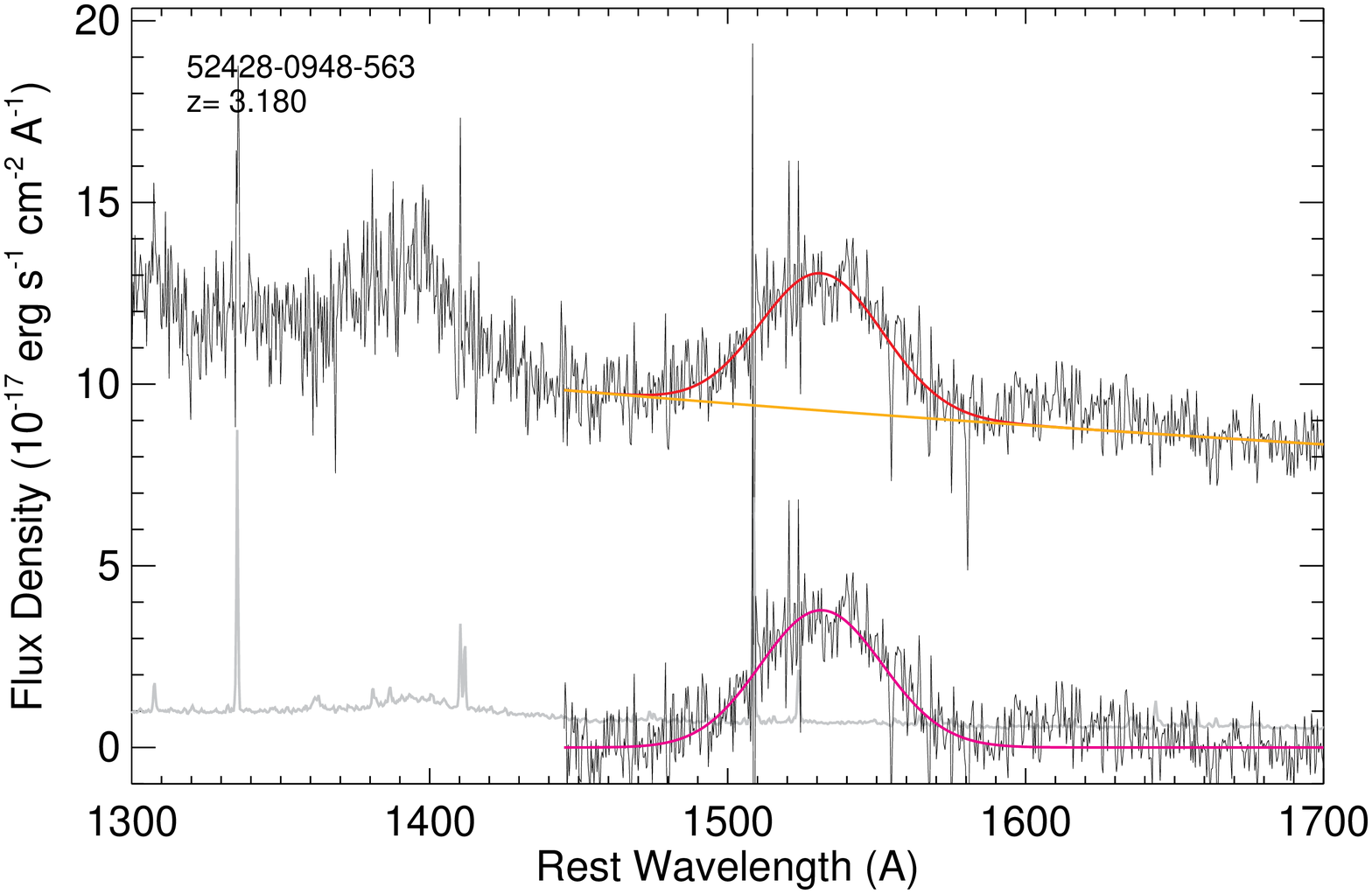} %{dai-civ-0948-563.eps}
\end{center}
\figcaption{Example of comparison of the \civ\ BEL fitting results with and without an F-test (smoothed over 2 pixels).
Top panel shows the SDSS results from \citet{shen11},
where no F-test was used and 
the emission line was fitted with 3 Gaussian components; 
Bottom panel shows the same object,
since the F-test shows a confidence level of 0.984,
we abandoned the additional Gaussian component
and kept only a single Gaussian for the BEL profile. 
Same color codes as in S11 are used to guide the eye. 
Upper and lower black lines in each panel
show the original and continuum-subtracted spectra.
The {\bf \color{gray}gray} line in the lower spectra is the flux-density errors. 
In {\bf \color{orange}orange} is the continuum,
covered by the composite spectra in {\bf \color{red}red} except for the emission line region.
In {\bf \color{green} green} are the broad Gaussians used for the BEL (covered by the composite spectra in the bottom panel).
The composite spectra of the emission line is in {\bf \color{magenta} magenta}.
The S11 has a dominant Gaussian FWHM of 9728 $\pm$ 506 \kms\ ,
which is consistent with our results of 9409 $\pm$ 282 \kms\ . %one fit result, 50 mock result is : 9161+-384
The equivalent width (EW) results are also consistent
(S11: 21.7 $\pm$ 1.4; this work: 20.3 $\pm$ 1.4).  %50 mock result used, one fit result: 20.8 +- 0.5
The additional Gaussian components in the SDSS fits 
are not necessary for this object.
\label{fig:ftest}}
\end{figure}

We wrote an IDL procedure that
first measures and subtracts the continuum,
and then fits one or more Gaussian profiless to the emission line.
The procedure is based on the code used for the SDSS quasar catalog (S11),
but includes more generality. 
In the cases where a single Gaussian is not a good fit to the line profile,
up to 3 Gaussian components are allowed. 
An F-test is used to evaluate the need for each
additional component.
The F-test is widely used to compare the best fits
of different models based on least squares comparison and the F distribution. 
The F value is computed as: \\
\begin{equation} \label{eqn:ftest}
F=\frac{\chi^{2} - \chi_{\rm new}^{2}}{\rm DOF - DOF_{new}}/\frac{\chi_{\rm new}^{2}}{\rm DOF_{new}}, 
\end{equation}
where DOF is the number of degrees of freedom for the variance \citep{numr}. 
We compute the F-test values using
the IDL mpftest program\footnote{http://cow.physics.wisc.edu/~craigm/idl/idl.html}. 
In each case, we allow up to 3 Gaussians for the BEL
and use an F-test confidence level of 0.999 as the threshold. 
Only in cases where the F-test threshold is met,
which means the new fit is significantly different from the old one, 
will the extra broad component be kept.
Fig.~\ref{fig:ftest} shows the fitting results of the same object with and without an F-test.
This procedure differs from the SDSS approach,
where as long as the new $\chi^2$ is smaller,
an additional Gaussian component is added.
Since the use of Gaussian profile(s) has no physical basis, 
we argue that the number of Gaussians should be minimized
except in special cases (BALs \& NALs, see Section.~\ref{sec:discussion}).

The introduction of an F-test significantly 
decreases the number of Gaussian components needed
for the emission line fits (Table~\ref{tab:ftest}). 
The percentage of objects that need 
more than one Gaussian component
drops significantly from 94\% to 6\% for \hbeta;
and from 74\% to 29\% for \mgii. 
However, for \civ, this percentage remains high at 76\%, 
partly due to the frequently observed asymmetry in the highly ionized \civ\ BELs. 

We measure the FWHMs in the quasar
optical spectra for the main BELs:  \hbeta, \mgii, and \civ.
First, the continuum is fitted with a power-law to
the emission line-free region (Table~\ref{tab:fitregion}).
\feii\ can be strong and broad due to many multiplets,
especially in the vicinity of \mgii\ and \hbeta\ lines. 
Therefore the \feii\ emission template is also used in the continuum fit
for \mgii\ and \hbeta. 
The continuum fit wavelength windows are
chosen such that there is no contamination from the tail of the BEL component.
We adopt the optical FeII template from \citet{bg92} for \hbeta\ , 
and the UV \feii\ \& FeIII templates from \citet{vandw01} for MgII.
No iron template is used for \civ, 
since the iron emission is generally weak in the \civ\ band. 
For \hbeta\ and \mgii, the continuum and iron removal could be S/N dependent. 
In cases where the S/N of the spectra is limited (average S/N per pixel$< 4$), 
the iron line removal is not feasible,
and for these objects we only fit a power-law continuum.
This affects only 3\% of the objects with a \mgii\ fit,
and 8\% of the objects with an \hbeta\ fit.

Up to 5 parameters are fitted simultaneously for the continuum: 
continuum normalization (A$_{\lambda}$) and continuum slope ($\alpha_{\lambda}$);  
for \hbeta\ and \mgii,  
\feii\ template normalization (A$_{Fe}$), \feii\ Gaussian line-width ($\sigma_{Fe}$),
and \feii\ velocity offset (${\rm v_{off}}$) relative to the redshift. 
We then fit up to 3 Gaussians to the emission lines
allowing for velocity offsets (BEL central wavelength), 
linewidth (FWHM \& $\sigma_{l}$), 
and equivalent width (EW) measurements.
Each Gaussian is fitted with 3 parameters: maximum value (factor), 
mean value (central $\lambda$), and standard deviation ($\sigma$).
In the case of broad or asymmetric emission lines where multiple 
Gaussian components are used, 
we provide two sets of linewidths:
the `dominant'  FWHM --- associated with the 
major component with the highest intensity;
and the `non-parametric' FWHM --- of the composite line profile.
The dominant FWHM increases by an average $\sim$30\% 
after introducing the F-test,
since fewer Gaussian components are used
to reconstruct the emission line profile --
this will increase the derived $\mbh$ (See \S~\ref{sec:virial}).
Yet the shift is usually within or around 1$\sigma$ of the 
FWHM error, and therefore the dominant FWHM after F-test
is in general consistent with the values without the test.

\begin{figure}[ht]
\begin{center}
\includegraphics[scale=0.4]{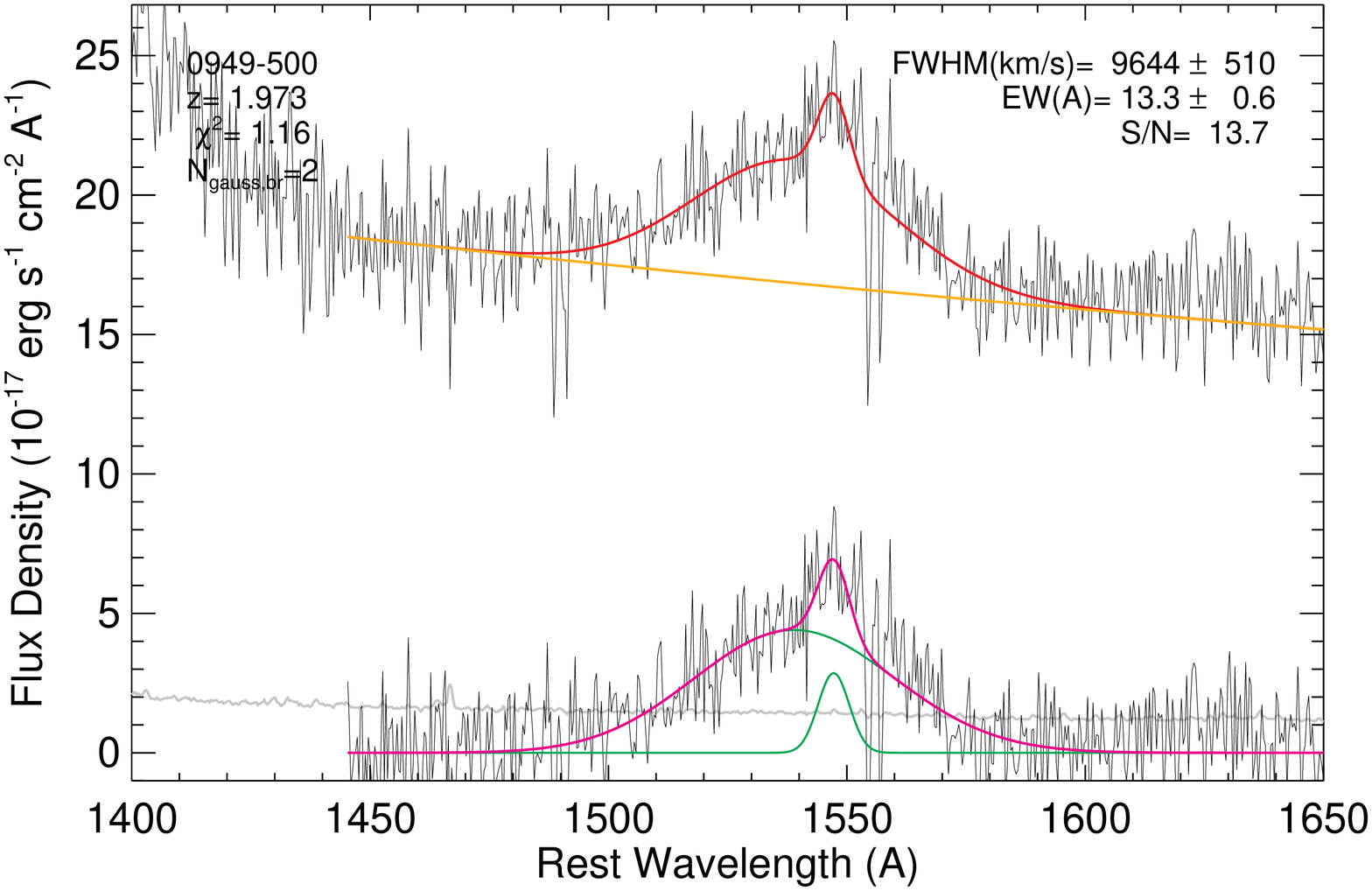} %{absold-Civ-0949-500.eps}  %fig15
\includegraphics[scale=0.4]{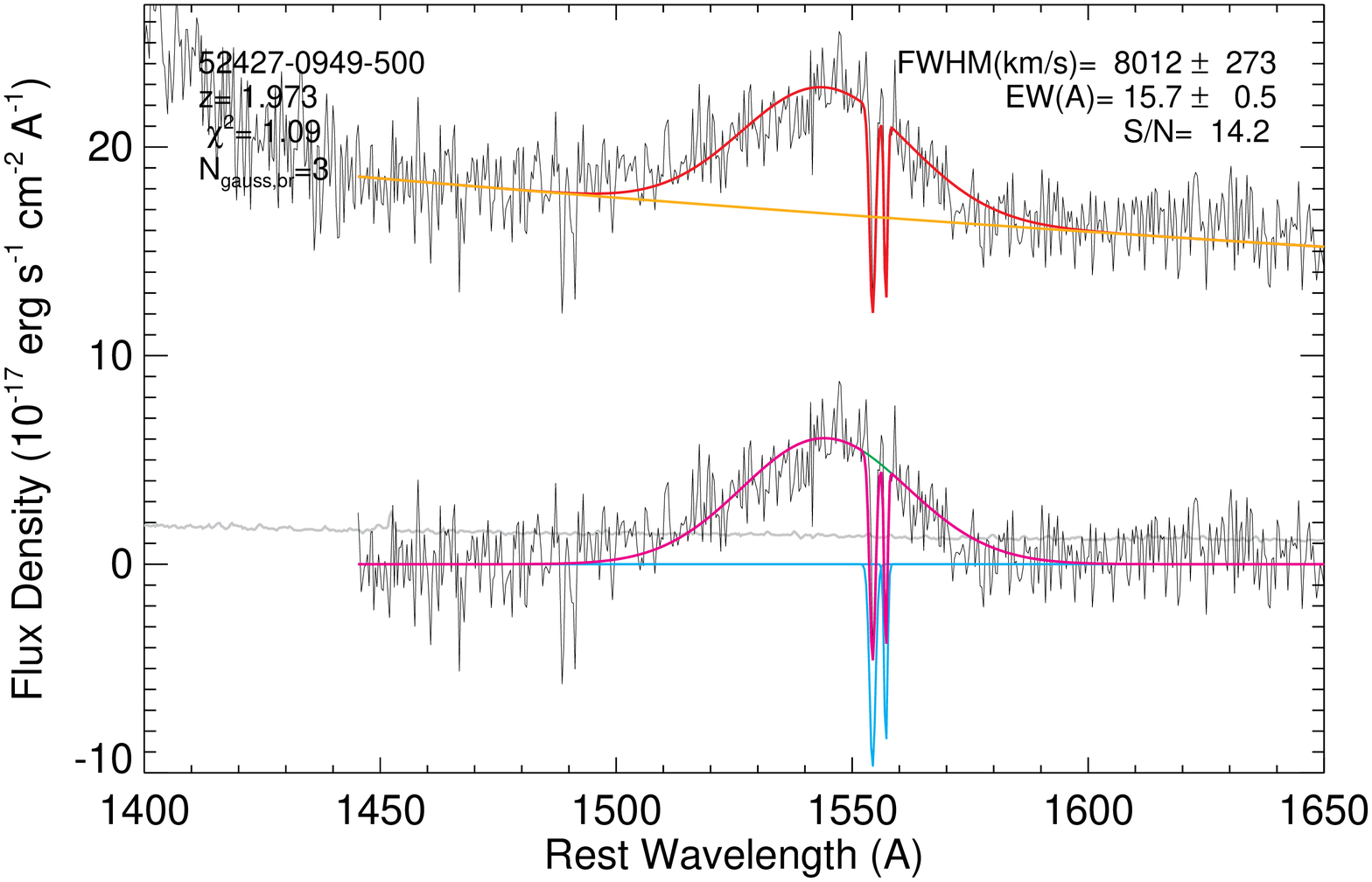} %{absnew-Civ-52427-0949-500.eps}
\end{center}
\figcaption{Example of a quasar with automatically (top) and manually (bottom) fitted
 \civ\ line profile (smoothed over 2 pixels).
The manual fit accounts for the absorption feature,
and better constrains the FWHM of the BEL.
Upper and lower black lines in each panel
show the original and continuum-subtracted spectra.
The {\bf \color{gray}gray} line in the lower spectra is the flux-density errors. 
In {\bf \color{orange}orange} is the continuum,
covered by the composite spectra in {\bf \color{red}red} except for the emission line region.
In {\bf \color{green} green} are the broad Gaussians used for the BEL,
and in {\bf \color{cyan}cyan} the absorption feature---a \civ\,$\lambda\lambda$4959,\,5007 doublet 
is clearly seen redshifted from the BEL peak.
The composite spectra of the emission line is in {\bf \color{magenta} magenta}.
\label{fig:egabs}}
\end{figure}

Both narrow absorption line (NAL) and broad absorption lines (BAL, FWHM $>$ 1000 $\kms$) 
are commonly observed in the \civ\ and \mgii\ 
BELs for MIR-selected quasars.
NALs and BALs can affect the standard multiple Gaussian fitting algorithm 
and therefore need to be treated separately.
If absorption features---NALs and BALs---are observed, 
the spectra are manually fit individually. 
This approach is adopted to retrieve as accurately as possible the 
line width measurement.
Fig.~\ref{fig:egabs} shows an example with absorption feature before and after the manual fit.
Since the FWHMs of the emission lines are manually measured
after subtracting the absorption features,
they lack error bars.
They will be used for $\mbh$ analysis but are flagged in the catalog. 
More discussion can be found in \S~\ref{sec:error} and \S~\ref{sec:discussion}, 
and 
in a forthcoming paper on the absorption features in MIR quasars (Dai \etal, in preparation).

\subsection{\civ}
\begin{figure}[ht]
\begin{center}
\includegraphics[scale=0.5]{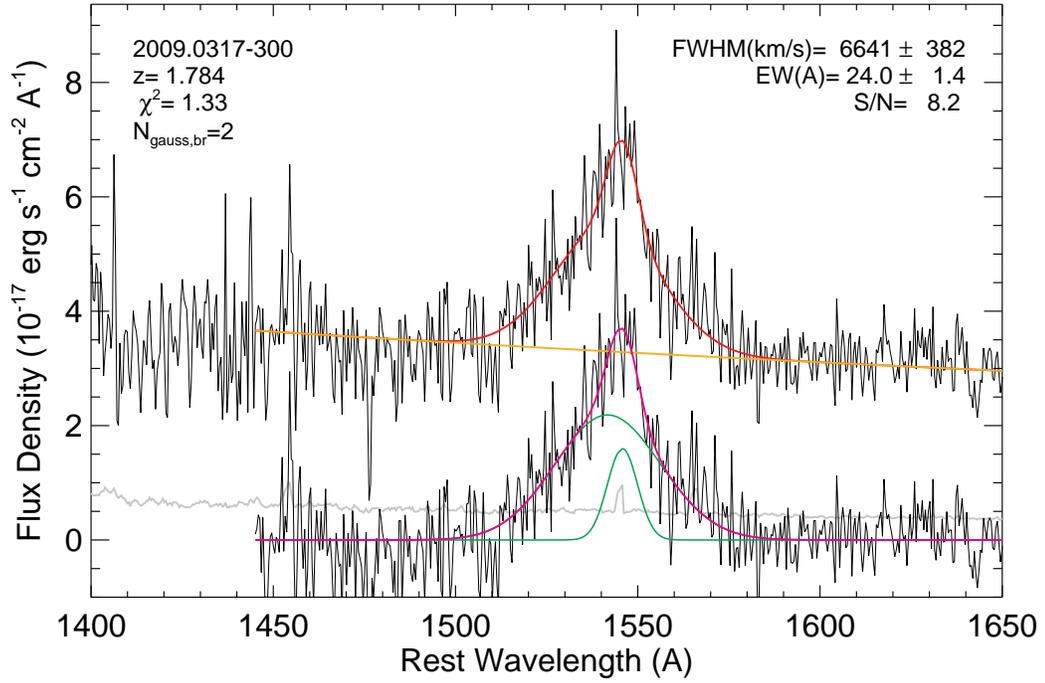} %{egciv-2009.0317-300.eps} %fig16
\end{center}
\figcaption{Example of the spectral fit for \civ\ BEL (smoothed over 2 pixels). 
Upper and lower black lines show the original and continuum-subtracted spectra.
Top left shows the redshift, $\chi ^2$ of the fit, 
and the number of Gaussians used in the broad line fits;
top right is the fitting results of the dominant FWHM, EW, and median S/N of the emission line region.
In {\bf \color{gray}gray} is the flux-density errors. 
In {\bf \color{orange}orange} is the continuum, %and {\bf \color{blue}blue} 
covered by the composite spectra in {\bf \color{red}red} except in the BEL region.
In {\bf \color{green} green} are the Gaussian components for the BEL.
The composite spectra of the emission line is in {\bf \color{magenta} magenta}.
The `dominant' FWHM is from the broader Gaussian in green,
while the `non-parametric' FWHM is from the composite line profile in magenta.
As shown in this case, 
the `dominant' FWHM is commonly broader than the `non-parametric' FWHM 
in 70\% of the targets with multiple Gaussians. 
\label{fig:egciv}}
\end{figure}

The \civ\ line is fitted for the 143 objects with $1.63 < z < 4.40$.
Iron contamination is not significant for \civ, 
hence, only a two parameter (A$_{\lambda}$, $\alpha_{\lambda}$) 
power-law continuum fit is used.
We subtract the continuum fit to the line-free regions, %[1445, 1465]\,$\AA$ and [1700, 1705]\,$\AA$ windows (S11), 
and then fit the \civ\ emission line (Table~\ref{tab:fitregion}). %in the [1500,1600]\,$\AA$ window.
We did not subtract a narrow \civ\ from the line profile
because it is still debated whether 
a narrow \civ\ component is present \citep{wills93, marziani96, sulentic07},
and to be comparable with other studies \citep[e.g.VP06, S11, ][]{assef11, park13}.
For the same reason, we did not fit the 1600\,$\AA$ feature, either \citep{laor94, fine10}.
It is common ($>$70\% ) that more than one Gaussian component
is required (Table~\ref{tab:ftest}) to fit the $\civ$ BEL profile
in each of the subsamples: 48/61 for SDSS,  56/75 for MMT09,
and 5/7 for MMT05b.
In $\sim$ 40\% of the $\civ$ emission lines,
NALs or BALs are seen in or adjacent 
to the BEL profile. 
Fig.~\ref{fig:egciv} shows an example of a typical \civ\ fit.

\subsection{\mgii}
\begin{figure}[ht]
\begin{center}
\includegraphics[scale=0.5]{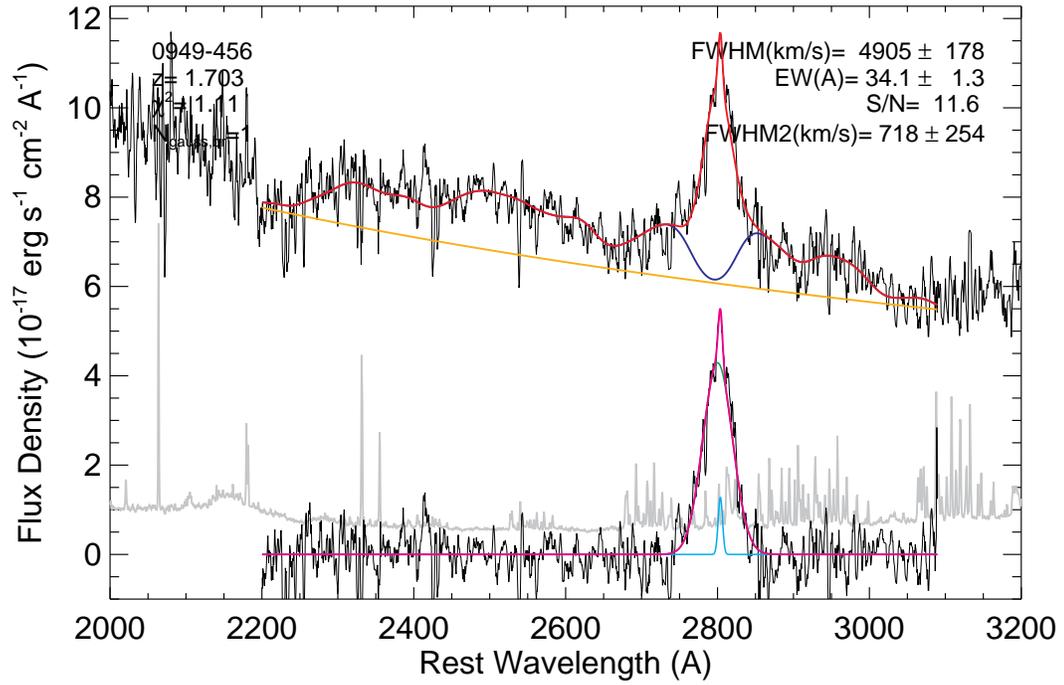} %{egmg.eps} %fig17
\end{center}
\figcaption{Example of the spectral fit for \mgii\ BEL (smoothed over 2 pixels). 
Upper and lower black lines show the original and continuum and $Fe$ template-subtracted spectra.
Colors and legends are explained in Fig.~\ref{fig:egciv},
with {\bf \color{violet}purple} curve showing the $Fe$ template,
mostly covered by the composite spectra in {\bf \color{red}red} except in the BEL region.
In {\bf \color{cyan}cyan} is the \mgii\ narrow emission component ($FWHM < 1200\,\kms$),
whose FWHM is marked by FWHM2 in the legend.
\label{fig:egmg}}
\end{figure}

The \mgii\ line is fitted for the 285 objects with $0.43 < z < 2.10$.
We adopt the iron template from \citet{vandw01} and fit the 
continuum plus iron template to the emission-line free region (Table~\ref{tab:fitregion}). 
In 9 sources with \mgii\ coverage,
the iron template is not constrained due to low spectra quality (S/N per pixel $<$ 4),
in which only power-law continuum was subtracted.
When the \mgii\ emission line is fit,
the \mgii\,2796, 2803 $\AA$ doublet ($\sim750 \kms$ at rest-frame) 
is not taken into account
given the much greater FWHM of the \mgii\ emission line in all cases.
As it is still debatable whether a narrow \mgii component should be removed from 
the BEL profile \citep[][]{md04, vo09, wang09},
we provide two sets of measurements, 
(1) with and (2) without a single Gaussian for the narrow component ($< 1200\,\kms$).
Objects that need multiple broad components are $\sim$ 30\% (Table~\ref{tab:ftest})
in each of the subsamples:
25/81 for SDSS,  53/183 for MMT09, and 6/21 for MMT05b.
NALs are seen in $\sim$ 8\% of the objects.
Fig.~\ref{fig:egmg} shows an example of a typical \mgii\ fit.

\subsection{\hbeta}
\begin{figure}[ht]
\begin{center}
\includegraphics[scale=0.5]{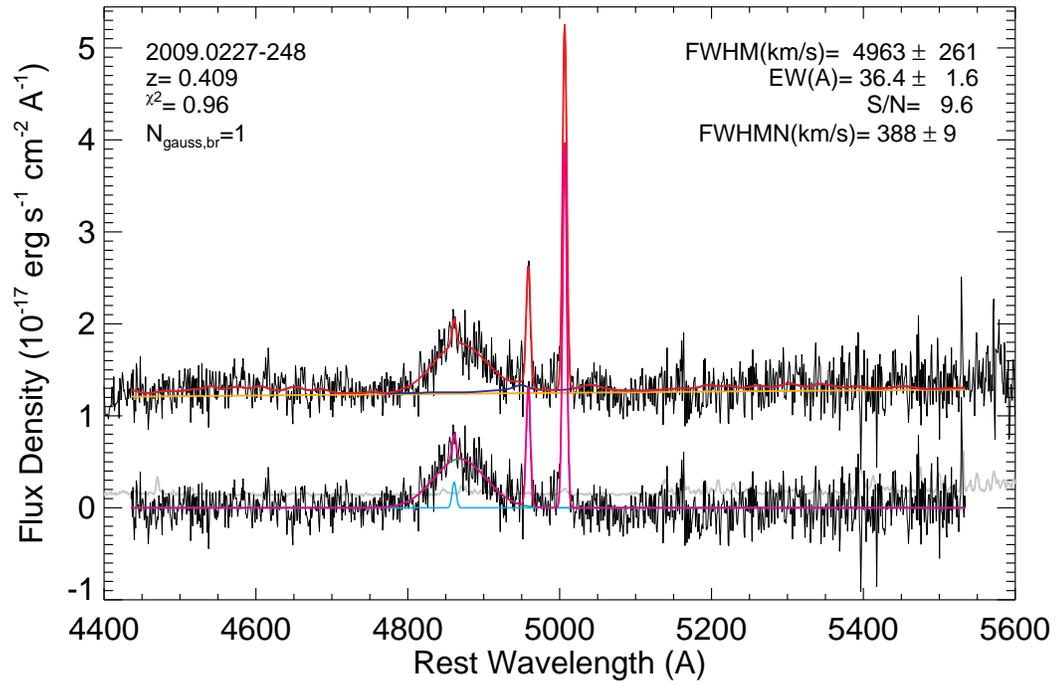} %{eghb.eps}  %fig18
\end{center}
\figcaption{Example of the spectral fit for \hbeta\ BEL (smoothed over 2 pixels)s. 
Upper and lower black lines show the original and continuum and $Fe$ template-subtracted spectra.
Colors and legends are explained in Fig.~\ref{fig:egciv} and Fig.~\ref{fig:egmg}.
In {\bf \color{cyan}cyan} is the \hbeta, \oiiiab\ narrow emission components,
whose FWHM is marked by FWHMN in the legend.
\label{fig:eghb}}
\end{figure}

The \hbeta\ line is fitted for the 75 objects with $z < 0.76$.
We adopt the iron template from \citet{bg92}  
and fit the continuum plus iron template
in the designated spectral windows (Table~\ref{tab:fitregion}).
In 4 objects with \hbeta\ coverage,
the iron template is not constrained due to the low quality of the spectra (S/N per pixel $<$ 4),
and only a power-law continuum was subtracted.
After subtracting the continuum and iron emission lines, 
we fit the \oiiiab\ doublets together with
the \hbeta\ component.
For the \hbeta\ components, 
we allow up to 3 Gaussians to fit the BEL,
and use a single 
Gaussian to account for each of the narrow \hbeta\ and \oiii\ emission lines. 
We require the narrow \hbeta\ component and the \oiii\ doublets
to have the same velocity shift and broadening, 
and constrained their FWHM to be $<$ 1200 km\,s$^{-1}$.
Only in $< 5\%$ cases do we need 
multiple Gaussians (Table~\ref{tab:ftest}) in each of the subsamples: 
3/38 for SDSS, 2/31 for MMT09, and 0/6 for MMT05b.
Fig.~\ref{fig:eghb} shows an example of a typical \hbeta\ fit.

\begin{table}[htdp]
\caption{Average FWHM uncertainties in our sample and comparison to literature.}
\begin{center}
\begin{tabular}{lcccccc}
\hline
Emission line  & This Work  & S11 & VO09 & VP06 & P13 \\
\hline
\civ & 26\% & 21\%  & 6\%& 9\% & 6\%\\
\hline
\mgii & 20\% & 27\% & 10\% &  &\\
\hline
\hbeta& 25\% & 27\% & 12\% & 10\%  & \\
\hline
\end{tabular}
\end{center} 
Notes: 
VP06, \citet{vp06}, 28 quasars (reverberation mapping, RM); 
VO09, \citet{vo09}, 34 SDSS quasars and 978 LBQS quasars (single-epoch spectra, SE); 
S11, \citet{shen11}, of 105,783 SDSS selected quasars (SE); 
P13, \citet{park13}, of 39 AGNs (RM). 
The uncertainty differences arise from spectral quality 
and the different methods used to measure them (\S\ref{sec:error}).
\label{tab:fwhmerr}
\end{table}

\subsection{Uncertainties of Spectral measurements and Error Estimates}
\label{sec:error}
The uncertainties in the spectral measurement arise from three main sources:
(1) the quality of the spectra and instrument errors;
(2) the adopted fitting process -- e.g. ambiguity introduced from using certain line profiles, 
and from using one or multiple components;
and (3) special features that could affect the algorithm
 -- in particular, a narrow line component, especially for \mgii\ and \civ\ (cf. S11);
instrumental broadening with BEL;
or strong NALs or BALs.

\begin{figure}[ht]
\begin{center}
\includegraphics[scale=0.5]{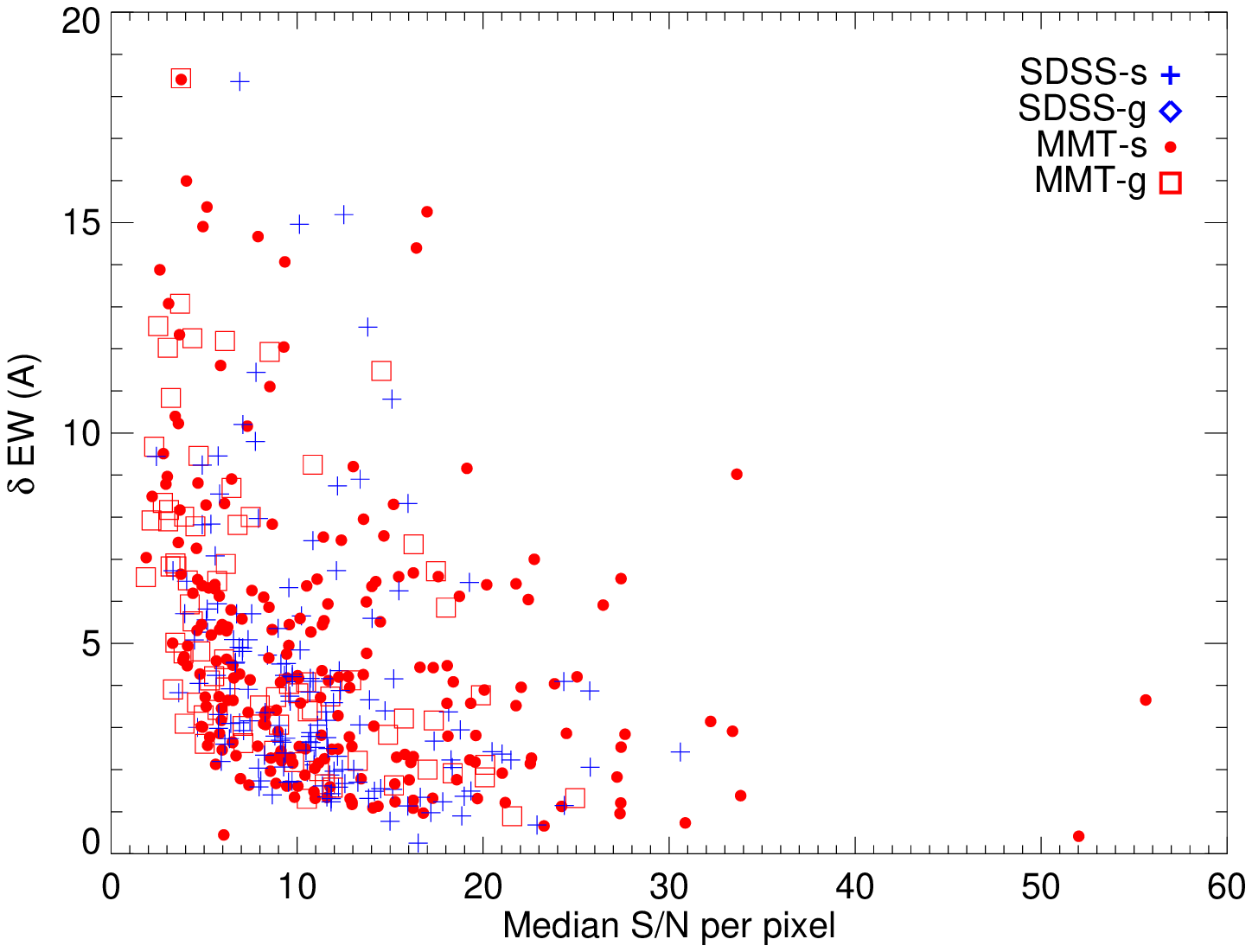} %{ew-sn.eps}   %fig19
\includegraphics[scale=0.5]{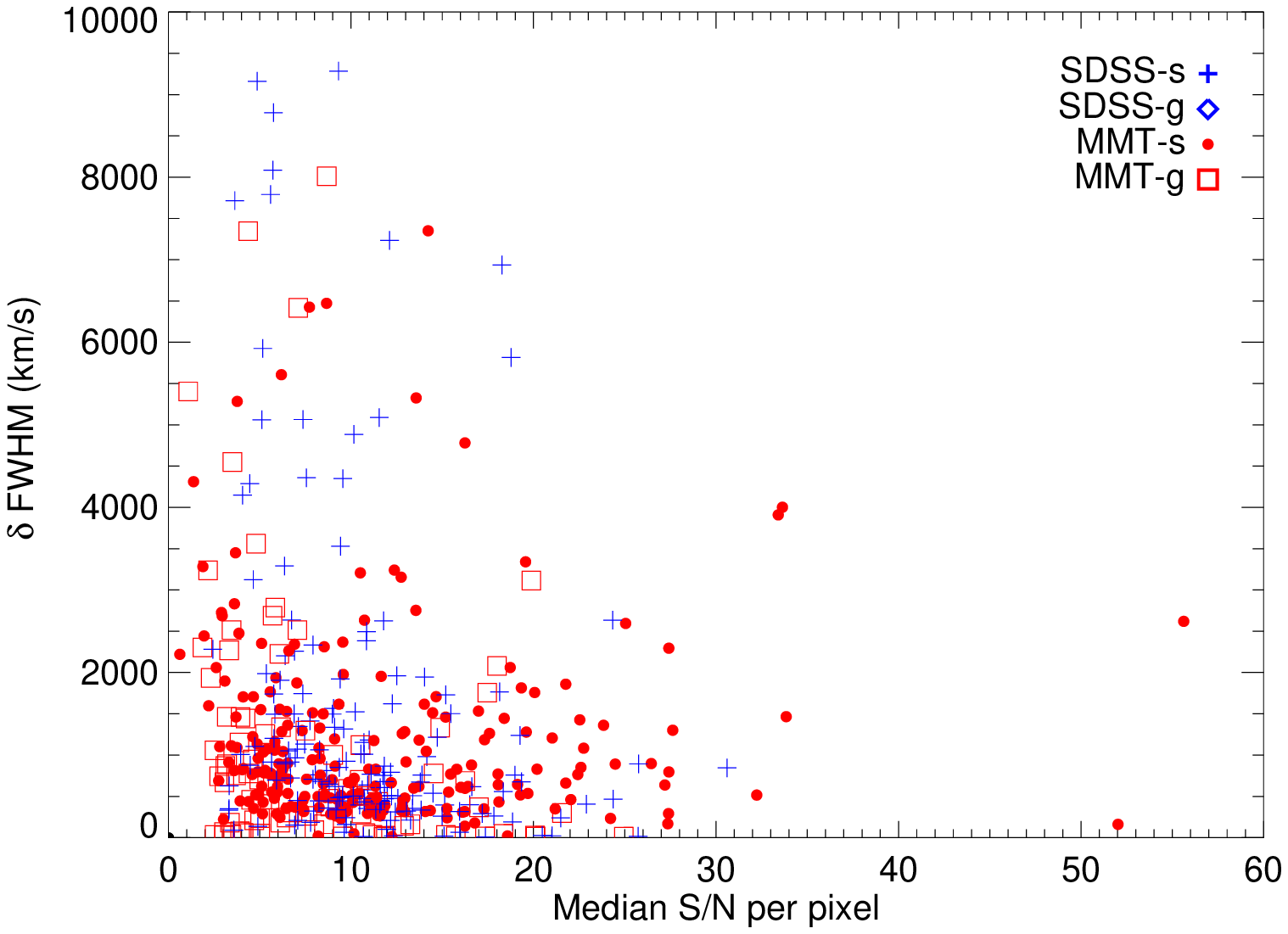} %{fwhm-sn.eps}
\includegraphics[scale=0.5]{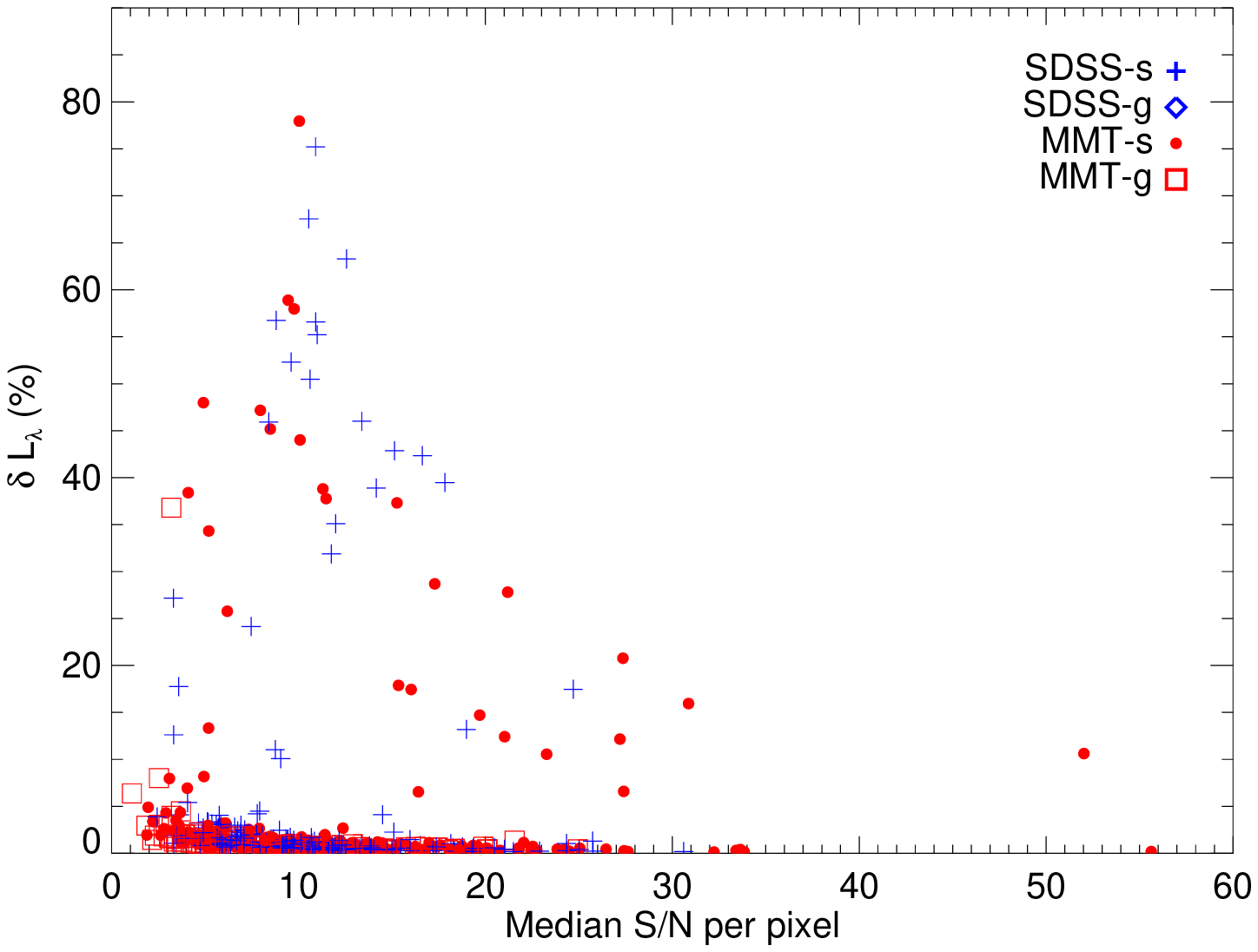} %{lcont-sn.eps}
\end{center}
\figcaption{Uncertainties in the EW and FWHM measurements (top),
and the continuum fitting results (bottom)
versus the median S/N per pixel of the fitting region.  
Color codes are explained in Fig.~\ref{fig:cov2}.
We observe decreasing uncertainties for EW and FWHM as the S/N of the spectrum
increases,
but the S/N influence on the continuum fitting is not obvious.
\label{fig:sn}}
\end{figure}

The fitting errors based on S/N are automatically accounted for through our IDL program
using the IDL program $mpfitfun.pro$\footnote{http://www.physics.wisc.edu/~craigm/idl/down/mpfitfun.pro}.
This program returns the 1\,$\sigma$ errors of each parameter
from the covariance matrix.
The quality of the spectra directly affects 
the fitting results.
We observed similar S/N dependences as in S11.
The uncertainty in the FWHM and EW measurements increases as 
the S/N in the line-fitting region decreases (Fig.~\ref{fig:sn}, top).
Little or no influence from the continuum S/N is 
found for the continuum fitting results  (Fig.~\ref{fig:sn}, bottom).

Instrumental broadening is not a problem for the BEL.
Hectospec has a spectral resolution of %$4.5 - 5.2 \,\AA$ FWHM, equivalence of 
170 - 380 $\kms$ at the redshifts ($0 < z < 4$) for the sample \citep{fab08}. 
The SDSS has a 1.5 $\sim$ 2 times higher resolution \citep{abazajian09}.
For the BELs,
$\sim 99\%$ have FWHM $> 2000\,\kms$,
so the instrumental resolution correction is negligible.
However, the instrument resolution is 
comparable to the NAL widths observed (a few hundred $\kms$), 
so that instrumental broadening must be removed. 
We used the formula: 
$FWHM_{\rm measured}^2 =  FWHM_{\rm intrinsic}^2 + FWHM_{\rm instrument}^2$
to correct the observed line-width for narrow absorption lines.
The non-Gaussian flat-topped fiber profile of MMT Hectospec \citep{fab08}
renders this correction imperfect, and will be discussed in 
the absorption paper (Dai \etal, 2014). 

We adopt the Monte Carlo flux randomization method
as in the SDSS routine (S11).  
This approach provides a more reasonable
estimate than from the program fit  alone,
as it also smoothes out the ambiguity of 
whether or not to subtract a narrow line for \civ\ or \mgii\ BELs. 
We generate 50 mock spectra with
the same wavelength and flux density error
arrays as the original spectrum, 
and randomly scatter the flux values 
with Gaussian noise (allowing negative values)
based on the original errors.
We then apply the same fitting procedure described in \S~\ref{sec:spec}.
The measurement uncertainties are defined as the standard deviation
 of the measured parameters in the 50 mock spectra. % (68\% range).  
This uncertainty is on average 2.1, 2.9, and 3.6 times larger 
than the fitting errors in FWHM for \hbeta\ , \mgii\ , and \civ\,, respectively.
The average FWHM uncertainties are summarized in Table~\ref{tab:fwhmerr}.
The uncertainties given in VP06 were adopted as the largest fitting error
from their five continuum settings and could be underestimated,
as the single fitting error is on average 2-3 times 
lower than using the Monte Carlo method. 
The average scaling factor between single fit and Monte Carlo uncertainties 
is then used to scale the uncertainties of FWHM and EW in $\sim$100 lines 
with strong absorption features.

The errors in FWHM and continuum measurements will directly affect the final SMBH mass (\S~\ref{sec:virial}). 
A 50\% uncertainty in FWHM translates to a 25\% uncertainty in SMBH mass. 
In general, the flux density and spectral measurement errors 
are in the range of 20$\sim$30\%.
For the SDSS subsample, our error estimates in general agree with the 
SDSS results. 

\begin{table} [hbdp]
\caption{Frequently Used Virial Black Hole Mass Estimators.}
\begin{center}
\begin{tabular}{lcccc}
\hline
\hline
Emission Line  & Continuum $\lambda\,(\AA)$ & a & b & Reference \\
\hline
 \hbeta & 5100 & 0.672 & 0.61 & \bf{MD04}\\
 ... & ... & 0.910 & 0.50 & VP06\\
 \hline
 \mgii  & 3000  & 0.505  & 0.62 & MD04 \\
...  & ...  & 0.860  & 0.50 & VO09 \\
...  & ...  & 0.740  & 0.62 & \bf{S11} \\
 \hline
\civ  & 1350 & 0.660 & 0.53 & \bf{VP06}\\
\hline
\hline
\end{tabular}
\end{center}
\label{tab:virial}
Notes: 
MD04: \citet{md04}; VP06: \citet{vp06}; VO09: \citet{vo09}; S11: \citet{shen11}.  
In bold fonts are the sets of estimators we used for the fiducial SMBH mass.
\end{table}

\section{Virial Black Hole Masses}
\label{sec:virial}
The SMBH mass is one key property in studying the SMBH-host connection. %and Eddington ratios 
Among the various $\mbh$ estimators \citep[e.g.][]{kormendy95, gebhardt00, marconihunt03}, 
the virial mass estimate is one of the simplest and most adopted \citep[e.g.][]{kaspi00, md04, vo09}.
The virial method is a powerful tool 
especially in the absence of host galaxy information,
where stellar velocity dispersion or bulge luminosity is missing.
The virial method is based on the assumption that the dynamics in the vicinity of the nucleus,
the `Broad-Emission-Line-Region' (BLR), is dominated by the gravity of the SMBH,
so that the mass of the central SMBH can be 
estimated from the virialized velocity of the line-emitting gas.
The virial method based on the emission lines are calibrated 
by reverberation mapping (RM) results,
which use time delays measured from the BEL variability \citep[e.g.][]{vp06, wang09, park13}. 
In the RM method, 
the BLR radius can be measured via the light travel time
delayed response of
the emission line flux to continuum variation. 
However,
only a few dozen objects have reliable RM masses
due to the demanding exposure and signal-to-noise (S/N) requirements \citep{denney13}. 
The virial method is more commonly used 
as it requires only single-epoch (SE) spectra.
For SE spectra, 
the BEL line-width is used as direct proxy for the SMBH mass,
based on the assumption that the BLR radius
is proportional to the luminosity---the observed R-L relationship \citep[VP06; ][]{collin06,bentz09}---
and the BEL line-width is proportional to the Keplerian velocity %rotational
of the accreting gas.

The virial mass estimators for SMBH based on SE spectra are usually expressed as:
\begin{equation} \label{eqn:virial}
 {\rm log} (\frac{\mbh}{\msun}) = a + b\,{\rm log} (\frac{\lambda L_{\lambda}}{10^{44} \ergs}) + c\,{\rm log} (\frac{\rm FWHM}{\kms})
\end{equation}
 where $\msun$ is the solar mass. 
The term $\lambda L_{\lambda}$ is the continuum luminosity, 
a proxy for the BLR radius \citep{kaspi00, bentz06, bentz13}.
They are measured from chosen wavelengths close to each BEL (Table~\ref{tab:virial}).
The coefficients $a$ and $b$ are empirical values based on 
the SMBH masses from RM and 
 comparison among different lines. 
$c$ normally has a fixed value of 2. 
Since the BEL line-width (FWHM) represents the virial velocity,
this 2 factor exemplifies the virial nature of the BLR ($\mbh \propto G v^2 R^{-1}$). 
Recently a few papers have suggested using 
other values for $c$ based on comparison of SE and RM results.  
For instance,
\citet{wang09} used 1.09 and 1.56
in front of the \hbeta\ and \mgii\ FWHMs, respectively.
\citet{park13} used 0.56 in 
front of the \civ\ FWHMs.
If a $< 2$ factor is adopted, 
the resulting SMBH mass estimate will be smaller accordingly.
Here we stick to the $c = 2$ value to be consistent with the SDSS quasar catalog (S11).

The \civ, \mgii, and \hbeta\ BELs %we choose,
are widely used as virial black hole mass calibrators \citep[e.g.][]{md04, vp06, vo09, shen11}.
We summarize the most frequently used virial estimators in Table~\ref{tab:virial}.
If multiple Gaussian components are used, 
in the catalog we provide both the dominant and the non-parametric $\mbh$ 
derived from the dominant and non-parametric FWHM.
In the following analysis of $\mbh$ properties, 
for the \mgii\ and \hbeta,
we use the $\mbh$ derived from the non-parametric FWHM to be consistent with the literature definitions.
This choice of non-parametric FWHM in general provides lower $\mbh$ estimates than from dominant FWHM,
and may underestimate the $\mbh$ for BELs if the emitting gas is in Keplerian motion.

\begin{figure}[ht]
%\begin{center}
\includegraphics[scale=0.7]{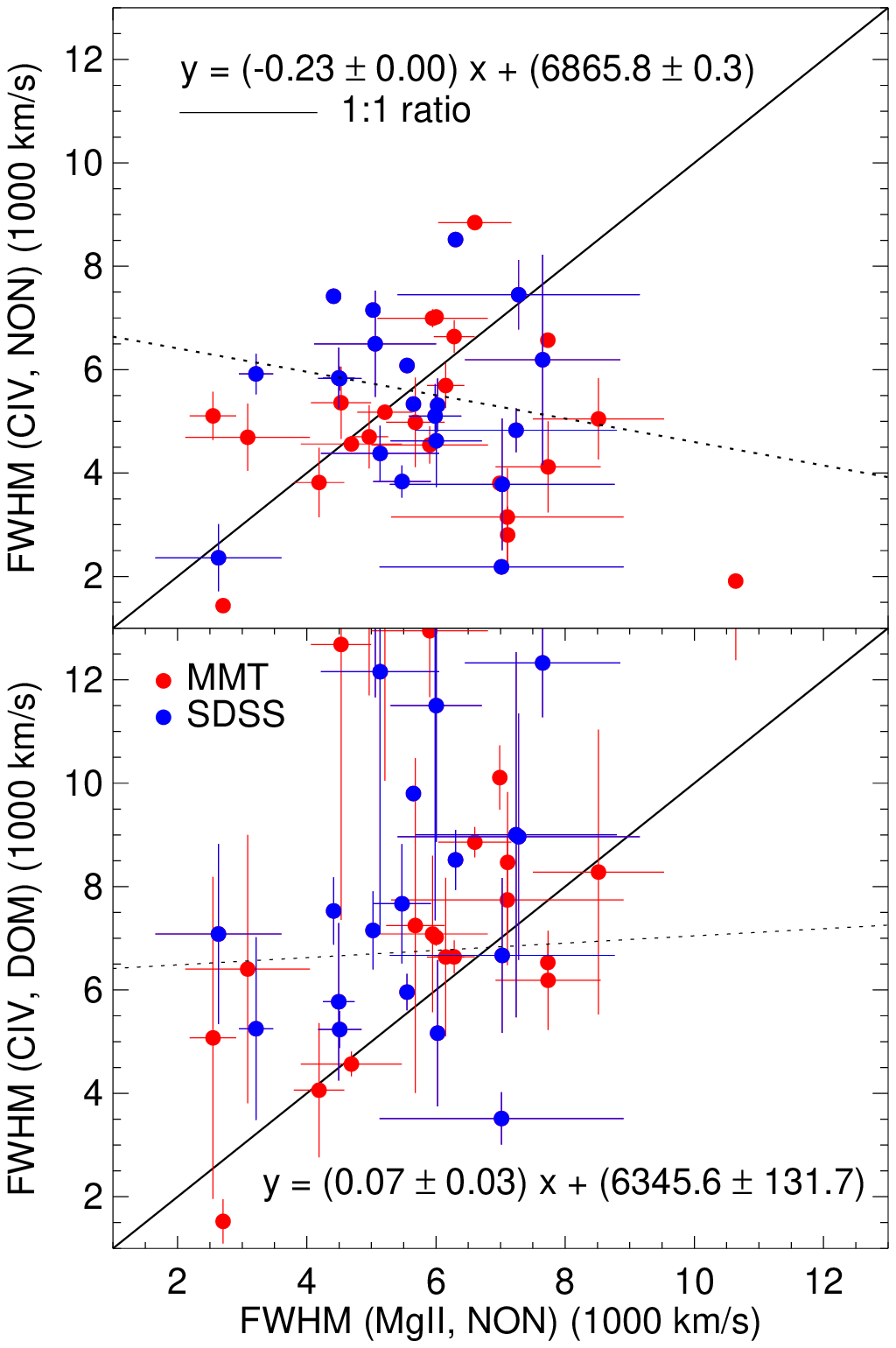} %{cf-mgiinon-civ-fwhm-5nl.eps}  %fig20
\includegraphics[scale=0.7]{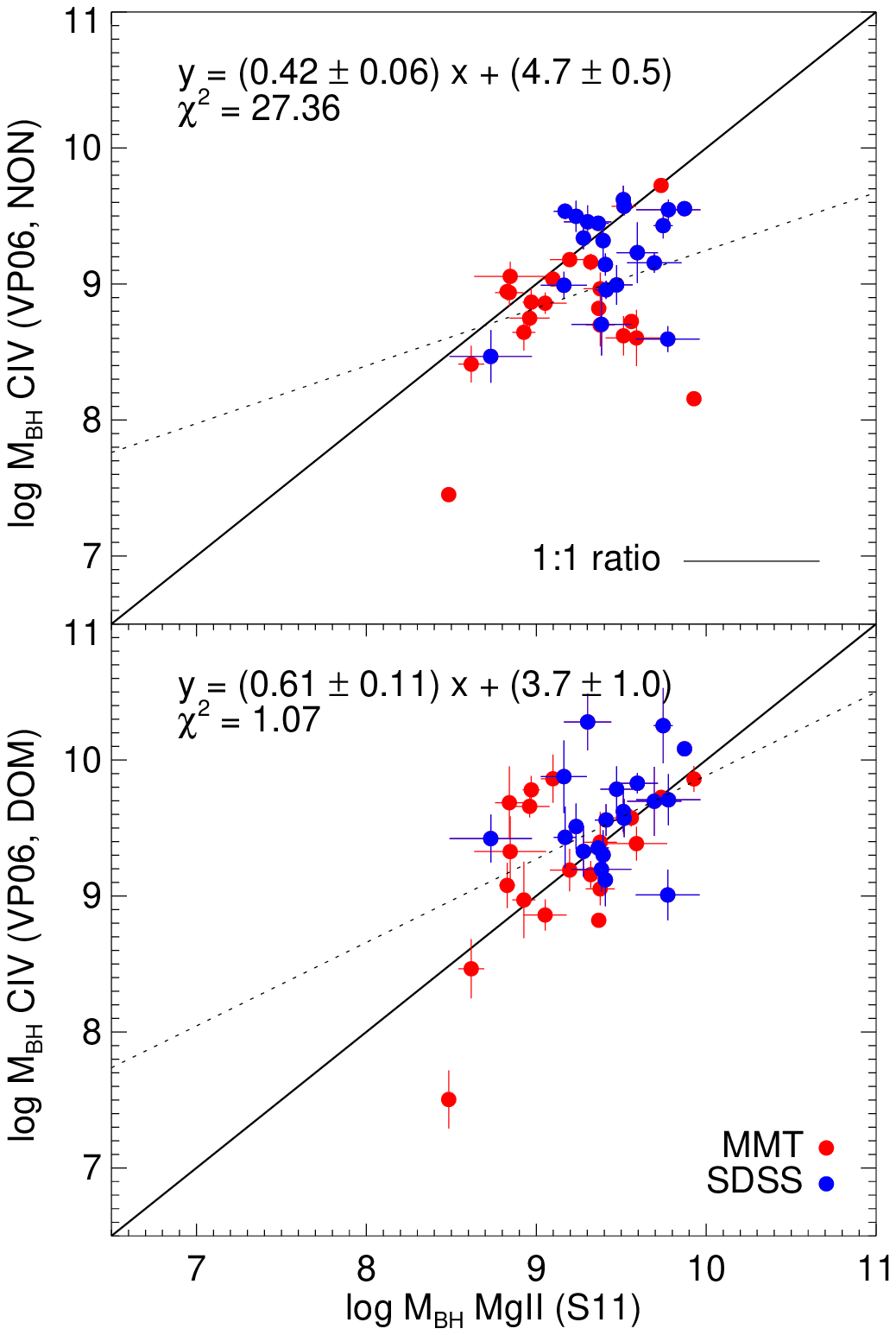} %{cf-mgiinon-civ-5nl.eps}
%\end{center}
\figcaption{ (left) Comparison of the non-parametric (top) and dominant (bottom) \civ\ FWHM
against the non-parametric \mgii\ emission line width. 
(right) SMBH mass ($\mbh$) in $\msun$ derived from the non-parametric (top) and dominant (bottom) \civ\ FWHM 
against the $\mbh$ from the non-parametric \mgii\ FWHM. 
In each pane, diagonal line marks the linear correlation (See \S~\ref{sec:virial}).
%and the deviation ($\chi^2$) from the 1:1 linear correlation 
%is given in the top left corner
The MIR-selected MMT targets are in red and the SDSS targets in blue.
The dominant \civ\ FWHMs are systematically higher than the 
non-parametric \civ\ FWHM 
in 70\% of the cases with multiple Gaussians
(See, e.g., Fig.~\ref{fig:egciv}),
and has a marginally smaller scatter from the1:1 linear correlation
with the \mgii\ non-parametric FWHM.
A better correlation to the \mgii\ derived $\mbh$ 
is also observed of the $\mbh$ from dominant \civ\ FWHM (linear fit slope: $0.61 \pm 0.11$) 
than from the non-parametric \civ\ FWHM (linear fit slope: $0.42 \pm 0.06$),
possibly indicating a non-virial component in the \civ\ BEL.
\label{fig:civmgiifwhm}}
\end{figure}

For the \civ\ calibrator, 
the line-width definition in literatures is 
also the same as the non-parametric FWHM \citep[VP06, see also][]{peterson04}. 
However,
it is debated as to whether it provides a reliable $\mbh$ estimate
due to the large scatter
between the generally consistent \civ\ and \hbeta\ derived $\mbh$ \citep{netzer07, assef11}.
This scatter may result from  
non-virial components from outflows or winds in the \civ\ BLR \citep[e.g.][]{richards11}.
For this MIR-selected quasar sample,
we find a marginally better correlation between 
the dominant \civ\ FWHM and 
the non-parametric \mgii\ FWHM (Figure.~\ref{fig:civmgiifwhm}, left).
Better consistency is also 
found between the $\mbh$ derived from the dominant \civ\ component
and \mgii\ BELs (Figure.~\ref{fig:civmgiifwhm}, right),
indicating a non-virial contribution in the non-parametric BEL profile.
Based on the correlation results,
we choose to use the dominant \civ\ FWHM for $\mbh$ estimates.
We will discuss the choice %of dominant \civ\ FWHM 
and its implications in \S~\ref{sec:discussion}.

\begin{figure}
\includegraphics[scale=0.7]{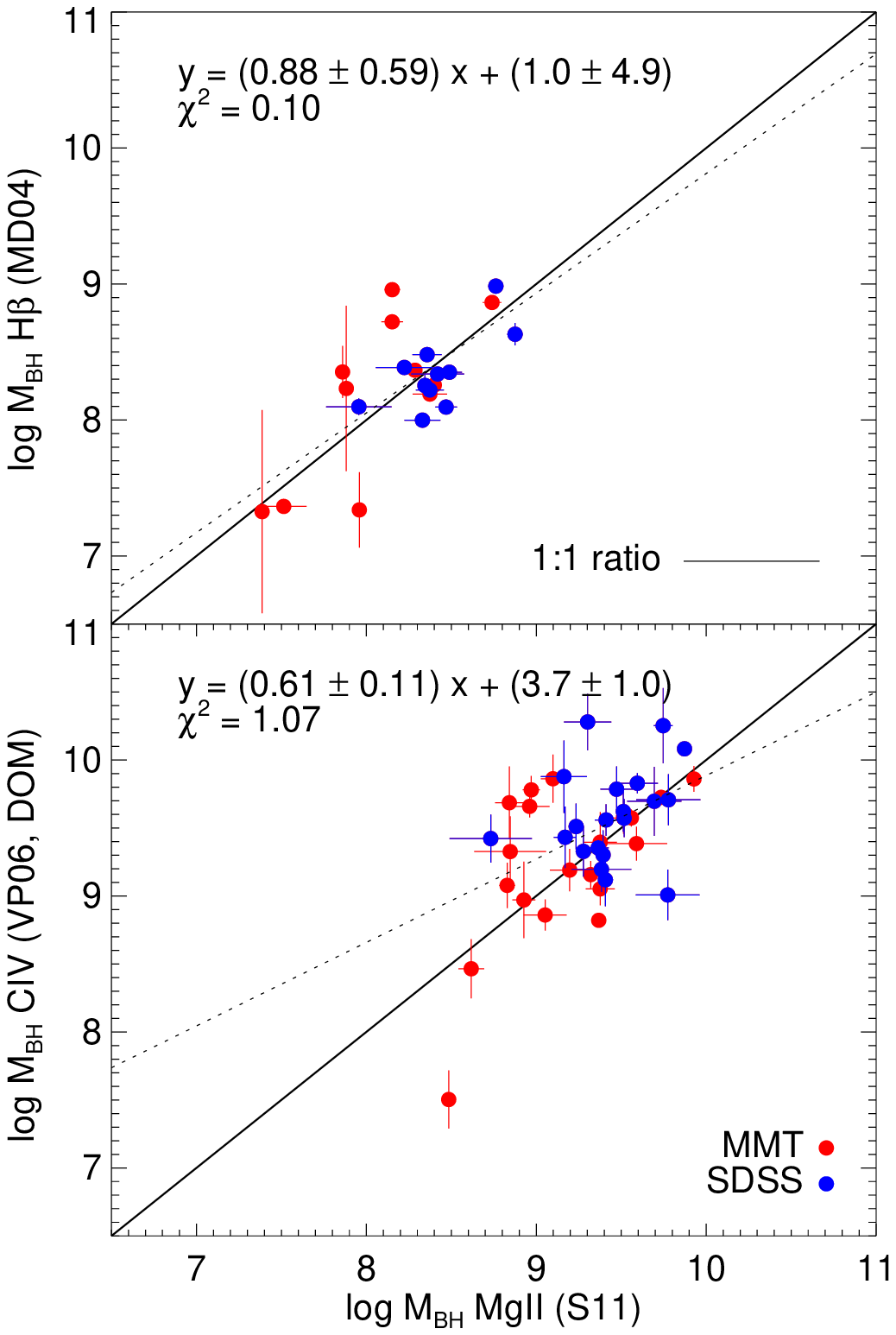}%{cfmbhline-s11.eps}  %fig21
\includegraphics[scale=0.7]{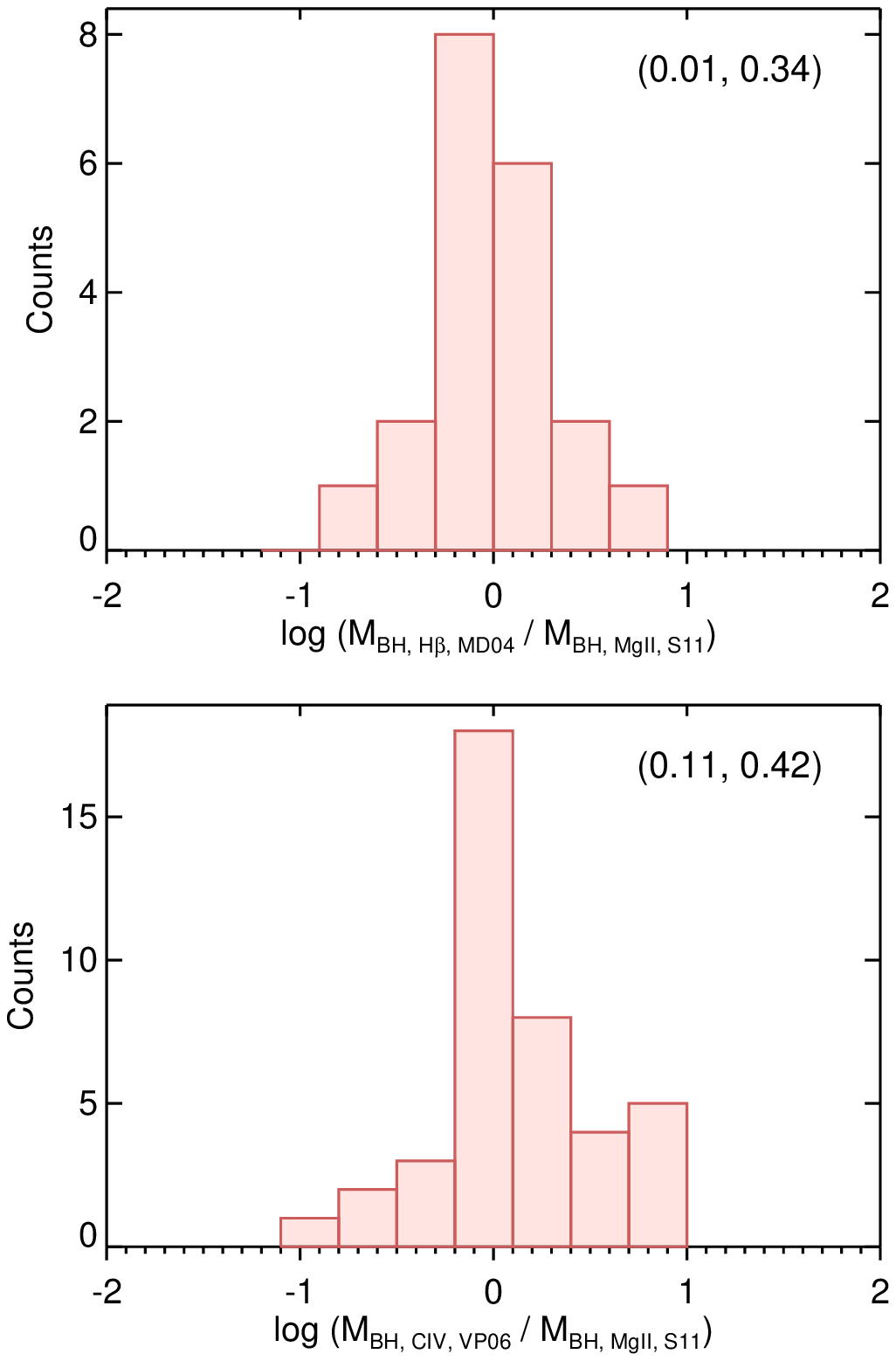}%{cfmbhline-hist.eps}
\figcaption{ {\it(left):} Comparison of SMBH masses in $\msun$ derived from different lines for the quasars with both
\hbeta\, and \mgii\ (top),  \mgii\ and \civ\ (bottom) BELs.
MMT sources are marked in red and SDSS sources in blue. 
A median line S/N per pixel $>$ 5 was required. 
Diagonal line marks the 1:1 correlation,
and dashed line marks the best fit linear correlation. 
The best-fit coefficient and associated errors are marked at the top left corner.
The $\mbh$ from S11 (\mgii) estimator shows a tight correlation 
with the $\mbh$ from MD04 (\hbeta), 
and is consistent with the $\mbh$ from VP06 (\civ).
{\it(right):} The mass ratio distributions for the two sets of estimates for the same object. 
A median line S/N per pixel $>$ 5 was required. 
The mean and 1$\sigma$ from a Gaussian fit to the distribution are plotted at the top right corner.
\label{fig:mbhcomp}}
\end{figure}

In our catalog, if applicable, 
we present multiple $\mbh$, using MD04 (\hbeta, \mgii), VP06 (\hbeta, \civ), 
VO09 (\mgii), and S11 (\mgii) estimators.
We attribute the $\mbh$ from MD04 (\hbeta), S11 (\mgii),
and VP06 (\civ) as the `fiducial' $\mbh$ to each object,
as the $\mbh$ from these parameters are best-correlated with each other (Fig.~\ref{fig:mbhcomp}, left).
We compare the different estimators 
based on the subsample of quasars that have 
two BELs with a median S/N per pixel of $>$ 5 and no BAL/NAL,
which leaves 20 objects with both \mgii\ and \hbeta\ BELs,
and 38 targets with both \civ\ and \mgii\ BELs.
%32 quasar spectra cover both \mgii\,and \hbeta\ BELs,
%and 41 spectra cover both \civ\, and \mgii\ BELs. 
The comparison of the $\mbh$ from different lines and 
estimators for quasars with two BELs
is achieved
by forcing a linear correlation and measuring the $\chi^2$ values 
to compare the sample scatter.

We first compare the three \mgii\ estimators (MD04, VO09, S11) with 
the \civ\ estimator (VP06), 
%We plotted the $\mbh$ derived from \civ\ and \mgii\ BELs,
and found a marginally smaller $\mbh$ 
scatter for VP06 (\civ) \& S11 (\mgii) ($\chi^2 = 1.07$) %(slope coefficient $=$ 0.6$\pm$0.1),
than for VP06 (\civ) \& VO09 (\mgii)  ($\chi^2 = 1.07$). %(slope coefficient $=$ 0.7$\pm$0.1),
Both have a $\chi^2$ value 
$\sim$1 dex better than VP06 (\civ) \& MD04 (\mgii).
The slope coefficient in all three sets of estimators 
agree with each other within errors at a value $\sim$0.6.
The scatter in log$(M_{\rm BH}, {\rm CIV, (VP06)} / M_{\rm BH}, {\rm MgII, (S11)})$ 
is similar to the scatter for the SDSS DR7 catalog (see Fig.10, S11). 
This small scatter between S11 and VP06 is by design, 
as the S11 coefficients were empirically adopted
to provide the best correlation between VP06 (\civ) and S11 (\mgii) results. 
For ease of comparison with the SDSS sample,
we assign the $\mbh$ from S11 as 
the fiducial $\mbh$ from \mgii\ BEL. 

We then make the same $\chi^2$ comparison
for the two \hbeta\ estimators (VP06, MD04) and 
the chosen \mgii\ estimator S11. 
For the same \hbeta\ BEL,
$\mbh$ from VP06 is systematically 0.2\,dex higher than from MD04,
since the VP06 `a' factor is $\sim$0.2 larger (Table~\ref{tab:virial}). 
S11 \& MD04 show a slightly smaller scatter ($\chi^2$ = 0.59)
than S11 \& VP06 ($\chi^2$ = 0.78),
so $\mbh$ from MD04 is chosen 
as the fiducial $\mbh$ in \hbeta\ BELs.
The scatter in log$(M_{\rm BH}, {\rm H\beta, (MD04)}$ / $M_{\rm BH}, {\rm MgII, (S11)})$ 
is also similar to that of
the SDSS DR7 catalog (see Fig.10, S11). 

In summary, for the MIR-selected sample,
we find that MD04 (\hbeta), S11 (\mgii), and VP06 (\civ)
show the best correlations and 
assign a fiducial $\mbh$ 
using these three estimators.
If $\mbh$ from \mgii\ and \hbeta\ BELs are both available, 
the $\mbh$ derived using \hbeta\ will be adopted as the 
fiducial $\mbh$ because of the robust SE mass scaling from \hbeta\ RM studies.
For targets with $\mbh$ from both \civ\ and \mgii\ BELs,
we attribute the  \mgii\  derived $\mbh$ given 
the possible complications of non-virial component from the \civ\ BELs.

In Fig.~\ref{fig:mbhcomp} (right), 
we plot the mass ratios distribution for the quasar subsample with 2 BELs (median S/N per pixel of $>$ 5). 
The mean and 1$\sigma$ from a Gaussian fit to the mass ratio distributions
are (0.01, 0.34) for  log $(M_{\rm BH}, {\rm H\beta, (MD04}) / M_{\rm BH}, {\rm MgII, (S11}))$
and (0.11, 0.42) for log $(M_{\rm BH}, {\rm CIV, (VP06}) / M_{\rm BH}, {\rm MgII, (S11}))$. 
The mean offsets are negligible since they are smaller than 
what a typical FWHM error would introduce:
a 30\% error in FWHM translates to an upper and lower uncertainty of 
$+$0.11 dex \& $-$0.15 dex in the log ($\mbh$) space,
and justifies the choice of these three estimators. 

\begin{figure}
\begin{center}
\includegraphics[scale=1.0]{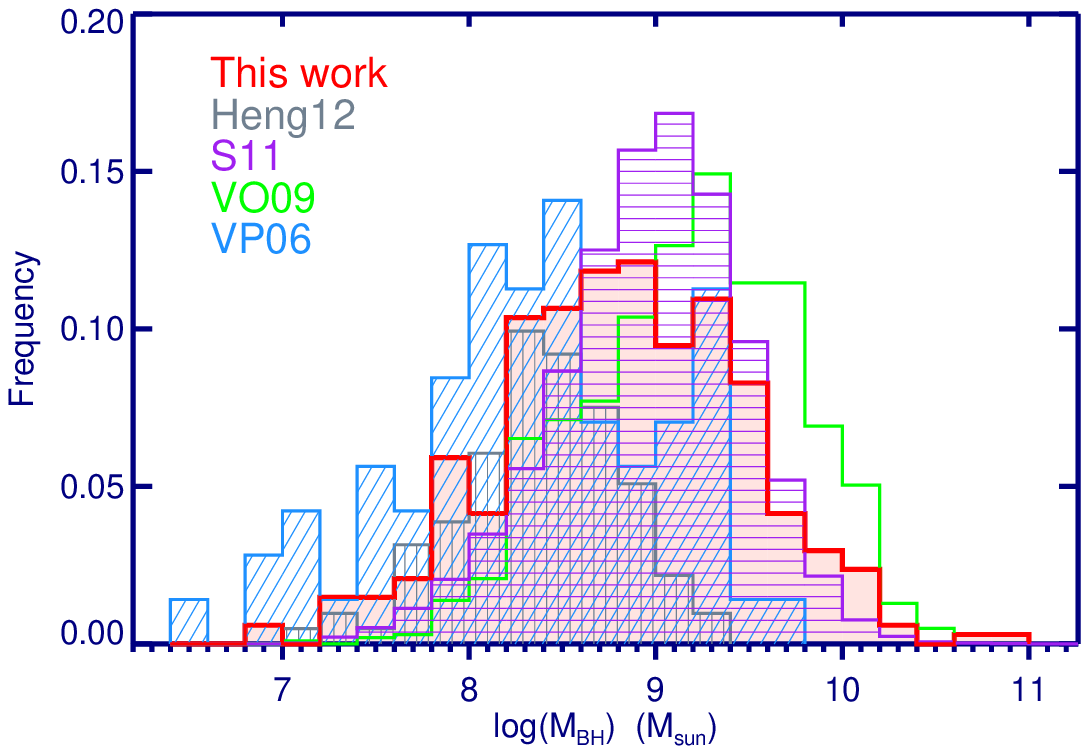}%{mbh-cf-r2.eps}  %fig22
\\
\includegraphics[scale=1.0]{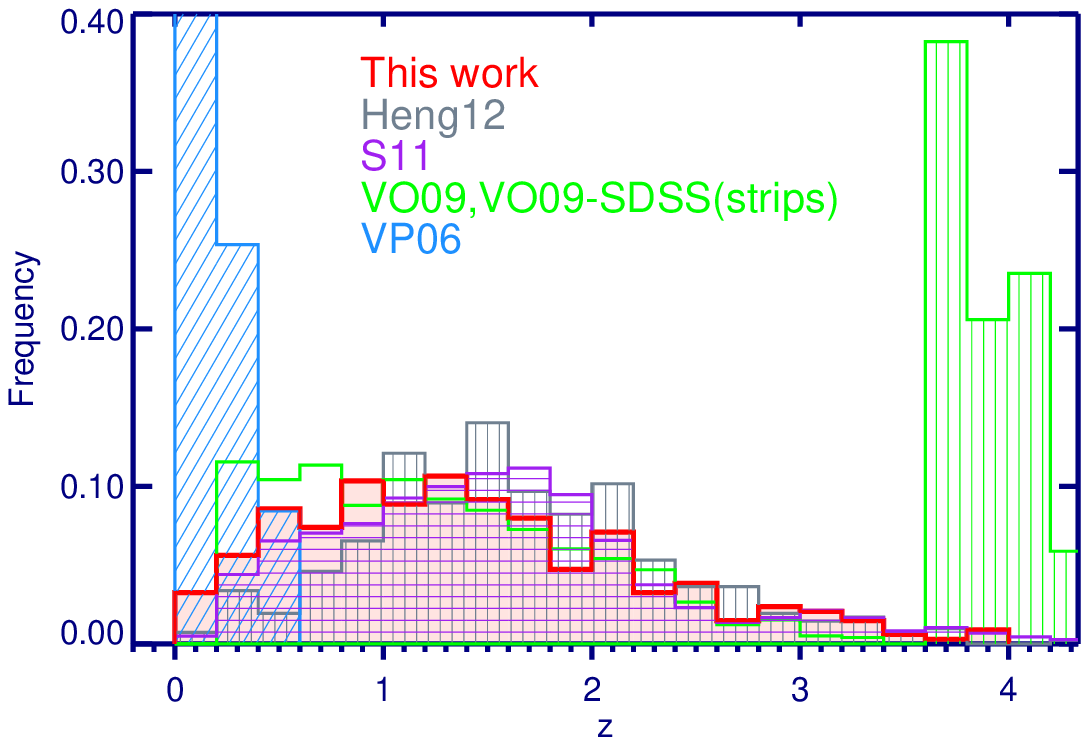}%{z-cf-r.eps}
\end{center}
\figcaption{{(top)} BH mass distribution of the relative frequency of
the 391 MIR-selected quasar sample ({\bf \color{red}red}).
Literature values from SDSS and other surveys are also plotted for comparison:
{\bf \color{gray}gray}, 413 X-ray selected quasars  (Heng et al., in prep),
{\bf \color{purple}purple},105,783 SDSS selected quasars \citep[][S11]{shen11},
{\bf \color{blue}blue}, 1,012 $z < 5$ quasars, including 34  SDSS quasars at 3.5$< z < $5 \citep[][VO09]{vo09},
{\bf \color{green}green}, 71 $z < 0.3$ quasars with reverberation mapping info \citep[][VP06]{vp06}.
{(bottom)} Redshift distribution of the MIR-selected quasars (red). 
Samples from the literature are color-coded in the same way as the top panel.
The MIR-selected quasar sample overlaps with Heng12, S11, VO09-BQS, LBQS quasars in redshifts,
and has a large overlap in $\mbh$ with the SDSS quasars. 
VP06 extends to the low mass end partly due to their relatively lower redshift from the RM constraint.  
\label{fig:mbh}}
\end{figure}

We show the SMBH mass and redshift distribution for 
the MIR-selected quasar sample in Fig.~\ref{fig:mbh},
and superpose samples from the literature for comparison.
The redshift distribution of the MIR-selected 
quasars is typical of an apparent-magnitude limited sample,
and has a large overlap with the SDSS, 
BQS, and LBQS catalogs. 
For $\mbh$,
the MIR-selected sample also overlaps with the
above mentioned samples,
but have a higher fraction of lower mass objects 
than the S11 sample---a direct result of the fainter 
magnitude limit applied.

\section{Bolometric Luminosity and Eddington Ratios}
\label{sec:lbol}
\begin{figure}
\begin{center}
\includegraphics[scale=0.8]{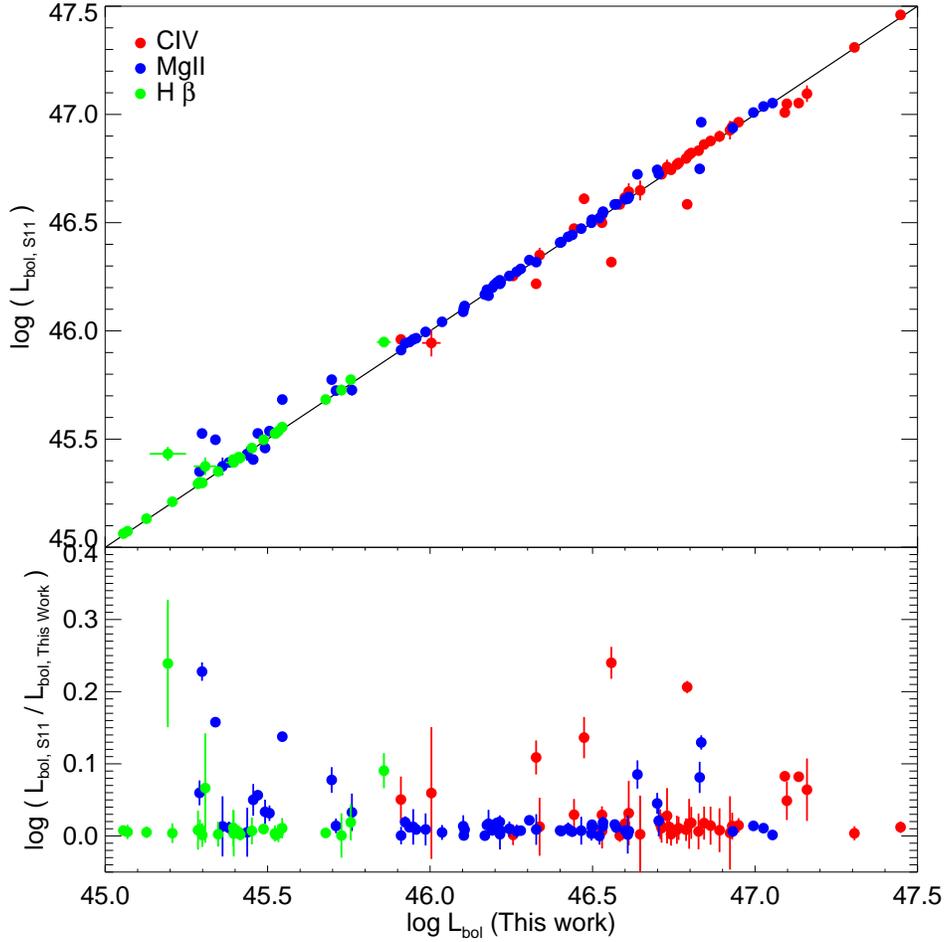}%{lbol-d.eps} %fig23
\end{center}
\figcaption{The bolometric luminosities ($\lbol$, top panel) and $\lbol$ comparisons for the same targets
between SDSS DR7 and this work (bottom, panel) as a function of $\lbol$.
Only objects with a median line S/N per pixel $> 3$ are included.
Targets with NALs/BALs are excluded. 
The two independently derived $\lbol$ are consistent with each other ($< 3 \sigma$)
for the majority ($\sim$80\%) of the MIR-selected SDSS subsample. 
\label{fig:lcont}}
\end{figure}

We measure the bolometric luminosity $\lbol$ from the fitted spectra continuum luminosities:
$\lbol = k \times L_{\lambda}$, 
where $L_{\lambda}$ are $L_{5100}(z < 0.76$, \hbeta), 
$L_{3000}(0.43 < z < 2.10$, \mgii), and $L_{1350} (1.63 < z < 3.18$, \civ) in $\ergs$;
and $k =$ 9.26, 5.15, and 3.81, respectively (cf. S11).
The coefficient $k$ values are from the composite SED from \citet[][R06]{rich06b},
a modified SED largely consistent with \citet{elvis94}. 
The R06 template should be applicable to at least 
the point source targets in this work,
since it is based on 259 $Sptizer$ detected SDSS type 1 (BEL) quasars,
and 96\% (248/259) of which also 
qualify the MIR-selection of $S_{\rm 24} > 400\mu$Jy for this sample.
Therefore, we caution the usage 
of the cataloged $\lbol$ and its derived parameters 
for extended objects.
We did not correct the spectra for intrinsic extinction (See also \S~\ref{sec:datareduction}). 
This may result in  
$\lbol$ being underestimated for systems with strong reddening;
or overestimated if there is significant host contamination.
A fourth estimator using $S_{\rm 24}$ flux shifted to the rest-frame 
is also introduced for comparison,
in which the $k$ values differ 
from redshift to redshift.   
Given the uncertainty in the quasar MIR SED shapes \citep{dai12},
we caution the use of the MIR flux-derived $\lbol$.
It is on average 0.5\,dex higher than the optical continuum-derived values, 
possibly from degenerate factors of reddening,
host contamination, and possible PAH emission contamination at $z > 2$.
For comparison, we will only discuss the continuum-derived $\lbol$ 
in the following discussion.
All MIR-selected quasars have $\lbol$ greater than $10^{44} {\rm erg\,s^{-1}}$,
confirming their quasar nature (Fig.~\ref{fig:lcont}).

For the MIR-selected SDSS subsample, 
a comparison with 
the SDSS DR7 quasar catalog (S11) shows 
consistency within 3$\sigma$
in continuum-derived $\lbol$ (Fig.~\ref{fig:lcont}) for over 80\% of the MIR-selected targets.
The MIR-selected quasars have an overall lower $\lbol$ distribution
than SDSS DR7 quasars,
since they include a large fraction (40\%) of objects fainter than the SDSS 
magnitude cut at $i > 20.2$.  
The median fitting errors for 
$\lbol$ are 2\%, 1\%, and 3\% for the \hbeta, \mgii, and \civ\ BELs,
respectively.
In objects that fall in $0.46 < z < 0.76$ or $1.63 < z < 2.10$, 
where two BELs are covered,
we find a $\sim$40\% consistency between the $\lbol$ from \civ\ and \mgii, 
and $\sim$15\% between \mgii\  and \hbeta,
evidence of reddening or host contribution.
In the following analysis, 
if two $\lbol$ are available for the same object, 
we use the $\lbol$ that corresponds 
to the chosen $\mbh$ (See \S~\ref{sec:virial}).

\begin{table} [hbdp]
\caption{Median SMBH mass, bolometric luminosity, and Eddington ratios of the MIR-selected quasars.}
\begin{center}
\begin{tabular}{lccccc}
\hline
\hline
Redshift & Subsample & $\#_{\rm obj}$ & $log(\mbh) (\msun)$  & $log(\lbol) (\ergs)$  & $log(\lbol/L_{\rm edd})$ \\
\hline 
$z < 1$  		& SDSS     &44 		&8.26 $\pm$ 0.53& 45.21 $\pm$ 0.40 &-1.07 $\pm$ 0.53\\
& MMT     &82 		&8.39 $\pm$ 0.56& 45.01 $\pm$ 0.46 &-1.33 $\pm$ 0.55\\
\hline 
& overall & 126 & 8.34 $\pm$ 0.55 & 45.06 $\pm$ 0.44 & -1.24 $\pm$ 0.55 \\
\hline 
$1 < z < 2$      		 & SDSS&	55 	&9.05 $\pm$ 0.47& 46.10 $\pm$ 0.61 &-1.05 $\pm$ 0.32\\
& MMT  &	126 	&8.85 $\pm$ 0.44& 45.72 $\pm$ 0.53 &-1.14 $\pm$ 0.34\\
\hline 
&overall & 181 &  8.91 $\pm$ 0.45 & 45.81 $\pm$ 0.56 & -1.10 $\pm$ 0.33\\
\hline 
$2 < z < 3$  			&SDSS  &	22 	&9.59 $\pm$ 0.24& 46.80 $\pm$ 0.36 &-0.98 $\pm$ 0.32\\
&MMT  & 43 	&9.29 $\pm$ 0.52& 46.27 $\pm$ 0.44 &-1.15 $\pm$ 0.38\\
\hline 
&overall & 65 &  9.40 $\pm$ 0.48 & 46.37 $\pm$ 0.50  &-1.05 $\pm$ 0.37\\
\hline 
$z > 3$ 		&SDSS     &17 		&9.92 $\pm$ 0.47& 46.86 $\pm$ 0.28 &-0.90 $\pm$ 0.44\\
&MMT      	& 2	   &10.78 $\pm$ 1.27& 47.69 $\pm$ 1.11 &-0.95 $\pm$ 0.16\\
\hline 
&overall &19 & 9.92 $\pm$ 0.54 &46.86 $\pm$ 0.38 &-0.91 $\pm$ 0.42\\
\hline
\hline
Redshift & Type & $\#_{\rm obj}$ & $log(\mbh) (\msun)$  & $log(\lbol) (\ergs)$  & $log(\lbol/L_{\rm edd})$ \\
\hline
$z < 1$      &point 	& 58& 8.34 $\pm$ 0.46 &45.29 $\pm$ 0.41 &-1.07 $\pm$ 0.46\\
		     &ext       & 68& 8.38 $\pm$ 0.61 &44.93 $\pm$ 0.42 &-1.34 $\pm$ 0.59\\
\hline 
$1 < z < 2$  &point 	&172& 8.91 $\pm$ 0.45 &45.81 $\pm$ 0.57 &-1.10 $\pm$ 0.34\\
			 &ext   	& 96& 8.90 $\pm$ 0.34 &45.95 $\pm$ 0.35 &-1.24 $\pm$ 0.24\\
\hline 
$2 < z < 3$  &point 	& 61& 9.44 $\pm$ 0.48 &46.40 $\pm$ 0.49 &-1.05 $\pm$ 0.38\\
			 &ext       &  4& 9.12 $\pm$ 0.26 &45.94 $\pm$ 0.49 &-1.25 $\pm$ 0.24\\
\hline 
$z > 3$      &point     &19 & 9.92 $\pm$ 0.54 &46.86 $\pm$ 0.38 &-0.91 $\pm$ 0.42\\  
             & ext &... &... &... & ...  \\  
\hline
\hline
\end{tabular}
\end{center}
\label{tab:ledd}
\end{table}

\begin{figure}
\begin{center}
\includegraphics[scale=0.7]{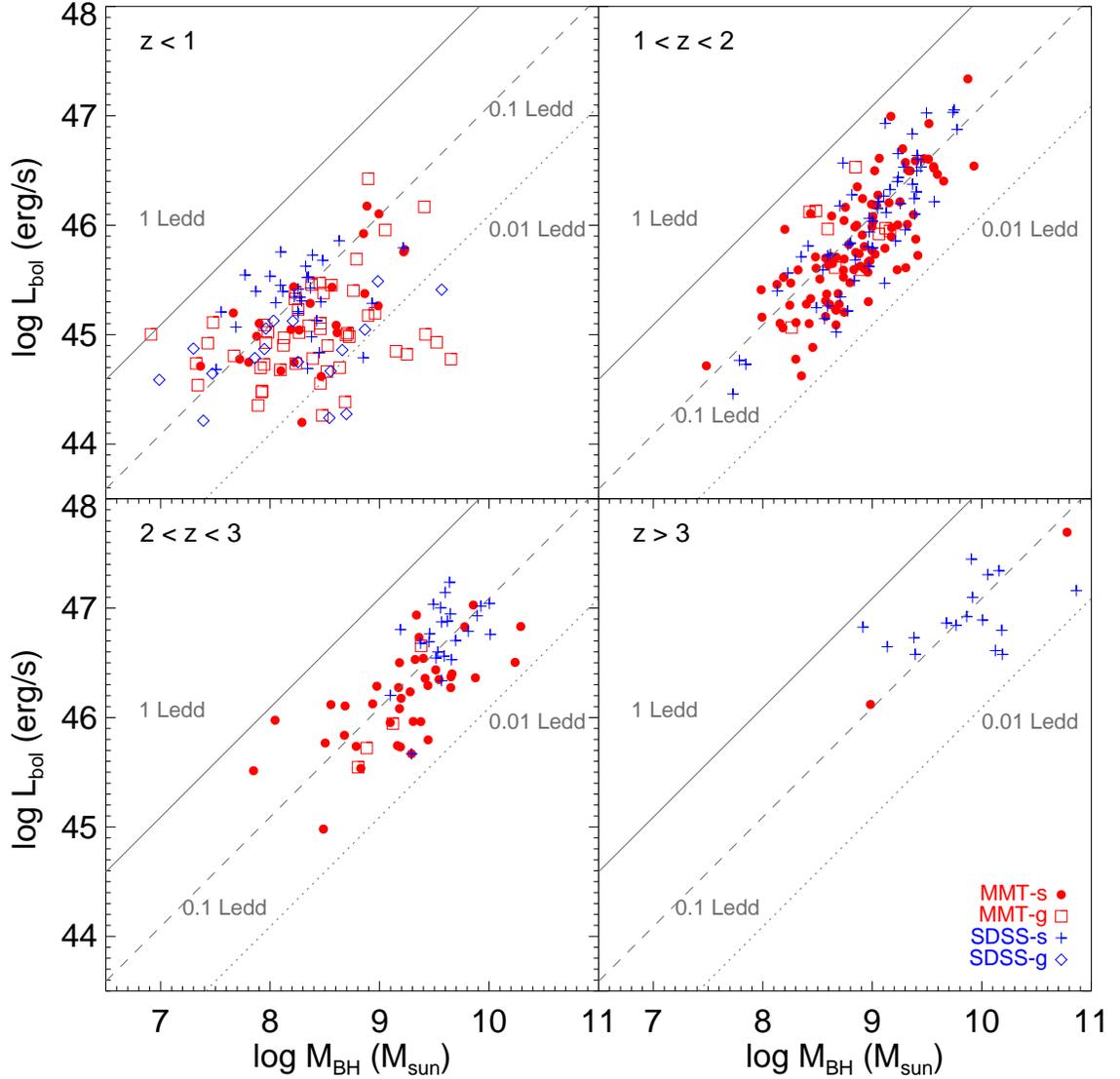}%{mbh-lbol-ext-s11-r.eps}  %fig24
\end{center}
\figcaption{Bolometric Luminosity ($\lbol$) of MIR-selected quasars as a function of SMBH mass ($\mbh$). 
The diagonal line marks the Eddington luminosity ($L_{\rm edd}$) of corresponding $\mbh$
at 1 (solid line), 0.1 (dashed line), and 0.01 (dotted line) $L_{\rm edd}$.
We separate the MMT (red) and SDSS (blue) subsamples by their morphologies,
%and the `extended' and 'point-like' sources by their symbols: 
pluses and filled circles for the `star-like' point sources, 
and open diamonds and squares for the `galaxy-like' extended sources (See \S~\ref{sec:completeness}).
The MMT quasars have lower $\lbol$ and $\mbh$
than their SDSS counterparts at $z < 3$. 
\label{fig:ledd1}}
\end{figure}

\begin{figure}
\begin{center}
\includegraphics[scale=0.7]{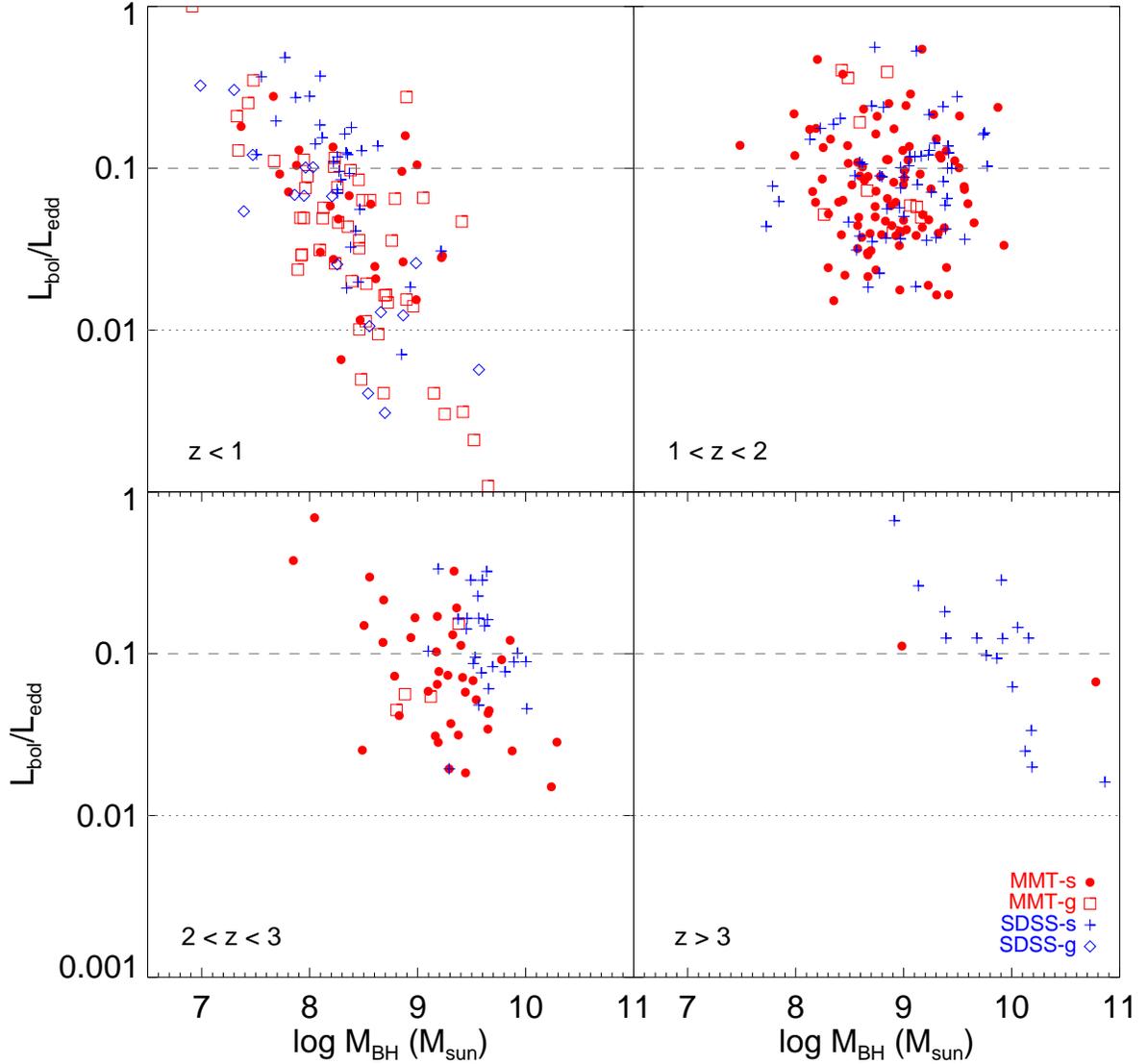}%{mbh-lbol-ledd-ext-s11-r.eps}  %fig25
\end{center}
\figcaption{Eddington Ratios ($\lbol/L_{\rm edd}$) of MIR-selected quasars as a function of SMBH mass ($\mbh$). 
Color codes and legends are the same as in Fig.~\ref{fig:ledd1}. 
The dashed and dotted lines mark the 0.1 and 0.01 Eddington ratios, respectively.
At $z < 1$, the extended sources show clearly lower $\lbol$ ($\sim$ 0.7 dex) 
and an average of $\sim$ 3 $\times$ lower Eddington ratios than the point sources;
at $1 < z < 2$ and $ 2< z < 3$, where limited extended sources are available, 
the point sources show a wide span of Eddington ratios and
scatter into the $\lbol/L_{\rm edd} < 0.1$ regime.
\label{fig:ledd2}}
\end{figure}

\begin{figure}
\begin{center}
\includegraphics[scale=0.7]{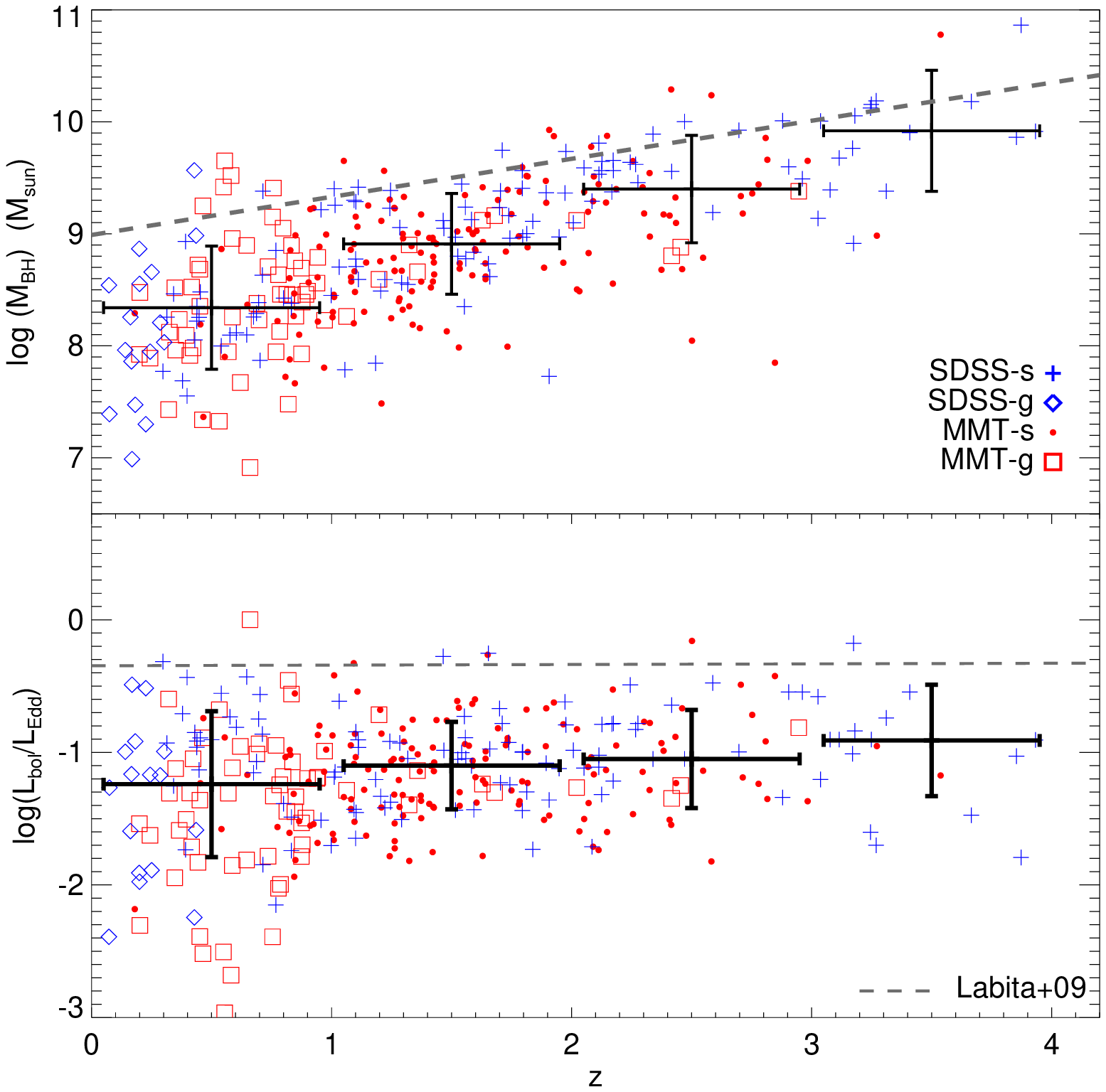}%{downsizer.eps}  %fig26
\end{center}
\figcaption{ SMBH mass ($\mbh$, top) and Eddington Ratios ($\lbol/L_{\rm Edd}$, bottom)
of MIR-selected quasars as a function of redshift.
Color codes and legends are the same as in Fig.~\ref{fig:ledd1}. 
In dashed lines are the proposed maximum mass values ($M_{\bullet(max)} = 0.34\,z + 8.99$)
and Eddington ratios ($\lbol/L_{\rm Edd} (max) = 0.005\,z + 0.45$) from \citet{labita09}.
The mean and standard deviation in each redshift bin is marked by black pluses.
We observe a downsizing effect in $\mbh$ but a more or less constant Eddington ratios
across the cosmic time.
\label{fig:downsize}}
\end{figure}

In Fig.~\ref{fig:ledd1}, we compare the $\mbh$ with $\lbol$. 
The diagonal line marks the Eddington luminosity 
for the corresponding SMBH mass.
Quasars rarely exceed $L_{\rm Edd}$ \citep{kollmeier06} 
and SDSS quasars tend to lie above $\sim$0.05 $L_{\rm Edd}$,
and below a `sub-Eddington boundary' \citep[][]{fkm04,labita09,sande10}.
Controversies exist as to whether the 
observed sub-Eddington limit is due to the incompleteness of SDSS sample
at low $\mbh$ ($\mbh < 3 \times 10^8 \msun$) 
and low Eddington ratio (ER, $L/L_{\rm Edd} < 0.07$)
 \citep{kellynshen13}.
For the MIR-selected sample, 
we do not observe a clear sub-Eddington limit (Fig.~\ref{fig:ledd2}).
The $\mbh$ for MIR-selected quasars
shows a trend of downsizing,
though the $\lbol/L_{\rm Edd}$ is 
relatively independent of redshift (Fig.~\ref{fig:downsize}).
These trends are similar to the results from the SDSS DR5 quasars \citep{labita09}.
Table~\ref{tab:ledd} summarizes 
the $\mbh$, $\lbol$, and $\lbol / L_{\rm Edd}$ differences between
the MMT and SDSS subsamples, 
and between point and extended sources.
At all redshift ranges,
the MMT identified quasars have a lower
median $\lbol/L_{\rm Edd}$ ratio than their SDSS counterparts,
possibly related to the inclusion of extended sources in the MMT sample,
since the mean $\lbol/L_{\rm Edd}$ ratio is
also lower for extended targets at all redshift.

At $z < 1$, the extended sources show lower $\lbol$ ($\sim$ 0.4 dex)
and lower ER (by a factor of 2) than the point sources (Fig.~\ref{fig:ledd2}). 
It is possible that the extended quasars %with extended photometric morphology 
reside in brighter or more massive host galaxies,
and at a less active evolutionary phase with lower $\lbol/L_{\rm Edd}$.
Of the 12 targets with rather low ERs ($\lbol / L_{\rm Edd} < 0.01$),
10 are extended sources.
Of the remaining 58 extended sources at $z < 1$,
16 have a $\lbol/L_{\rm Edd} > 0.1$,
and 42 are at $0.01 < \lbol/L_{\rm Edd} < 0.1$.
The $\lbol/L_{\rm Edd}$ may be underestimated
as quasars may contribute significantly in the rest-frame FIR
as suggested by \citet{kuras03} and \citet{dai12}.
On the other hand, the ER
may also be overestimated
because of the possible host contribution to the $\lbol$
at $z < 1$;
though the reddening correction of the spectra
will counteract that effect.
In the spectrum of at least a few MIR-selected SDSS sources 
with extended photometry,
stellar absorption and sometimes a Balmer break is observed.
For example, the 6 newly-identified SDSS quasars with extended morphology
all show signatures of host galaxy (e.g. Fig.~\ref{fig:hb6}): 
all have CaII H\&K absorption,
and four (4) also show the G band in absorption. 

At $1 < z < 3$, 
the MMT identified subsample has systematically lower $\lbol$ 
and $\mbh$ than their SDSS counterparts (Fig.~\ref{fig:ledd1}, see also Table~\ref{tab:ledd}).
The MMT sources extend the SDSS selection
to fainter magnitudes (Fig.~\ref{fig:Mi}),
so at a given redshift, 
they must either have lower $\lbol/L_{\rm Edd}$, 
or of smaller $\mbh$. 
\citet{kellynshen13} suggested that the
sub-Eddington boundary found for SDSS quasars
was a magnitude-limit effect, 
and there was a large population of 
low $\lbol/L_{\rm Edd}$ quasars down to $ \mbh\,\sim\,5\,\times 10^8 \msun$ 
(log$(\mbh) = 8.7\,\msun$)
and $\lbol/L_{\rm Edd} \sim 0.07$ (log$(\lbol/L_{\rm Edd}) = -1.15$). 
These do not appear in the MIR-quasar population for 1$< z < $3.
Instead of a shift of the $\mbh$ and $\lbol/L_{\rm Edd}$ to smaller values,
comparable mean and scatter of ERs and $\mbh$ are observed
at 1$< z < $2 and 2$< z < $3 (Fig.~\ref{fig:ledd2}). 
At $1 < z < 3$, the point sources also scatter 
into the $< 0.1\,\lbol/L_{\rm edd}$ regime.
However, given the small numbers of extended objects at $z > 1$---possibly 
due to the resolution restrictions of the telescope---it is 
difficult to tell whether there is any systematic difference
in the SMBH accretion rate between 
extended and point-like quasars at earlier cosmic time.

\begin{table} [hbdp]
\caption{Sample Entry of the basic parameters for the MIR-selected quasar sample.}
\begin{center}
\begin{tabular}{lcccccc}
\hline
\hline
ID & RA & DEC & redshift  & $S_{\rm 24}$  & $\delta S_{\rm 24}$ & SDSS photometry\\
   & (J2000) & (J2000) & & $\mu Jy$ & $\mu Jy$ & (u, g, r, i, z)\\
 \hline
 2009.0131-005    &     162.3289  &  59.4024 & 1.650 & 2282.43 &19.42 & (21.73, 21.07, 20.69, 20.22, 19.89) \\
\hline
\end{tabular}
\end{center}
(This table is available in its entirety in a machine-readable form in the online journal. 
A portion is shown here for guidance regarding its form and content. 
Detailed catalog format can be found in Table~\ref{tab:properties}.
SDSS photometry errors are not shown here due to space limitation.)
\label{tab:cat1}
\end{table}

\begin{table} [hbdp]
\caption{Luminosities and SMBH mass of the MIR-selected quasar sample.}
\begin{center}
\begin{tabular}{lccccccccc}
\hline
\hline
ID & Flag\_EXT & Flag\_ABS & Flag\_FAINT  & log$\mbh$  & $\delta \mbh$ & log$\lbol$ & $\delta\lbol$ & $\lbol / L_{\rm edd}$ & ... \\
   & & & &$\msun$ & in \% & $\ergs$ & in \% & \\
 \hline
 2009.0131-005    & 6  &  1  &0 &  9.17  &   0.17  &   46.99 & 0.01 & 0.50 & ...\\
\hline
\end{tabular}
\end{center}
(This table is truncated for viewing convenience. 
It is available in its entirety in a machine-readable form in the online journal. 
A portion is shown here for guidance regarding its form and content. 
Detailed catalog format can be found in Table~\ref{tab:results}.)
\label{tab:cat2}
\end{table}

\begin{table} [hbdp]
\caption{Fitting parameters of the MIR-selected quasar sample.}
\begin{center}
\begin{tabular}{lccccccc}
\hline
\hline
ID & ... & \civ\_ dom\_ FWHM &... &  \mgii\_ dom\_ FWHM & ... &  \hbeta\_ dom\_ FWHM  & ... \\
 & & $\kms$ & & \kms & & \kms&  \\
 \hline
 2009.0131-005    &... & 2056.6 $\pm$ 221.2 &... & 4773.9 $\pm$  363.0 &... & ... & ...   \\
\hline
\end{tabular}
\end{center}
(This table is truncated for viewing convenience, 
only the dominant FWHM for each line is listed.
Additional columns and
format information can be found in Table~\ref{tab:param}.
It is available in its entirety in a machine-readable form in the online journal. 
A portion is shown here for guidance regarding its form and content. )
\label{tab:cat3}
\end{table}

\section{The Spectral Catalog}
\label{sec:cata}
We have included all the measured properties from line fitting, and the derived properties 
in the online master catalogs. 
This catalog will be available in its entirety in a machine-readable form in the online journal.  
Objects are arranged in increasing RA order, 
and the ID reflects the spectroscopic subsamples: MMT09, MMT05b and SDSS. 
The MMT05f faint objects are then appended to the end of each table for
comparison.
Table~\ref{tab:cat1},\ref{tab:cat2}, and \ref{tab:cat3} show sample entries 
of the three master tables.
Table~\ref{tab:cat1} lists all the basic parameters, including the object ID, position, redshift, SDSS 
and MIPS 24$\mu m$ photometries of the quasar sample;
Table~\ref{tab:cat2} shows a sample entry the results, including flags, luminosities, SMBH mass, and ERs;
Table~\ref{tab:cat3} includes the fitting parameters: continuum normalization and slope, 
iron template normalization and broadening, 
wavelength, S/N, FWHM, line area, and EW of each emission line.
The catalog format can be found in Table~\ref{tab:properties}, Table~\ref{tab:results},
and Table~\ref{tab:param}.
Unless otherwise stated, a null value is given if no measurements are available. 
%and a null value for its associated uncertainty.

\section{Discussion}
\label{sec:discussion}
The catalog of MIR-selected quasars can be used to study the statistics of 
type 1 quasars and their physical properties.

We find that a significant and constant fraction (20\%) of MIR-selected quasars
have extended optical photometry at $z < 1.5$, 
indicating luminous host galaxies (Table~\ref{tab:completeness}). 
%This redshift limit is mostly due to resolution limits. 
The MMT-recovered quasars include a small population of redder targets
than the SDSS quasars (Fig.~\ref{fig:f24-r}). 
The MMT quasars share similar distributions with the SDSS 
quasars in all colors at $i \leq 20.2$,
and cover fainter objects than SDSS did not cover at $i > 20.2$ 
(Fig.~\ref{fig:colors1},\ref{fig:colors2},\ref{fig:colors3}).
The SDSS quasar algorithm is biased towards point sources at $i > 19.1$
and is therefore missing quasars residing in extended hosts. 
Unresolved quasars comprise about 94\% of all SDSS quasars. 
SDSS did not include extended objects in their target selection 
based on the assumption that the expected yield of quasars would be low.
The MIR flux limit used in this sample is more inclusive
and recovers the otherwise rejected extended sources. 
%The 24\micron\ selection favors hot dust rich hosts.
The extended population consists of 20\% of the total MIR quasar population,
and calls for re-examination and updated simulations for 
quasar distributions at all redshifts. 

Although the SDSS algorithm completeness was 
simulated and found to be consistent with MIR color-selected quasar samples, 
e.g. \citet{lacy13},
we discovered additional quasars using the flux-limited MIR-selection.
At $i > 19.1$, 9 additional MIR quasars that meet
the SDSS selection were recovered with the MMT spectroscopy,
resulting in an updated SDSS completeness of 70\%. 
At $i < 20.2$ and $z > 2.9$, we only found 1 additional MIR quasar which is consistent with
the SDSS completeness of 90\%. 
This completeness difference arises from the different selection criteria,
as both optical and MIR color selections restrict the sample to power-law like SEDs,
whereas the MIR flux selection adopted here 
includes everything that meet the apparent magnitude requirement.
At $z < 3$ and $i > 19.1$,
the observed quasar number densities per square degree are higher 
than at the SDSS covered $i < 19.1$ region.

%derived from simulation based on 
%a uniform redshift distribution R02,
%and it was pointed out that the completeness level may be 
%overestimated for $2.4 < z < 3.0$ due to selection bias \citep{vandenberk05}, 
%a modified $> 80\%$ completeness at all other $z$ ranges
%is still optimistic, 
%given our discovery of additional quasars, especially at $i > 19.1$. 
%only 5 MMT sources are at $2.4 < z < 3.0$,
%which would not significantly change the overall completeness. 
%In the MIR-selected sample, 
%the fraction of objects at 2.4 $< z <$ 3.0 is 5.2\%, 
%almost identical to the 5.3\% in SDSS DR7 quasar catalog.

In Fig.~\ref{fig:sp}, the MIR-selected quasars
show a redshift distribution peaking at $z\sim$1.4, consistent
with previous studies of the cosmic evolution of
AGN number densities \citep{hms05,silverman08}.
We see evidence of downsizing in the MIR-selected targets,
with the most massive SMBHs appearing at earlier times;
though the ER remains almost constant at $1 < z < 4$ with large scatters.
Objects with low $\lbol/L_{\rm Edd} < 0.01$ are also observed at $z < 1$.

\begin{figure}
\begin{center}
\includegraphics[scale=1.0]{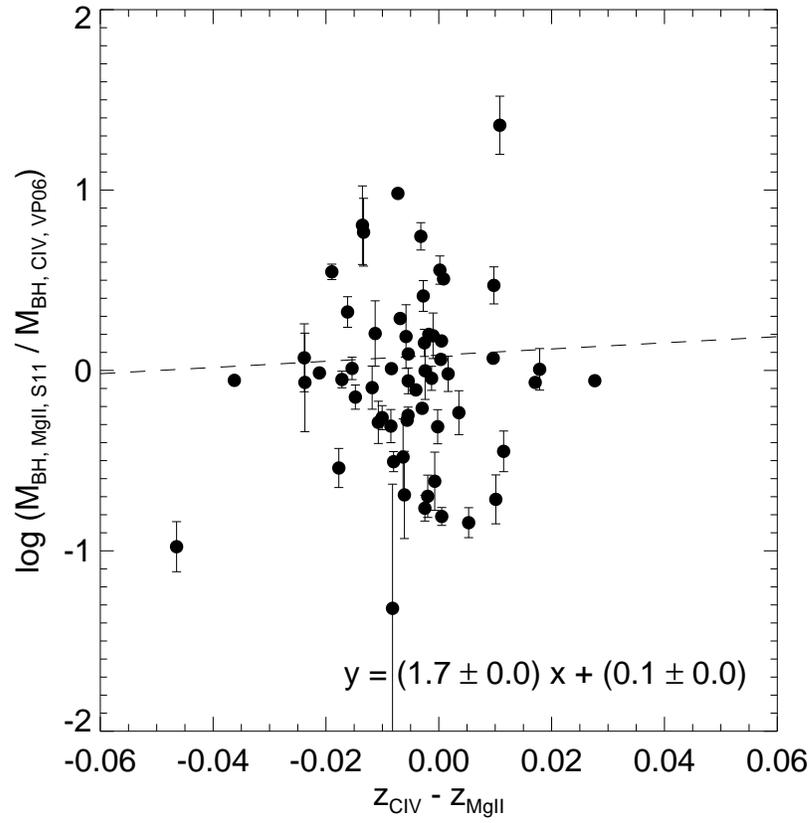}%{voff.eps} %civmgiiratio-s11-fit-non.eps}  %fig27
\end{center}
\figcaption{Comparison of the \civ$-$\mgii\ redshift  and $\mbh$ differences %log $(M_{\rm BH, CIV, VP06} / M_{\rm BH, MgII, S11})$
in the same objects covering both \civ\ and \mgii\ BELs.
Blueshifted \civ\ corresponds to negative values.
The $\mbh$ ratios indicate a loose to null correlation 
with the redshift difference ($z_{\rm \civ}-z_{\rm \mgii}$).
%:  log$(M_{\rm BH, MgII} / M_{\rm BH, CIV}) = (1.7 \pm 0.0) \times (z_{\civ}-z_{\mgii}) + (0.1 \pm 0.0)$. 
%The large scatter may be related to the fluctuation due to 
%non-virial component in the \civ\ BEL profile.  
\label{fig:voff}}
\end{figure}

Controversies exist as to whether \civ\  
line-widths are attributed solely to gravity,
or are affected by outflows or jets,
and as a result, whether the \civ\ emission 
derived masses are as reliable as 
\mgii\ and \hbeta\  derived masses \citep[VP06,][]{shen08, assef11}.
This concern arises from 
both
the typically 
blueshifted \civ\ BEL peak compared to other quasars BELs \citep[][S11]{gaskell82, rich02b},
the commonly observed BAL/NALs \citep[][W08]{weymann81} within the \civ\ emission line profiles, 
and the strong line asymmetries \citep[][See also \S~\ref{sec:spec}]{wilkes84, rich02b}. 
The blueshift of the \civ\ BEL peak 
relative \mgii\
is observed in $\sim$80\% of the MIR-quasars
whose spectra
covers both \civ\ and \mgii\ BELs. 
In the MIR-selected quasar sample,
there is no strong correlation between 
the \civ\ and \mgii\ FWHMs (Figure.~\ref{fig:civmgiifwhm}, left).
There is also no strong trend of 
decreasing ratios of log$(M_{\rm BH, (MgII)} / M_{\rm BH, (CIV)})$ 
with increasing \civ$-$\mgii\, blueshifts,
in contrast to the correlation reported in S11 \& \citet{richards11},
although the scatter is large for both
$\mbh$ ratios and \civ$-$\mgii\, blueshifts (Fig.~\ref{fig:voff}).

A non-virial \civ\ emission component 
can be used to explain the large scatter observed
between \civ\ and other BEL derived $\mbh$ \citep[S11][]{richards11, denney12}.
\citet{denney12}
found a `non-variable, largely core' emission component
in the \civ\ BEL
by comparing the SE spectra to the RM spectra.
After removing this non-variable component,
the \civ\ derived $\mbh$ shows a better 
correlation with the \hbeta\ derived $\mbh$. 
In this MIR-selected quasar sample,
we found that the $\mbh$ derived from the dominant \civ\ FWHM
shows a marginally better correlation with the $\mbh$ from \mgii\ BEL (slope coefficient $=$ 0.61 $\pm$ 0.11)
than that from the non-parametric \civ\ FWHM (slope coefficient $=$ 0.42 $\pm$ 0.07, 
Figure.~\ref{fig:civmgiifwhm}, right) and has smaller scatter.
If non-parametric \civ\  FWHM is used instead,
a sudden jump in the $\mbh$ distribution at $z \sim$ 1.6
would appear, where the $\mbh$ starts to be derived from the \civ\  BELs.
This sudden increase is not physical 
and supports our choice of the dominant \civ\ FWHM.
In 70\% of the \civ\ BEL with multiple Gaussians,
the non-parametric \civ\ FWHMs are smaller than the 
dominant \civ\ FWHMs,
due to contributions from narrower Gaussians
that fit the line core
(\eg, Fig.~\ref{fig:egciv}).
These narrower additional Gaussian component resembles 
the non-virial emission component found in \citet[][Fig.3]{denney12}.
The marginally better correlation of dominant \civ\ $\mbh$ to \mgii\ derived $\mbh$
suggests contamination from non-virial \civ\ components
to the non-parametric \civ\ FWHM. 
The choice of dominant \civ\ FWHM instead of the 
conventional non-parametric FWHM 
for $\mbh$ estimates may
provide a way to tackle this problem.

We find a high fraction of objects with 
absorption features in the MIR-selected sample.
For \civ, 
$\sim$40\% of the BELs quasars show absorption, NALs or BALs;
and this fraction is $\sim$ 20 \% for \mgii\ objects. 
The fraction of BALs in MIR-selected quasars, 
is 17$\pm\ 3\%$ in \civ, % (1.63 $< z < $4.40),
and 10$\pm$2\% in \mgii. % (0.43 $< z < $ 2.10).
The \civ\ numbers agree with the overall fraction of 10-15\% found for SDSS quasars \citep[e.g.][]{trump06,knigge08}.
The \mgii\ quasars show a $> 3\sigma$ higher fraction of BALs than the 1.31\% in \citet{trump06}. 
A velocity offset ($v_{\rm off}$) of $|v_{off}| \leq 3000\,\kms$
 between the NAL/BALs and the system redshift
is commonly used to define the associated NALs \citep[e.g.][]{wild08, shen12}.
%though earlier papers %(FIND!)  used 5000-6000\,\kms.  
The boundary between NAL and BAL widths also differs from paper to paper.
\citet[][W08]{wild08} used an upper boundary of 700 \kms\ for associated NALs,
and \citet{shen12} used 500 \kms. 
Here we adopt the 700 \kms\ limit
and find the fraction of NALs to be 17\% (27 objects) for \civ\  and 13\% (40 objects) for \mgii.
These fractions are $\sim 3\sigma$ higher 
for both \civ\ and \mgii\ NALs than in the SDSS-color-selected quasars \citep{wild08}.
They are consistent with the SDSS quasars
within errors for high velocity ($> 3000 \kms$) narrow absorbers \citep{vest03, wild08}.
About 25\% and 20\% of the \civ\  and \mgii\ absorbers show absorptions
redshifted from the emission line peak, 
indicating possible inflows towards the SMBH in the BLR. 
We will present the subsample of quasars 
with redshifted \civ\ absorption and 
explore the possibilities in a forthcoming paper (Dai \etal, 2014, in preparation).

\section{Summary}
\label{sec:sum}
We construct a catalog of MIR-selected quasars in the Lockman Hole-SWIRE field
and present their SMBH mass 
and Eddington ratios in this paper.
This broad-emission-line, type 1 quasar sample is MIPS 24\micron\ selected and
optically identified in three spectroscopic surveys: MMT09, MMT05b, and SDSS. 
In the catalog we compiled their photometries, continuum and emission line properties,
and luminosities and virial SMBH mass ($\mbh$) derived from the spectral measurements. 

We find a significant population of quasars with extended photometric morphologies.
%which was overlooked in the SDSS quasar survey.
A constant fraction of 20\% extended objects are 
observed in the MIR-selected quasars across the magnitude ranges.
We then compare and estimate the completeness of the SDSS quasar selection algorithm
to be $\sim 70\%$ at i $<$ 19.1,
about 20\% lower than the reported 90\% completeness (R02).
At 19.1 $< i < $20.2 and $z > 3$, 
our result is consistent with the reported SDSS completeness.
%On the other hand, the SDSS rejection of $z< 3$ objects at 19.1 $< i \leq $20.2 
%due to efficiency concerns is too conservative, 
At $z< 3$, SDSS only covered the $i < 19.1$ region.
In this redshift range, we observe a significantly higher quasar number density at 
19.1 $< i < $20.2  ($\sim$24\,deg$^{-2}$) than at $i < 19.1$  ($\sim$10\,deg$^{-2}$). 
The number density at $i > 20.2$ is even higher, reaching $\sim$45\,deg$^{-2}$. 
The MIR selection used here efficiently extends
the magnitude limit of the quasar population to the low z sources. 
Compared to color selections, 
the MIR selection recovers a high fraction of extended objects,
and provides a more complete sample 
to study the total quasar population. 
%Follow-up studies of the extended and $z< 3$, 19.1 $< i \leq $20.2 objects 
%will be of great value
%for our understanding of the total quasar population.} 
%for both point and extended sources,
%and (13$\pm$7\%) at 19.1 $< i < $20.2 for point sources. 
%These values are significantly lower than the reported 

We measure the linewidth and calculated the virial SMBH mass ($\mbh$), 
bolometric luminosity ($\lbol$),
and the ERs ($\lbol / L_{\rm edd}$) for the MIR-selected quasars. 
The consistency between the $\mbh$ estimated by the \civ\, \mgii\, and \hbeta\ emission lines is also tested.
We found a better correlation between \civ\ and \mgii\ derived $\mbh$ using
the dominant \civ\ FWHM instead of the conventional non-parametric \civ\ FWHM,
indicating contribution from non-virial component to the latter. 

The $\log \mbh (\msun)$ derived from emission line-width has an 
average error of $\sim$30\%,
with a distribution from 7-11, 
peaking at $\log \mbh (\msun) = 8.8 $. 
The MMT identified quasars supplement the SDSS quasars at lower $\mbh$, lower $\lbol$, 
and in the SDSS exclusion zones.
A systematic offset in Eddington ratios 
is found between extended and point sources at $z < 1$,
indicating a less active AGN phase for the extended objects.
Similar large scatter of $\lbol/L_{\rm Edd}$ from 0.01 to 1 is observed 
at $1< z < 3$.
The $\mbh$ for MIR-selected quasars shows a strong trend of downsizing, 
but the Eddington ratio remains relatively independent of redshift.

We also find a high fraction of quasars with absorption features
in this MIR-selected sample, 
which will be presented in a forthcoming paper. 

The complete catalog is being made publicly available online along with the MMT-Hectospec spectra. 
A similar 24\micron\ flux-limited redshift survey by the authors
is underway in XMM-LSS, FLS (First Look Survey), and the EGS (Extended Groth Strip) fields.

\acknowledgments 
%We thank the referee for insightful comments that helped improve the 
%clarity of this manuscript. 
%,  XXX for assistance with XXX ......
%acknowledges support from the . %We thank the referee for helpful comments.
Y.S.D acknowledges support from the 
Smithsonian Astrophysical Observatory (SAO) through the SAO Predoctoral Fellowship.
We thank the anonymous referee for suggestions that led to the
improvement of the manuscript.
We thank Yue Shen for sharing the spectral measurement code, 
and Richard Cool for sharing and supporting of the HSRED reduction code.

Observations reported here were obtained at the MMT Observatory, a
joint facility of the Smithsonian Institution and the University of Arizona. 
This work is also based partly on observations made with the Spitzer Space Telescope,
operated by the Jet Propulsion Laboratory, California Institute of Technology under a contract with NASA.
Funding for the SDSS and SDSS-II has been provided by
the Alfred P. Sloan Foundation, the Participating Institutions,
the National Science Foundation, the U.S. Department of
Energy, the National Aeronautics and Space Administration,
the Japanese Monbukagakusho, the Max Planck Society, and
the Higher Education Funding Council for England. The SDSS
Web site is http://www.sdss.org/.

{\it Facility:}  \facility{MMT},  \facility{Spitzer Space Telescope}, \facility{Sloan}

\clearpage

\clearpage

\begin{landscape}
\begin{longtable}{ccc}
\caption{The MIR-selected Quasar Catalog 1. Properties}\label{tab:properties}\\
\hline \hline \\[-2ex]
   \multicolumn{1}{c}{\textbf{Column}} &
   \multicolumn{1}{c}{\textbf{Format}} &
   \multicolumn{1}{c}{\textbf{Description}} \\[0.5ex] \hline
   \\[-1.8ex]
\endfirsthead
%This is the header for the remaining page(s) of the table...
\multicolumn{3}{c}{{\tablename} \thetable{} -- Continued} \\[0.5ex]
  \hline \hline \\[-2ex]
  \multicolumn{1}{c}{\textbf{Column}} &
  \multicolumn{1}{c}{\textbf{Format}} &
  \multicolumn{1}{c}{\textbf{Description}} \\[0.5ex] \hline
  \\[-1.8ex]
\endhead
%This is the footer for all pages except the last page of the table...
  \hline
  \multicolumn{3}{l}{{Continued on Next Page\ldots}} \\
\endfoot
%This is the footer for the last page of the table...
  \\[-1.8ex] \hline \hline
\endlastfoot
1\dotfill  & STRING   & MMT designation of observation date-fiber number,  \\
\dotfill & \dotfill         &  or SDSS DR7 designation of spectroscopic MJD$+$plate number$+$ fiber number\\%hhmmss.ss$+$ddmmss.s (J2000.0)\\
2\dotfill   & DOUBLE & Right ascension in decimal degrees (J2000.0) \\
3\dotfill   & DOUBLE & Declination in decimal degrees (J2000.0) \\
4\dotfill   & DOUBLE & Redshift (See \S~\ref{sec:datareduction})\\
5\dotfill & DOUBLE   & 24 \micron\ flux density ($S_{24}$) from SWIRE photometry\\
6\dotfill & DOUBLE   & Uncertainty in 24 \micron\ flux density ($S_{24-ERR}$) from SWIRE photometry\\
7--11\dotfill & DOUBLE   & SDSS photometry in AB magnitude (p.u, p.g, p.r, p.i, p.z) \\
12--16\dotfill & DOUBLE   & Uncertainty in SDSS photometry (p.Err\_u, p.Err\_g, p.Err\_r, p.Err\_i, p.Err\_z) \\
%\end{deluxetable*}
\end{longtable}
\tablecomments{}
\normalsize

%\clearpage
\begin{longtable}{ccc}
\caption{The MIR-selected Quasar Catalog 2. Results}\label{tab:results}\\
\hline \hline \\[-2ex]
   \multicolumn{1}{c}{\textbf{Column}} &
   \multicolumn{1}{c}{\textbf{Format}} &
   \multicolumn{1}{c}{\textbf{Description}} \\[0.5ex] \hline
   \\[-1.8ex]
\endfirsthead
%This is the header for the remaining page(s) of the table...
\multicolumn{3}{c}{{\tablename} \thetable{} -- Continued} \\[0.5ex]
  \hline \hline \\[-2ex]
  \multicolumn{1}{c}{\textbf{Column}} &
  \multicolumn{1}{c}{\textbf{Format}} &
  \multicolumn{1}{c}{\textbf{Description}} \\[0.5ex] \hline
  \\[-1.8ex]
\endhead
%This is the footer for all pages except the last page of the table...
  \hline
  \multicolumn{3}{l}{{Continued on Next Page\ldots}} \\
\endfoot
%This is the footer for the last page of the table...
  \\[-1.8ex] \hline \hline
\endlastfoot
1\dotfill  & STRING   & MMT designation of observation date-fiber number\hfill \\
\dotfill & \dotfill         &  or SDSS DR7 designation of spectroscopic MJD$+$plate number$+$ fiber number\\
2\dotfill   &  INTEGER & extended source flag (p.type), '3' for extended object, '6' for point source \\ %Flag\_EXT, 
3\dotfill   & INTEGER &  absorption flag, '1' for sources with absorption, '0' for targets without absorption \\ %Flag\_ABS, 
4\dotfill   & INTEGER &  faint object flag, '1' for sources with $S_{\rm 24} < 400\mu$Jy--- the MMT05f subsample \\ %Flag\_FAINT, 
% BAD_FLAG: 
5\dotfill & DOUBLE   & Fiducial Virial SMBH mass $\log\mbh$ in $\msun$ (\S~\ref{sec:virial}) \\
6\dotfill & DOUBLE   & Measurement uncertainty of the fiducial $\log\mbh$ in percentage\\
7\dotfill & DOUBLE  & Bolometric luminosity $\log\lbol$ in $\ergs$\\
8\dotfill & DOUBLE  & Uncertainty in $\log\lbol$ in percentage \\
9\dotfill & DOUBLE   & Eddington ratio ($\lbol/L_{\rm edd}$) based on the fiducial $\mbh$\\
10\dotfill & DOUBLE   & Virial SMBH mass based on dominant \civ\ , $\log M_{BH, \civ\, VP06}$ in $\msun$ \\
11\dotfill & DOUBLE   & Measurement uncertainty in dominant \civ\ ,  $\log M_{BH, \civ\, VP06}$ in percentage \\
12\dotfill & DOUBLE   & Virial SMBH mass based on non-parametric \civ\, $\log M_{BH, \civ\ non, VP06}$ in $\msun$ \\
13\dotfill & DOUBLE   & Measurement uncertainty in non-parametric \civ\ ,  $\log M_{BH, \civ\ non, VP06}$ in percentage\\
14\dotfill & DOUBLE   & Virial SMBH mass based on dominant \mgii\ , $\log M_{BH, \mgii\, MD04}$ in $\msun$ \\
15\dotfill & DOUBLE   & Measurement uncertainty in dominant \mgii\ ,  $\log M_{BH, \mgii\, MD04}$ in percentage\\
16\dotfill & DOUBLE   & Virial SMBH mass based on non-parametric \mgii\ , $\log M_{BH, \mgii\ non, MD04}$ in $\msun$\\
17\dotfill & DOUBLE   & Measurement uncertainty in non-parametric \mgii\ ,  $\log M_{BH, \mgii\ non, MD04}$ in percentage\\
18\dotfill & DOUBLE   & Virial SMBH mass based on dominant \mgii\ , $\log M_{BH, \mgii\, VO09}$ in $\msun$ \\
19\dotfill & DOUBLE   & Measurement uncertainty in dominant \mgii\ ,  $\log M_{BH, \mgii\, VO09}$ in percentage\\
20\dotfill & DOUBLE   & Virial SMBH mass based on non-parametric \mgii\ , $\log M_{BH, \mgii\ non, VO09}$ in $\msun$\\
21\dotfill & DOUBLE   & Measurement uncertainty in non-parametric \mgii\ ,  $\log M_{BH, \mgii\ non, VO09}$ in percentage\\
22\dotfill & DOUBLE   & Virial SMBH mass based on dominant \mgii\ , $\log M_{BH, \mgii\, S11}$ in $\msun$ \\
23\dotfill & DOUBLE   & Measurement uncertainty in dominant \mgii\ ,  $\log M_{BH, \mgii\, S11}$ in percentage\\
24\dotfill & DOUBLE   & Virial SMBH mass based on non-parametric \mgii\ , $\log M_{BH, \mgii\ non, S11}$ in $\msun$\\
25\dotfill & DOUBLE   & Measurement uncertainty in non-parametric \mgii\ ,  $\log M_{BH, \mgii\ non, S11}$ in percentage\\
26\dotfill & DOUBLE   & Virial SMBH mass based on dominant \hbeta\ , $\log M_{BH, H\beta, VP06}$ in $\msun$ \\
27\dotfill & DOUBLE   & Measurement uncertainty in dominant \hbeta\ ,  $\log M_{BH, H\beta, VP06}$ in percentage\\
%28\dotfill & DOUBLE   & Virial SMBH mass based on non-parametric \hbeta\ , $\log M_{BH, H\beta non, VP06}$ in $\msun$\\
%29\dotfill & DOUBLE   & Measurement uncertainty in non-parametric \hbeta\ ,  $\log M_{BH,  H\beta non, VP06}$ in percentage\\
%
28\dotfill & DOUBLE   & Virial SMBH mass based on dominant \hbeta\ , $\log M_{BH, H\beta, MD04}$ in $\msun$ \\
29\dotfill & DOUBLE   & Measurement uncertainty in dominant \hbeta\ ,  $\log M_{BH, H\beta, MD04}$ in percentage\\
%32\dotfill & DOUBLE   & Virial SMBH mass based on non-parametric \hbeta\ , $\log M_{BH, H\beta non, MD04}$ in $\msun$\\
%33\dotfill & DOUBLE   & Measurement uncertainty in non-parametric \hbeta\ ,  $\log M_{BH, H\beta, non, MD04}$ in percentage\\
%
30\dotfill & DOUBLE   & Monochromatic line luminosity at 1350\AA\ $\log L_{\rm 1350}$ in $\ergs$ \\
31\dotfill & DOUBLE   & Uncertainty in $\log L_{\rm 1350}$ in percentage \\
32\dotfill & DOUBLE   & Monochromatic line luminosity at 3000\AA\ $\log L_{\rm 3000}$ in $\ergs$\\
33\dotfill & DOUBLE   & Uncertainty in $\log L_{\rm 3000}$ in percentage \\
34\dotfill & DOUBLE   & Monochromatic line luminosity at 5100\AA\ $\log L_{\rm 5100}$ in $\ergs$\\
35\dotfill & DOUBLE   & Uncertainty in $\log L_{\rm 5100}$ in percentage \\
36\dotfill & DOUBLE  & Bolometric luminosity $\log\lbol\_\civ$ in $\ergs$ (0.580925$+ \log L_{\rm 1350}$)\\
37\dotfill & DOUBLE  & Uncertainty in $\log\lbol\_\civ$ in percentage\\
38\dotfill & DOUBLE  & Bolometric luminosity $\log\lbol\_\mgii$ in $\ergs$ (0.711807$+ \log L_{\rm 3000}$)\\
39\dotfill & DOUBLE  & Uncertainty in $\log\lbol\_\mgii$ in percentage\\
40\dotfill & DOUBLE  & Bolometric luminosity $\log\lbol\_H\beta$ in $\ergs$ (0.96661$+ \log L_{\rm 5100}$)\\
41\dotfill & DOUBLE  & Uncertainty in $\log\lbol\_H\beta$ in percentage\\
42\dotfill & DOUBLE  & Bolometric luminosity $\log\lbol\_{\rm MIR}$ in $\ergs$ (conversion factor is redshift dependent) \\
43\dotfill & DOUBLE  & Uncertainty in $\log\lbol\_{\rm MIR}$ in percentage\\
%\enddata
%\end{deluxetable*}
\end{longtable}
\tablecomments{
Notes: Unless otherwise stated, null value is given if not measured, and $-1$ for its associated error. }
%(1) Bolometric luminosities computed using bolometric corrections in
%\citet{rich06} using one of the $5100$\AA, $3000$\AA, or
%$1350$\AA\ monochromatic luminosities depending on redshift; 
%(2) Uncertainties are measurement errors only; 
\normalsize

%\end{landscape}
%\begin{landscape}
\begin{longtable}{ccc}
\caption{The MIR-selected Quasar Catalog --3. Parameters}\label{tab:param}\\
%\scriptsize
\hline \hline \\[-2ex]
   \multicolumn{1}{c}{\textbf{Column}} &
   \multicolumn{1}{c}{\textbf{Format}} &
   \multicolumn{1}{c}{\textbf{Description}} \\[0.5ex] \hline
   \\[-1.8ex]
\endfirsthead
%This is the header for the remaining page(s) of the table...
\multicolumn{3}{c}{{\tablename} \thetable{} -- Continued} \\[0.5ex]
  \hline \hline \\[-2ex]
  \multicolumn{1}{c}{\textbf{Column}} &
  \multicolumn{1}{c}{\textbf{Format}} &
  \multicolumn{1}{c}{\textbf{Description}} \\[0.5ex] \hline
  \\[-1.8ex]
\endhead
%This is the footer for all pages except the last page of the table...
  \hline
  \multicolumn{3}{l}{{Continued on Next Page\ldots}} \\
\endfoot
%This is the footer for the last page of the table...
  \\[-1.8ex] \hline \hline
\endlastfoot
1\dotfill  & STRING   & MMT designation of observation date-fiber number, \\
\dotfill & \dotfill         &  or SDSS DR7 designation of spectroscopic MJD$+$plate number$+$ fiber number\\%hhmmss.ss$+$ddmmss.s (J2000.0)\\
%
%%%%%%%%%%%%%%%    CIV   %%%%%%%%%%%%%%%%
2\dotfill & DOUBLE   & Power-law normalization for \civ\ continuum fit at 3000\AA \\
3\dotfill & DOUBLE   & Uncertainty in Power-law normalization\\
4\dotfill & DOUBLE   & Power-law slope $\alpha_{\rm \civ}$ for the continuum fit \\
5\dotfill & DOUBLE   & Uncertainty in $\alpha_{\rm \civ}$\\
6\dotfill & DOUBLE   & Central wavelength of the dominant \civ component\\
7\dotfill & DOUBLE   & Central wavelength of the second \civ component \\
8\dotfill & DOUBLE   & Central wavelength of the third \civ component \\
9\dotfill & DOUBLE   & Central wavelength of the non-parametric \civ component \\
10\dotfill & DOUBLE   & Line dispersion ($\sigma_l$) of the dominant \civ\ component in $\kms$ \\
11\dotfill & DOUBLE   & Uncertainty in $\sigma_l$ of the dominant \civ\ component in $\kms$\\
12\dotfill & DOUBLE    & Full-width-half-maximum (FWHM) of the dominant \civ\ in $\kms$ \\
13\dotfill & DOUBLE   & Uncertainty in the dominant \civ\ FWHM in $\kms$ \\
14\dotfill & DOUBLE    & Integrated line area of the dominant \civ\ \\
15\dotfill & DOUBLE    & Uncertainty in the integrated line area of the dominant \civ\ \\
16\dotfill & DOUBLE   & Restframe equivalent width (EW) of the dominant \civ\ (\AA)\\
17\dotfill & DOUBLE   & Uncertainty in EW of the dominant \civ\ \\
18\dotfill & DOUBLE   & Line dispersion ($\sigma_l$) of the secondary \civ\ component in $\kms$ \\
19\dotfill & DOUBLE   & Uncertainty in $\sigma_l$ of the secondary \civ\ component in $\kms$\\
20\dotfill & DOUBLE    & Full-width-half-maximum (FWHM) of the secondary \civ\ in $\kms$ \\
21\dotfill & DOUBLE   & Uncertainty in the secondary \civ\ FWHM in $\kms$\\
22\dotfill & DOUBLE    & Integrated line area of the secondary \civ\ \\
23\dotfill & DOUBLE    & Uncertainty in the integrated line area of the secondary \civ\ \\
24\dotfill & DOUBLE   & Restframe equivalent width (EW) of the secondary \civ\ (\AA)\\
25\dotfill & DOUBLE   & Uncertainty in EW of the secondary \civ\ \\
26\dotfill & DOUBLE   & Line dispersion ($\sigma_l$) of the third \civ\ component in $\kms$ \\
27\dotfill & DOUBLE   & Uncertainty in $\sigma_l$ of the third \civ\ component in $\kms$\\
28\dotfill & DOUBLE    & Full-width-half-maximum (FWHM) of the third \civ\ in $\kms$ \\
29\dotfill & DOUBLE   & Uncertainty in the third \civ\ FWHM in $\kms$\\
30\dotfill & DOUBLE    & Integrated line area of the third \civ\ \\
31\dotfill & DOUBLE    & Uncertainty in the integrated line area of the third \civ\ \\
32\dotfill & DOUBLE   & Restframe equivalent width (EW) of the third \civ\ (\AA)\\
33\dotfill & DOUBLE   & Uncertainty in EW of the third \civ\ \\
%
%34\dotfill & DOUBLE   & Line dispersion ($\sigma_l$) of the non-parametric \civ\ component in $\kms$ \\
%35\dotfill & DOUBLE   & Uncertainty in $\sigma_l$ of the non-parametric \civ\ component in $\kms$\\
34\dotfill & DOUBLE    & Full-width-half-maximum (FWHM) of the non-parametric \civ\ in $\kms$ \\
35\dotfill & DOUBLE   & Uncertainty in the non-parametric \civ\ FWHM in $\kms$\\
36\dotfill & DOUBLE    & Integrated line area of the non-parametric \civ\ \\
37\dotfill & DOUBLE    & Uncertainty in the integrated line area of the non-parametric \civ\ \\
38\dotfill & DOUBLE   & Restframe equivalent width (EW) of the non-parametric \civ\ (\AA)\\
39\dotfill & DOUBLE   & Uncertainty in EW of the non-parametric \civ\ \\
40\dotfill & DOUBLE   & Reduced $\chi^2$ for the \civ\  continuum fit \\
41\dotfill & DOUBLE   & Reduced $\chi^2$ for the \civ\  emission line fit\\
42\dotfill & SHORT  & Status code for the \civ\  continuum fit (See IDL program `mpfitfun.pro') \\
43\dotfill & SHORT  & Status code for the \civ\  emission line fit\\
44\dotfill & SHORT  & Number of good pixels for the \civ\ emission line fitting region (1500-1600 \AA)\\
45\dotfill & DOUBLE  & Median S/N per pixel for the \civ\  emission line fitting region \\
%%%%%%%%%%%%%%%    Mgii  %%%%%%%%%%%%%%%%
46\dotfill & DOUBLE   & Power-law normalization for \mgii\ continuum fit at 3000\AA \\
47\dotfill & DOUBLE   & Uncertainty in Power-law normalization\\
48\dotfill & DOUBLE   & Power-law slope $\alpha_{\rm \mgii}$ for the continuum fit \\
49\dotfill & DOUBLE   & Uncertainty in $\alpha_{\rm \mgii}$\\
50\dotfill & DOUBLE   & Normalization of the \feii\ template \\
51\dotfill & DOUBLE   & Uncertainty in \feii\ normalization \\
52\dotfill & DOUBLE   & FWHM of the \feii\ component for \mgii\ continuum fit \\
53\dotfill & DOUBLE   & Uncertainty in $FWHM_{\rm Fe}$\\
54\dotfill & DOUBLE   & Central wavelength of the dominant \mgii\ component\\
55\dotfill & DOUBLE   & Central wavelength of the second \mgii\ component \\
56\dotfill & DOUBLE   & Central wavelength of the third \mgii\ component \\
57\dotfill & DOUBLE   & Central wavelength of the narrow \mgii\ component \\
58\dotfill & DOUBLE   & Central wavelength of the non-parametric \mgii\ component \\
59\dotfill & DOUBLE   & Line dispersion ($\sigma_l$) of the dominant \mgii\ component in $\kms$ \\
60\dotfill & DOUBLE   & Uncertainty in $\sigma_l$ of the dominant \mgii\ component in $\kms$\\
61\dotfill & DOUBLE    & Full-width-half-maximum (FWHM) of the dominant \mgii\ in $\kms$ \\
62\dotfill & DOUBLE   & Uncertainty in the dominant \mgii\ FWHM in $\kms$ \\
63\dotfill & DOUBLE    & Integrated line area of the dominant \mgii\ \\
64\dotfill & DOUBLE    & Uncertainty in the integrated line area of the dominant \mgii \\
65\dotfill & DOUBLE   & Restframe equivalent width (EW) of the dominant \mgii\ (\AA) \\
66\dotfill & DOUBLE   & Uncertainty in EW of the dominant \mgii\ \\
67\dotfill & DOUBLE   & Line dispersion ($\sigma_l$) of the secondary \mgii\ component in $\kms$ \\
68\dotfill & DOUBLE   & Uncertainty in $\sigma_l$ of the secondary \mgii\ component in $\kms$\\
69\dotfill & DOUBLE    & Full-width-half-maximum (FWHM) of the secondary \mgii\ in $\kms$ \\
70\dotfill & DOUBLE   & Uncertainty in the secondary \mgii\ FWHM in $\kms$\\
71\dotfill & DOUBLE    & Integrated line area of the secondary \mgii\ \\
72\dotfill & DOUBLE    & Uncertainty in the integrated line area of the secondary \mgii\ \\
73\dotfill & DOUBLE   & Restframe equivalent width (EW) of the secondary \mgii\ (\AA)\\
74\dotfill & DOUBLE   & Uncertainty in EW of the secondary \mgii\ \\
75\dotfill & DOUBLE   & Line dispersion ($\sigma_l$) of the third \mgii\  component in $\kms$ \\
76\dotfill & DOUBLE   & Uncertainty in $\sigma_l$ of the third \mgii\ component in $\kms$\\
77\dotfill & DOUBLE    & Full-width-half-maximum (FWHM) of the third \mgii\ in $\kms$ \\
78\dotfill & DOUBLE   & Uncertainty in the third \mgii\ FWHM in $\kms$\\
79\dotfill & DOUBLE    & Integrated line area of the third \mgii\ \\
80\dotfill & DOUBLE    & Uncertainty in the integrated line area of the third \mgii\ \\
81\dotfill & DOUBLE   & Restframe equivalent width (EW) of the third \mgii\ (\AA)\\
82\dotfill & DOUBLE   & Uncertainty in EW of the third \mgii\ \\
%
%83\dotfill & DOUBLE   & Line dispersion ($\sigma_l$) of the non-parametric \mgii\ component in $\kms$ \\
%84\dotfill & DOUBLE   & Uncertainty in $\sigma_l$ of the non-parametric \mgii\ component in $\kms$\\
83\dotfill & DOUBLE    & Full-width-half-maximum (FWHM) of the non-parametric \mgii\ in $\kms$ \\
84\dotfill & DOUBLE   & Uncertainty in the non-parametric \mgii\ FWHM in $\kms$\\
85\dotfill & DOUBLE    & Integrated line area of the non-parametric \mgii\\\
86\dotfill & DOUBLE    & Uncertainty in the integrated line area of the non-parametric \mgii\ \\
87\dotfill & DOUBLE   & Restframe equivalent width (EW) of the non-parametric \mgii\ (\AA)\\
88\dotfill & DOUBLE   & Uncertainty in EW of the non-parametric \mgii\ \\
89\dotfill & DOUBLE   & Line dispersion ($\sigma_l$) of the narrow \mgii\ component in $\kms$ \\
90\dotfill & DOUBLE   & Uncertainty in $\sigma_l$ of the narrow \mgii\ component in $\kms$\\
91\dotfill & DOUBLE    & Full-width-half-maximum (FWHM) of the narrow \mgii\ in $\kms$ \\
92\dotfill & DOUBLE   & Uncertainty in the narrow \mgii\ FWHM in $\kms$\\
93\dotfill & DOUBLE    & Integrated line area of the narrow \mgii\\\
94\dotfill & DOUBLE    & Uncertainty in the integrated line area of the narrow \mgii\ \\
95\dotfill & DOUBLE   & Restframe equivalent width (EW) of the narrow \mgii\ (\AA)\\
96\dotfill & DOUBLE   & Uncertainty in EW of the narrow \mgii\ \\
97\dotfill & DOUBLE   & Reduced $\chi^2$ for the \mgii\  continuum fit \\
98\dotfill & DOUBLE   & Reduced $\chi^2$ for the \mgii\ emission line fit\\
99\dotfill & SHORT  & Status code for the \mgii\  continuum fit (See IDL program `mpfitfun.pro') \\
100\dotfill & SHORT  & Status code for the \mgii\ emission line fit\\
101\dotfill & SHORT  & Number of good pixels for the \mgii\ emission line fitting region (2700-2900 \AA)\\
102\dotfill & DOUBLE  & Median S/N per pixel for the \mgii\  emission line fitting region \\
%%%%%%%%%%%%%%%    Hbeta    %%%%%%%%%%%%%%%%
103\dotfill & DOUBLE   & Power-law normalization for \hbeta\ continuum fit at 3000\AA \\
104\dotfill & DOUBLE   & Uncertainty in Power-law normalization\\
105\dotfill & DOUBLE   & Power-law slope $\alpha_{\rm H\beta}$ for the continuum fit \\
106\dotfill & DOUBLE   & Uncertainty in $\alpha_{\rm H\beta}$\\
107\dotfill & DOUBLE   & Normalization of the \feii$+$\oiii\ template \\
108\dotfill & DOUBLE   & Uncertainty in \feii$+$\oiii\ normalization \\
109\dotfill & DOUBLE   & FWHM of the \feii$+$\oiii\ component for \hbeta\ continuum fit \\
110\dotfill & DOUBLE   & Uncertainty in $FWHM_{\rm Fe}$\\
111\dotfill & DOUBLE   & Central wavelength of the dominant \hbeta\ component\\
112\dotfill & DOUBLE   & Central wavelength of the second \hbeta\ component \\
113\dotfill & DOUBLE   & Central wavelength of the third \hbeta\ component \\
114\dotfill & DOUBLE   & Central wavelength of the narrow \hbeta\ component\\
115\dotfill & DOUBLE   & Central wavelength of the \oiiia\ component\\
116\dotfill & DOUBLE   & Central wavelength of the \oiiib\ component\\
117\dotfill & DOUBLE   & Line dispersion ($\sigma_l$) of the dominant \hbeta\  component in $\kms$ \\
118\dotfill & DOUBLE   & Uncertainty in $\sigma_l$ of the dominant \hbeta\ component in $\kms$\\
119\dotfill & DOUBLE    & Full-width-half-maximum (FWHM) of the dominant \hbeta\ in $\kms$ \\
120\dotfill & DOUBLE   & Uncertainty in the dominant \hbeta\ FWHM in $\kms$ \\
121\dotfill & DOUBLE    & Integrated line area of the dominant \hbeta\ \\
122\dotfill & DOUBLE    & Uncertainty in the integrated line area of the dominant \hbeta \\
123\dotfill & DOUBLE   & Restframe equivalent width (EW) of the dominant \hbeta\ (\AA)\\
124\dotfill & DOUBLE   & Uncertainty in EW of the dominant \hbeta \\
125\dotfill & DOUBLE   & Line dispersion ($\sigma_l$) of the secondary \hbeta\ component in $\kms$ \\
126\dotfill & DOUBLE   & Uncertainty in $\sigma_l$ of the secondary \hbeta\ component in $\kms$\\
127\dotfill & DOUBLE    & Full-width-half-maximum (FWHM) of the secondary \hbeta\ in $\kms$ \\
128\dotfill & DOUBLE   & Uncertainty in the secondary \hbeta\ FWHM in $\kms$\\
129\dotfill & DOUBLE    & Integrated line area of the secondary \hbeta\ \\
130\dotfill & DOUBLE    & Uncertainty in the integrated line area of the secondary \hbeta\  \\
131\dotfill & DOUBLE   & Restframe equivalent width (EW) of the secondary \hbeta\  (\AA)\\
132\dotfill & DOUBLE   & Uncertainty in EW of the secondary \hbeta\  \\
133\dotfill & DOUBLE   & Line dispersion ($\sigma_l$) of the third \hbeta\  component in $\kms$ \\
134\dotfill & DOUBLE   & Uncertainty in $\sigma_l$ of the third \hbeta\  component in $\kms$\\
135\dotfill & DOUBLE    & Full-width-half-maximum (FWHM) of the third \hbeta\ in $\kms$ \\
136\dotfill & DOUBLE   & Uncertainty in the third \hbeta\ FWHM in $\kms$\\
137\dotfill & DOUBLE    & Integrated line area of the third \hbeta\ \\
138\dotfill & DOUBLE    & Uncertainty in the integrated line area of the third \hbeta\  \\
139\dotfill & DOUBLE   & Restframe equivalent width (EW) of the third \hbeta\ (\AA)\\
140\dotfill & DOUBLE   & Uncertainty in EW of the third \hbeta\  \\
141\dotfill & DOUBLE   & Line dispersion ($\sigma_l$) of the narrow \hbeta\ component in $\kms$ \\
142\dotfill & DOUBLE   & Uncertainty in $\sigma_l$ of the narrow \hbeta\ component in $\kms$\\
143\dotfill & DOUBLE    & Full-width-half-maximum (FWHM) of the narrow \hbeta\ in $\kms$ \\
144\dotfill & DOUBLE   & Uncertainty in the narrow \hbeta\ FWHM in $\kms$\\
145\dotfill & DOUBLE    & Integrated line area of the narrow \hbeta\ \\
146\dotfill & DOUBLE    & Uncertainty in the integrated line area of the narrow \hbeta\ \\
147\dotfill & DOUBLE   & Restframe equivalent width (EW) of the narrow \hbeta\ (\AA)\\
148\dotfill & DOUBLE   & Uncertainty in EW of the narrow \hbeta\ \\
149\dotfill & DOUBLE   & Line dispersion ($\sigma_l$) of the \oiiia\ component in $\kms$ \\
150\dotfill & DOUBLE   & Uncertainty in $\sigma_l$ of the \oiiia\ component in $\kms$\\
151\dotfill & DOUBLE    & Full-width-half-maximum (FWHM) of the \oiiia\ in $\kms$ \\
152\dotfill & DOUBLE   & Uncertainty in the \oiiia\  FWHM in $\kms$\\
153\dotfill & DOUBLE    & Integrated line area of the \oiiia\ \\
154\dotfill & DOUBLE    & Uncertainty in the integrated line area of the  \oiiia\ \\
155\dotfill & DOUBLE   & Restframe equivalent width (EW) of the \oiiia (\AA) \\
156\dotfill & DOUBLE   & Uncertainty in EW of the  \oiiia\  \\
157\dotfill & DOUBLE   & Line dispersion ($\sigma_l$) of the \oiiib\ component in $\kms$ \\
158\dotfill & DOUBLE   & Uncertainty in $\sigma_l$ of the \oiiib\ component in $\kms$\\
159\dotfill & DOUBLE    & Full-width-half-maximum (FWHM) of the  \oiiib\ in $\kms$ \\
160\dotfill & DOUBLE   & Uncertainty in the  \oiiib\  FWHM in $\kms$\\
161\dotfill & DOUBLE    & Integrated line area of the  \oiiib\ \\
162\dotfill & DOUBLE    & Uncertainty in the integrated line area of the \oiiib\ \\
163\dotfill & DOUBLE   & Restframe equivalent width (EW) of the \oiiib\ (\AA)\\
164\dotfill & DOUBLE   & Uncertainty in EW of the  \oiiib\  \\
165\dotfill & DOUBLE   & Ratio of  (\oiiib / \oiiia)  \\
166\dotfill & DOUBLE   & Reduced $\chi^2$ for the  \hbeta\ continuum fit \\
167\dotfill & DOUBLE   & Reduced $\chi^2$ for the  \hbeta\ emission line fit\\
168\dotfill & SHORT  & Status code for the \hbeta\ continuum fit (See IDL program `mpfitfun.pro') \\
169\dotfill & SHORT  & Status code for the  \hbeta\ emission line fit\\
170\dotfill & SHORT  & Number of good pixels for the  \hbeta\ emission line fitting region (4700-5100 \AA)\\
171\dotfill & DOUBLE  & Median S/N per pixel for the  \hbeta\ emission line fitting region \\
%\enddata
%\end{deluxetable*}
\end{longtable}
\tablecomments{
Notes: (1) Unless otherwise stated, null value is given if not measured, and $-1$ for its associated error. }
%(1) Status code from `mpfitfun.pro': `-99' is assigned if force input.}
\normalsize

\end{landscape}
\end{CJK}

\begin{thebibliography}

\expandafter\ifx\csname natexlab\endcsname\relax\def\natexlab#1{#1}\fi

\bibitem[Abazajian et al.(2009)]{abazajian09} Abazajian, K.~N., 
Adelman-McCarthy, J.~K., Ag{\"u}eros, M.~A., et al.\ 2009, \apjs, 182, 543 

\bibitem[Anderson et al.(2003)]{anderson03} Anderson, S.~F., 
Voges, W., Margon, B., et al.\ 2003, \aj, 126, 2209 
\bibitem[Antonucci(1993)]{antonucci93} Antonucci, R.\ 1993, \araa, 31, 473 


\bibitem[Assef et al.(2011)]{assef11} Assef, R.~J., Denney, 
K.~D., Kochanek, C.~S., et al.\ 2011, \apj, 742, 93 


\bibitem[Bentz et al.(2006)]{bentz06} Bentz, M.~C., Peterson, 
B.~M., Pogge, R.~W., Vestergaard, M., \& Onken, C.~A.\ 2006, \apj, 644, 133 
\bibitem[Bentz et al.(2009)]{bentz09} Bentz, M.~C., Peterson, 
B.~M., Netzer, H., Pogge, R.~W., \& Vestergaard, M.\ 2009, \apj, 697, 160 
\bibitem[Bentz et al.(2013)]{bentz13} Bentz, M.~C., Denney, 
K.~D., Grier, C.~J., et al.\ 2013, \apj, 767, 149 


\bibitem[Bergeron(1986)]{bergeron86} Bergeron, J.\ 1986, \aap, 155, L8 

\bibitem[Bournaud et al.(2011)]{bournaud11} Bournaud, F., Chapon, 
D., Teyssier, R., et al.\ 2011, \apj, 730, 4 


\bibitem[Boroson 
\& Green(1992)]{bg92} Boroson, T.~A., \& Green, R.~F.\ 1992, \apjs, 80, 109 

\bibitem[{Calzetti et al.(2000)}]{calzetti00} Calzetti, D., Armus,  L., Bohlin, R.~C., Kinney, A.~L., Koornneef, J.,\& Storchi-Bergmann, T.\ 2000, \apj, 533, 682
\bibitem[Caputi et al.(2007)]{caputi07} Caputi, K.~I., Lagache, 
G., Yan, L., et al.\ 2007, \apj, 660, 97 


\bibitem[{Chary\& Elbaz(2001)}]{ce01} Chary, R., \& Elbaz, D.\ 2001, \apj, 556, 562
\bibitem[Cool et al.(2008)]{cool08} Cool, R.~J., et al.\ 2008, \apj, 682, 919 

\bibitem[Collin et 
al.(2006)]{collin06} Collin, S., Kawaguchi, T., Peterson, B.~M., \& Vestergaard, M.\ 2006, \aap, 456, 75 

\bibitem[Dai et al.(2012)]{dai12} Dai, Y.~S., Bergeron, J., 
Elvis, M., et al.\ 2012, \apj, 753, 33
 
 \bibitem[Dekel et al.(2009)]{dekel09} Dekel, A., Birnboim, Y., 
Engel, G., et al.\ 2009, \nat, 457, 451 

\bibitem[Denney(2012)]{denney12} Denney, K.~D.\ 2012, \apj, 759,  44 


\bibitem[Denney et al.(2013)]{denney13} Denney, K.~D., Pogge, 
R.~W., Assef, R.~J., et al.\ 2013, \apj, 775, 60 


\bibitem[Di Matteo et al.(2012)]{dimatteo12} Di Matteo, T., 
Khandai, N., DeGraf, C., et al.\ 2012, \apjl, 745, L29 


\bibitem[Dole et al.(2004)]{dole04} Dole, H., Rieke, G.~H., 
Lagache, G., et al.\ 2004, \apjs, 154, 93 

\bibitem[{Donley et al.(2008)}]{don08} Donley, J.~L., Rieke, G.~H., P{\'e}rez-Gonz{\'a}lez, P.~G., \& Barro, G.\ 2008, \apj, 687, 111 
\bibitem[Donley et al.(2012)]{donley12} Donley, J.~L., 
Koekemoer, A.~M., Brusa, M., et al.\ 2012, \apj, 748, 142 


\bibitem[Eales et al.(2010)]{eales10} Eales, S.~A., Raymond, G., Roseboom, I.~G., et al.\ 2010, \aap, 518, L23 
\bibitem[Evans et al.(2006)]{evans06} Evans, A.~S., Solomon, P.~M., Tacconi, L.~J., Vavilkin, T., \& Downes, D.\ 2006, \aj, 132, 2398 
\bibitem[Elbaz et al.(2010)]{elbaz10} Elbaz, D., Hwang, H.~S., Magnelli, B., et al.\ 2010, \aap, 518, L29 
\bibitem[{Elvis et al.(1994)}]{elvis94} Elvis, M., et al. \ 1994, \apjs, 95, 1 
\bibitem[Elvis(2000)]{elvis00} Elvis, M.\ 2000, \apj, 545, 63 
\bibitem[Elvis et al.(2002)]{elvis02} Elvis, M., Marengo, M., \& Karovska, M.\ 2002, \apjl, 567, L107 
\bibitem[Fabricant et al.(2005)]{fab05} Fabricant, D., et al.\ 2005, \pasp, 117, 1411 
\bibitem[Fabricant et al.(2008)]{fab08} Fabricant, D.~G., 
Kurtz, M.~J., Geller, M.~J., et al.\ 2008, \pasp, 120, 1222 
\bibitem[Fazio et al. (2004)]{fazio04} Fazio, G.~G., Hora, 
J.~L., Allen, L.~E., et al.\ 2004, \apjs, 154, 10 

\bibitem[Ferrarese 
\& Merritt(2000)]{ferrarese00} Ferrarese, L., \& Merritt, D.\ 2000, \apjl, 539, L9 

\bibitem[Fine et al.(2006)]{fine06} Fine, S., Croom, S.~M., 
Miller, L., et al.\ 2006, \mnras, 373, 613 

\bibitem[Fine et al.(2010)]{fine10} Fine, S., Croom, S.~M., 
Bland-Hawthorn, J., et al.\ 2010, \mnras, 409, 591 

\bibitem[Falcke et 
al.(2004)]{fkm04} Falcke, H., K{\"o}rding, E., \& Markoff, S.\ 2004, \aap, 414, 895 

\bibitem[Fukugita et al.(1996)]{fukugita96} Fukugita, M., 
  Ichikawa, T., Gunn, J.~E., Doi, M., Shimasaku, K., \& Schneider, D.~P.\ 
  1996, \aj, 111, 1748 


\bibitem[Gandhi et 
al.(2009)]{2009A&A...502..457G} Gandhi, P., Horst, H., Smette, A., et al.\ 2009, \aap, 502, 457 

\bibitem[Gaskell(1982)]{gaskell82} Gaskell, C.~M.\ 1982, \apj, 
263, 79 

\bibitem[Gebhardt et al.(2000)]{gebhardt00} Gebhardt, K., Bender, 
R., Bower, G., et al.\ 2000, \apjl, 539, L13 

\bibitem[Griffin et al.(2010)]{grif10} Griffin, M.~J., et al.\ 2010, \aap, 518, L3 

\bibitem[Genzel et al.(1998)]{genzel98} Genzel, R., et al.\ 
1998, \apj, 498, 579  
           
\bibitem[Glikman et al.(2004)]{glikman04} Glikman, E., Gregg, M.~D., Lacy, M., et al.\ 2004, \apj, 607, 60 
\bibitem[Glikman et al.(2007)]{glikman07} Glikman, E., Helfand, D.~J., White, R.~L., et al.\ 2007, \apj, 667, 673 
\bibitem[Glikman et al.(2012)]{glikman12} Glikman, E., Urrutia, 
T., Lacy, M., et al.\ 2012, \apj, 757, 51 

\bibitem[Goulding et al.(2014)]{goulding14} Goulding, A.~D., 
Forman, W.~R., Hickox, R.~C., et al.\ 2014, \apj, 783, 40 




\bibitem[{Haas et al.(2003)}]{haas03} Haas, M., et al. \ 2003, \aap, 402, 87 

%\bibitem[Hamann et al.(1997)]{1997ApJ...488..155H} Hamann, F., Beaver, 
%E.~A., Cohen, R.~D., et al.\ 1997, \apj, 488, 155 

\bibitem[Hamann et al.(1995)]{hamann95} Hamann, F., Barlow, 
T.~A., Beaver, E.~A., et al.\ 1995, \apj, 443, 606 

\bibitem[Hamann et al.(2001)]{hamann01} Hamann, F.~W., Barlow, 
T.~A., Chaffee, F.~C., Foltz, C.~B., \& Weymann, R.~J.\ 2001, \apj, 550, 142   %AAL

\bibitem[Hasinger et 
al.(2005)]{hms05} Hasinger, G., Miyaji, T., \& Schmidt, M.\ 2005, \aap, 441, 417 


\bibitem[{Hatziminaoglou et al.(2008)}]{hatzi08} Hatziminaoglou, E., et al.\ 2008, \mnras, 386, 1252
\bibitem[Hatziminaoglou et al.(2010)]{hatzi10} Hatziminaoglou, E., et al.\ 2010, \aap, 518, L33 
\bibitem[Hao et al.(2010)]{hao10} Hao, H., Elvis, M., Civano, F., et al.\ 2010, \apjl, 724, L59 

\bibitem[Heckman et al.(1991)]{heckman91} Heckman, T.~M., Miley, 
G.~K., Lehnert, M.~D., \& van Breugel, W.\ 1991, \apj, 370, 78 


%\bibitem[Heckman et al.(1990)]{heckman90} Heckman, T.~M., Armus, 
%L., \& Miley, G.~K.\ 1990, \apjs, 74, 833 


\bibitem[Hopkins et al.(2006)]{hopkins06} Hopkins, P.~F., 
Hernquist, L., Cox, T.~J., et al.\ 2006, \apjs, 163, 1 
 
\bibitem[Huang et al., in preparation]{hjs10} Huang, J-S, et al., in preparation
\bibitem[{Iono et al.(2007)}]{iono07}Iono, D. et al. \ 2007, \apj, 659, 283

\bibitem[Jahnke 
\& Macci{\`o}(2011)]{jahnke11} Jahnke, K., \& Macci{\`o}, A.~V.\ 2011, \apj, 734, 92 

\bibitem[Juneau et al.(2013)]{juneau13} Juneau, S., Dickinson, 
M., Bournaud, F., et al.\ 2013, \apj, 764, 176 

\bibitem[Lacy et al.(2013)]{2013ApJS..208...24L} Lacy, M., Ridgway, S.~E., 
Gates, E.~L., et al.\ 2013, \apjs, 208, 24 

\bibitem[Kaspi et al.(2000)]{kaspi00} Kaspi, S., Smith, P.~S., 
Netzer, H., et al.\ 2000, \apj, 533, 631 

\bibitem[Kauffmann et al.(2003)]{kauffmann03} Kauffmann, G., 
Heckman, T.~M., Tremonti, C., et al.\ 2003, \mnras, 346, 1055 

\bibitem[{Kaviani, Haehnelt, \& Kauffmann(2003)}]{kaviani03}Kaviani, A., Haehnelt, M. G., \& Kauffmann, G. \ 2003, \mnras, 340, 739

\bibitem[Kelly \& Shen(2013)]{kellynshen13} Kelly, B.~C., \& Shen, Y.\ 2013, \apj, 764, 45 
\bibitem[Klaas et al.(2001)]{klaas01} Klaas, U., et al.\ 2001, \aap, 379, 823 
\bibitem[Knigge et al.(2008)]{knigge08} Knigge, C., Scaringi, S., Goad, M.~R., \& Cottis, C.~E.\ 2008, \mnras, 386, 1426 
\bibitem[Kollmeier et al.(2006)]{kollmeier06} Kollmeier, J.~A., Onken, C.~A., Kochanek, C.~S., et al.\ 2006, \apj, 648, 128 
\bibitem[Kormendy \& Richstone(1995)]{kormendy95} Kormendy, J., \& Richstone, D.\ 1995, \araa, 33, 581 
\bibitem[Kormendy \& Ho(2013)]{kormendy13} Kormendy, J., \& Ho, L.~C.\ 2013, \araa, 51, 511 
\bibitem[{Kov{\'a}cs et al.(2010)}]{kovac10} Kov{\'a}cs, A., Omont, A., Beelen, A., et al.\ 2010, \apj, 717, 29 
\bibitem[{Krips et al.(2007)}]{krips07}Krips, M. et al. \ 2007, \apjl, 671, L5
\bibitem[Kuraszkiewicz et al.(2003)]{kuras03} Kuraszkiewicz, 
J.~K., Wilkes, B.~J., Hooper, E.~J., et al.\ 2003, \apj, 590, 128 

\bibitem[Labita et al.(2009)]{labita09} Labita, M., Decarli, R., Treves, A., \& Falomo, R.\ 2009, \mnras, 396, 1537 
\bibitem[Lacy et al.(2002)]{lacy02} Lacy, M., Gregg, M., Becker, R.~H., et al.\ 2002, \aj, 123, 2925 
\bibitem[{Lacy et al.(2004)}]{lacy04} Lacy, M., et al., 2004, \apjs, 154, 166 
\bibitem[Lacy et al.(2013)]{lacy13} Lacy, M., Ridgway, S.~E., Gates, E.~L., et al.\ 2013, \apjs, 208, 24 
\bibitem[{Lagache et al.(2005)}]{lagache05}Lagache, G., et al. \ 2005, ARAA, 43, 727
\bibitem[Laor et al.(1994)]{laor94} Laor, A., Bahcall, J.~N., Jannuzi, B.~T., et al.\ 1994, \apj, 420, 110 
\bibitem[Lawrence et al.(2007)]{lawrence07} Lawrence, A., et al.\ 2007, \mnras, 379, 159
\bibitem[{LeFloc{\'h} et al.(2005)}]{lefloch05}LeFloc{\'h}, E. et al. \ 2005, \apj, 632, 169
\bibitem[{Leipski et al.(2010)}]{leip10} Leipski, C., Meisenheimer, K., Klaas, U., et al.\ 2010, \aap, 518, L34 
\bibitem[Li et al.(2008)]{li08} Li, Y., et al.\ 2008, \apj, 678, 41  
\bibitem[Lilly et al.(2013)]{lilly13} Lilly, S.~J., Carollo, C.~M., Pipino, A., Renzini, A., \& Peng, Y.\ 2013, \apj, 772, 119 
\bibitem[Lonsdale et al.(2003)]{lonsdale03} Lonsdale, C.~J., et al.\ 2003, \pasp, 115, 897 
\bibitem[Lusso et al.(2012)]{2012MNRAS.425..623L} Lusso, E., Comastri, A., Simmons, B.~D., et al.\ 2012, \mnras, 425, 623 
\bibitem[Lusso et al.(2010)]{2010A&A...512A..34L} Lusso, E., Comastri, A., Vignali, C., et al.\ 2010, \aap, 512, A34 
\bibitem[Lutz et al.(2008)]{lutz08} Lutz, D., et al.\ 2008, \apj, 684, 853


\bibitem[Marconi \& Hunt(2003)]{marconihunt03} Marconi, A., \& Hunt, L.~K.\ 2003, \apjl, 589, L21 
\bibitem[Marziani et al.(1996)]{marziani96} Marziani, P., Sulentic, J.~W., Dultzin-Hacyan, D., Calvani, M., 
\& Moles, M.\ 1996, \apjs, 104, 37 
\bibitem[{{McLure} \& {Dunlop}(2004)}]{md04}{McLure}, R.~J., \& {Dunlop}, J.~S. 2004, \mnras, 352, 1390 (MD04)
\bibitem[McLure \& Dunlop(2002)]{md02} McLure, R.~J., \& Dunlop, J.~S.\ 2002, \mnras, 331, 795 
\bibitem[{Mihos et al.(1994)}]{mihos94}Mihos, J.C. \& Hernquist, L. \ 1994, \apjl, 431, L9
\bibitem[Mink et al.(2007)]{mink07} Mink, D.~J., Wyatt, W.~F., Caldwell, N., et al.\ 2007, Astronomical Data Analysis Software and Systems XVI, 376, 249 

\bibitem[Mullaney et al.(2011)]{2011MNRAS.414.1082M} Mullaney, J.~R., 
Alexander, D.~M., Goulding, A.~D., 
\& Hickox, R.~C.\ 2011, \mnras, 414, 1082 

\bibitem[Newman et al.(2013)]{newman13} Newman, J.~A., Cooper, 
M.~C., Davis, M., et al.\ 2013, \apjs, 208, 5 

\bibitem[{Netzer et al.(2007)}]{netzer07} Netzer, H., et al.\ 2007, \apj, 666, 806 

\bibitem[{``Numerical Recipes'', Second Edition, (1992)}]{numr}``Numerical recipes 
in FORTRAN (Second Edition)", Teukolsky, S.~A., Vetterling, W.~T.,~\& Flannery, B.~P., William H. Press, 613, 1992



\bibitem[Ogle et al.(2006)]{ogle06} Ogle, P., Whysong, D., 
\& Antonucci, R.\ 2006, \apj, 647, 161 
\bibitem[{Oliver et al.(2010)}]{oliver10} Oliver, S.~J., Wang, L., Smith, A.~J., et al.\ 2010, \aap, 518, L21

\bibitem[Oke \& Gunn(1983)]{oke83} Oke, J.~B.~\& Gunn, J.~E.\ 
  1983, \apj, 266, 713 

\bibitem[Papovich et al.(2006)]{papovich06} Papovich, C., Cool, 
R., Eisenstein, D., et al.\ 2006, \aj, 132, 231 
\bibitem[Park et al.(2013)]{park13} Park, D., Woo, J.-H., 
Denney, K.~D., \& Shin, J.\ 2013, \apj, 770, 87 
\bibitem[{Pei et al.(1999)}]{pei99}Pei, Y. C. et al. \ 1999, \apj, 522, 604
\bibitem[Peng(2007)]{peng07} Peng, C.~Y.\ 2007, \apj, 671, 
1098 
\bibitem[Peterson et al.(2004)]{peterson04} Peterson, B.~M., 
Ferrarese, L., Gilbert, K.~M., et al.\ 2004, \apj, 613, 682 
\bibitem[Pilbratt et al.(2010)]{pilbratt10} Pilbratt, G.~L., et al.\ 2010, \aap, 518, L1 
\bibitem[Polletta et al.(2006)]{polletta06} Polletta, M.~d.~C., 
Wilkes, B.~J., Siana, B., et al.\ 2006, \apj, 642, 673 
\bibitem[{Puget et al.(1996)}]{puget96}Puget, J.-L., et al., \ 1996, \aap, 308,5

\bibitem[{Richards et al.(2002a), R02}]{rich02a} Richards, G.~T., et al., 2002a, \aj, 123, 2945
\bibitem[Richards et al.(2002b)]{rich02b} Richards, G.~T., Vanden Berk, D.~E., Reichard, T.~A., Hall, P.~B., Schneider, D.~P., 
SubbaRao, M., Thakar, A.~R., \& York, D.~G.\ 2002b, \aj, 124, 1 
\bibitem[Richards et al.(2003)]{rich03} Richards, G.~T., Hall, 
P.~B., Vanden Berk, D.~E., et al.\ 2003, \aj, 126, 1131 
\bibitem[Richards et al.(2006a)]{rich06a} Richards, G.~T., et al.\ 2006a, \aj, 131, 2766 
\bibitem[Richards et al.(2006b), R06]{rich06b} Richards, G.~T., et al.\ 2006b, \apjs, 166, 470 
\bibitem[{Richards et al.(2009)}]{rich09} Richards, G.~T., et al., 2009, \aj, 137, 3884
\bibitem[Richards et al.(2011)]{richards11} Richards, G.~T., 
Kruczek, N.~E., Gallagher, S.~C., et al.\ 2011, \aj, 141, 167 

\bibitem[Rieke et al.(2004)]{rieke04} Rieke, G.~H., Young, 
E.~T., Engelbracht, C.~W., et al.\ 2004, \apjs, 154, 25 

\bibitem[Rieke et al.(2009)]{rieke09} Rieke, G.~H., Alonso-Herrero, A., Weiner, B.~J., P{\'e}rez-Gonz{\'a}lez, P.~G., Blaylock, M., Donley, J.~L., \& Marcillac, D.\ 2009, \apj, 692, 556 
\bibitem[Roseboom et al.(2010)]{roseboom10} Roseboom, I.~G., et al.\ 2010, \mnras, 409, 48
\bibitem[Sanders et al.(1988)]{sanders88} Sanders, D.~B., Soifer, B.~T., Elias, J.~H., Madore, B.~F., Matthews, K., Neugebauer, G., 
           \& Scoville, N.~Z.\ 1988, \apj, 325, 74 

\bibitem[Sajina et al. (2005)]{sajina05} Sajina, A., Lacy, M., 
\& Scott, D.\ 2005, \apj, 621, 256 

\bibitem[{Sakamoto et al.(2006)}]{saka06}Sakamoto, K., Ho, P. T. P., \& Peck, A. B. \ 2006, \apj, 644, 862
\bibitem[{Salvato et al.(2009)}]{salva09} Salvato, M., et al., 2009, \apj, 690, 1250 
\bibitem[Sanders et al.(1989)]{sanders89} Sanders, D.~B., 
	Phinney, E.~S., Neugebauer, G., Soifer, B.~T., \& Matthews, K.\ 1989, \apj, 347, 29

\bibitem[Schneider et al.(2007)]{schneider07} Schneider, D.~P., Hall, P.~B., Richards, G.~T., et al.\ 2007, \aj, 134, 102 
  
\bibitem[Serjeant et al.(2010)]{serjeant10} Serjeant, S., et al.\ 2010, \aap, 518, L7 

\bibitem[Schlegel, Finkbeiner, \& Davis(1998)]{schlegel98} 
  Schlegel, D.~J., Finkbeiner, D.~P., \& Davis, M.\ 1998, \apj, 500, 525 


\bibitem[Schmidt 
\& Green(1983)]{sg83} Schmidt, M., \& Green, R.~F.\ 1983, \apj, 269, 352 

\bibitem[Schweitzer et al.(2006)]{schweitzer06} Schweitzer, M., et al.\ 2006, \apj, 649, 79

\bibitem[Schneider et al.(2010)]{schneider10} Schneider, D.~P., 
Richards, G.~T., Hall, P.~B., et al.\ 2010, \aj, 139, 2360 

\bibitem[Shen et al.(2008)]{shen08} Shen, Y., Greene, J.~E., 
Strauss, M.~A., Richards, G.~T., \& Schneider, D.~P.\ 2008, \apj, 680, 169 


\bibitem[{{Shen} \& {Kelly}(2010)}]{shen10}
{Shen}, Y., \& {Kelly}, B.~C. 2010, \apj, 713, 41

\bibitem[Shen et al.(2011), hereafter S11]{shen11} Shen, Y., Richards, G.~T., 
Strauss, M.~A., et al.\ 2011, \apjs, 194, 45 (S11) 

\bibitem[Shen 
\& M{\'e}nard(2012)]{shen12} Shen, Y., \& M{\'e}nard, B.\ 2012, \apj, 748, 131 


\bibitem[Shen(2013)]{shen13} Shen, Y.\ 2013, Bulletin of the 
Astronomical Society of India, 41, 61 


\bibitem[Shi et al.(2009)]{shi09} Shi, Y., Rieke, G.~H., Ogle, P., Jiang, L., \& Diamond-Stanic, A.~M.\ 2009, \apj, 703, 1107 

\bibitem[Shupe et al.(2008)]{shupe08} Shupe, D.~L., Rowan-Robinson, M., Lonsdale, C.~J., et al.\ 2008, \aj, 135, 1050 


\bibitem[Silverman et al.(2008)]{silverman08} Silverman, J.~D., 
Green, P.~J., Barkhouse, W.~A., et al.\ 2008, \apj, 679, 118 



\bibitem[Smith et al.(2002)]{smith02} Smith, J.~A.~et al.\ 
  2002, \aj, 123, 2121 
\bibitem[Steinhardt 
\& Elvis(2010)]{sande10} Steinhardt, C.~L., \& Elvis, M.\ 2010, \mnras, 402, 2637 
\bibitem[{Stern et al.(2005)}]{stern05} Stern, D., et al., 2005, \apj, 631, 163 
\bibitem[Stern et al. (2012)]{stern12} Stern, D., Assef, R.~J., 
Benford, D.~J., et al.\ 2012, \apj, 753, 30 
\bibitem[Stevens et al., in preparation]{steve10} Stevens, J. A., et al., in preparation
\bibitem[Stoughton et al.(2002)]{stoughton02} Stoughton, C., Lupton, R.~H., Bernardi, M., et al.\ 2002, \aj, 123, 485 
\bibitem[Sturm et al.(2006)]{sturm06} Sturm, E., Rupke, D., Contursi, A., et al.\ 2006, \apjl, 653, L13 
\bibitem[Sulentic et al.(2007)]{sulentic07} Sulentic, J.~W., Bachev, R., Marziani, P., Negrete, C.~A., 
\& Dultzin, D.\ 2007, \apj, 666, 757 


\bibitem[Trump et al.(2006)]{trump06} Trump, J.~R., Hall, 
P.~B., Reichard, T.~A., et al.\ 2006, \apjs, 165, 1 
\bibitem[Tremaine et al.(2002)]{tremaine02} Tremaine, S., 
Gebhardt, K., Bender, R., et al.\ 2002, \apj, 574, 740 

\bibitem[Urrutia et al.(2009)]{urrutia09} Urrutia, T., Becker, R.~H., White, R.~L., et al.\ 2009, \apj, 698, 1095 
%\bibitem[Urrutia et al.(2012)]{urrutia12} Urrutia, T., Lacy, M., Spoon, H., et al.\ 2012, \apj, 757, 125 
\bibitem[Urry \& Padovani(1995)]{urry95} Urry, C.~M., \& Padovani, P.\ 1995, \pasp, 107, 803 


\bibitem[Vanden Berk et al.(2001)]{vandenberk01} Vanden Berk, D.~E., Richards, G.~T., Bauer, A., et al.\ 2001, \aj, 122, 549 
\bibitem[Vanden Berk et al.(2005)]{vandenberk05} Vanden Berk, D.~E., 
Schneider, D.~P., Richards, G.~T., et al.\ 2005, \aj, 129, 2047 
\bibitem[Veilleux et al.(2009)]{veilleux09} Veilleux, S., et al.\ 2009, \apjs, 182, 628 
\bibitem[{{Vestergaard}(2002)}]{vest02}{Vestergaard}, M. 2002, \apj, 571, 733
\bibitem[{{Vestergaard} \& {Osmer}(2009)}]{vo09}{Vestergaard}, M., \& {Osmer}, P.~S. 2009, \apj, 699, 800 (VO09)
\bibitem[Vestergaard \& Wilkes(2001)]{vandw01} Vestergaard, M., \& Wilkes, B.~J.\ 2001, \apjs, 134, 1 
\bibitem[Vestergaard(2003)]{vest03} Vestergaard, M.\ 2003, \apj, 599, 116 
\bibitem[{Vestergaard \& Peterson(2006)}]{vp06} Vestergaard, M., \& Peterson, B.~M.\ 2006, \apj, 641, 689 (VP06)

\bibitem[Wang et al.(2009)]{wang09} Wang, J.-G., Dong, X.-B., 
Wang, T.-G., et al.\ 2009, \apj, 707, 1334 
\bibitem[Weymann et al.(1979)]{weymann79} Weymann, R.~J., 
Williams, R.~E., Peterson, B.~M., \& Turnshek, D.~A.\ 1979, \apj, 234, 33 
\bibitem[Weymann et 
al.(1981)]{weymann81} Weymann, R.~J., Carswell, R.~F., \& Smith, M.~G.\ 1981, \araa, 19, 41 
\bibitem[Weymann et al.(1991)]{weymann91} Weymann, R.~J., Morris, 
S.~L., Foltz, C.~B., \& Hewett, P.~C.\ 1991, \apj, 373, 23 
\bibitem[Wilkes(1984)]{wilkes84} Wilkes, B.~J.\ 1984, \mnras, 
207, 73 
\bibitem[Willmer et al.(2006)]{willmer06} Willmer, C.~N.~A., et al.\ 2006, \apj, 647, 853
\bibitem[White et al.(1997)]{white97} White, R.~L., Becker, 
R.~H., Helfand, D.~J., \& Gregg, M.~D.\ 1997, \apj, 475, 479 
\bibitem[Wills et al.(1993)]{wills93} Wills, B.~J., Netzer, H., 
Brotherton, M.~S., et al.\ 1993, \apj, 410, 534 
\bibitem[Wild et al.(2008), hereafter W08]{wild08} Wild, V., Kauffmann, G., 
White, S., et al.\ 2008, \mnras, 388, 227 
\bibitem[Wright et al. (2010)]{wright10} Wright, E.~L., 
Eisenhardt, P.~R.~M., Mainzer, A.~K., et al.\ 2010, \aj, 140, 1868 
\bibitem[Woo 
\& Urry(2002)]{2002ApJ...579..530W} Woo, J.-H., \& Urry, C.~M.\ 2002, \apj, 579, 530 

\bibitem[York et al.(2000)]{york00} York, D.~G., Adelman, J., 
Anderson, J.~E., Jr., et al.\ 2000, \aj, 120, 1579    %SDSS technical summary
\bibitem[York et al.(2006)]{york06} York, D.~G., Khare, P., 
Vanden Berk, D., et al.\ 2006, \mnras, 367, 945   %SDSS AAL


\end{thebibliography}
\end{document}